\documentclass[preprint,onecolumn,12pt]{aastex}
\usepackage{natbib}
\usepackage{graphicx}
\usepackage{ramya}
\usepackage{rotating}
\usepackage{lscape}
\usepackage{subfig}
\usepackage{epstopdf}

\def\specchar#1{\sc{#1}\ }
\def\hi{\mbox{H$\,$\specchar{i}}}
\def\hii{\mbox{H$\,$\specchar{ii}}} 
\def\heii{\mbox{He$\,$\specchar{ii}}}
\def\oiii{\mbox{[O$\,$\specchar{iii}]}}
\def\nev{\mbox{[Ne$\,$\specchar{v}]}}
\def\fev{\mbox{[Fe$\,$\specchar{v}]}}
\def\fevii{\mbox{[Fe$\,$\specchar{vii}]}}
\def\kms{\hbox{km$\;$s$^{-1}$}}
\def\msyr{\hbox{$M_{\odot}\;$yr$^{-1}$}}
\def\msun{\hbox{$M_{\odot}$}}

\begin{document}

\shortauthors{S. Ramya et al.}
\shorttitle{Radio continuum \& \hi in BCDs}   

\title{Radio continuum and \hi study of Blue Compact Dwarf Galaxies}
\author{S. Ramya}   
\affil{Indian Institute of Astrophysics, Bengaluru, India}    
\author{N. G. Kantharia}
\affil{National Centre for Radio Astrophysics (TIFR), Pune, India}
\author{T. P. Prabhu}   
\affil{Indian Institute of Astrophysics, Bengaluru, India}    

\date{Received / Accepted}



\begin{abstract}
The multifrequency radio continuum and 21cm \hi observations of five blue compact dwarf (BCD) galaxies, Mrk 104, Mrk 108, Mrk 1039, Mrk 1069 and I Zw 97 using the Giant Meterwave Radio Telescope (GMRT) are presented here. Radio continuum emission at 610 MHz and 325 MHz is detected from all the observed galaxies whereas only a few are detected at 240 MHz. In our sample, three galaxies (Mrk 104, Mrk 108 and Mrk 1039) are members of groups and two galaxies (Mrk 1069 and I Zw 97) are isolated galaxies. The radio emission from Mrk 104 and Mrk 108 is seen to encompass the entire optical galaxy whereas the radio emission from Mrk 1039, Mrk 1069, I Zw 97 is confined to massive \hii regions. This, we suggest, indicates that the star formation in the latter group of galaxies has recently been triggered and that the environment in which the galaxy is evolving plays a role. Star formation rates (SFR) calculated from 610 MHz emission is in the range $0.01-0.1\ M_\odot$~yr$^{-1}$; this is similar to the SFR obtained for individual star forming regions in BCDs. The integrated radio spectra of four galaxies are modelled over the frequency range where data is available.  We find that two of the galaxies Mrk 1069 and Mrk 1039, show a turnover at low frequencies which is well fitted by free-free absorption whereas the other two galaxies, Mrk 104 and Mrk 108, show a power law at the lowest GMRT frequencies. The flatter spectrum, localized star formation and radio continuum in isolated galaxies lend support to stochastic self-propagating star formation (SSPSF). The \hi observations of four galaxies Mrk 104, Mrk 108, Mrk 1039 and Mrk 1069 show extended disks as large as $\sim1.1-6$ times the optical size. All the observed BCDs (except Mrk 104) show rotating disk with a half power width of $\sim50-124$ \kms. Solid body rotation is common in our sample. We note that the tidal dwarf (TD) origin is possible for two of the BCDs in our sample.

\end{abstract} 


\section{\textsc{Introduction}}
\label{sec:intro_radio}

Blue compact dwarf galaxies (BCDs) are star forming dwarf galaxies whose bluer colours are attributed to ongoing star formation. They are gas-rich, compact, low luminosity ($M_{B}= -17$ to $-14$) objects with low metal abundances ($\frac{1}{50}Z_{\odot}< \ Z <\frac{1}{2}Z_{\odot}$: \citealt{izo06}). They are not forming stars for the first time, as was predicted earlier (\citealt{ssb73}; \citealt{ss72}). All BCDs possess a faint low surface brightness component that is detected both in the optical and in the IR (\citealt{caon05} and references therein; \citealt{ramya09}), implying the presence of old red stars. By analyzing colour maps and surface brightness profiles of the low surface brightness (LSB) component, the ages and chemical abundances of the underlying host galaxies have been determined (\citealt{noeske00}; \citealt{pap02}; \citealt{ramya09}). The enrichment of the ISM in dwarf galaxies mainly occurs during these short starburst events \citep{leg00}. \cite{kun86} proposed that if the metals produced during a starburst are immediately mixed with the surrounding \hii regions, the metallicity will rise very quickly to values of the order of 1/50th of the solar value which explains why no galaxy with metallicity lower than I Zw 18 (1/50th $Z_\odot$), a nearby BCD, has ever been found \citep{leg00}. \\

A search for quiescent BCDs (QBCDs), carried out by \citet{alm08} indicates that after their bursting phase of a few 10 Myr to a few 100 Myr, BCDs enter the quiescent stage. BCDs spend about 30 times more time in the quiescent phase. However, QBCDs are found to be more metal rich than BCDs \citep{alm08} which is yet to be understood. It is noticed that, none of the dwarfs or low surface brightness (LSB) galaxies show a SFR equal to zero (\cite{leg00} and references therein) which implies even during quiescence star formation occurs at a very low rate. \cite{leg00} has concluded through his modelling that the observed oxygen and carbon abundances in  I Zw 18 can be reproduced by a continuous SFR of $10^{-4}$ \msyr after 14 Gyr. However to reproduce the present colours, they had to include a bursting episode. All this suggests the existence of a weak but continuous regime of star formation in these galaxies. A study of extremely isolated BCDs by \cite{zitrin09} concludes that the galaxy colours are better explained by the combination of a continuous star formation process with a recent instantaneous star burst, than by a combination of several instantaneous bursts as suggested previously. \

A few BCDs also emit high ionization lines of \heii $\lambda4686$ \AA, \nev\ $\lambda\lambda3346,3426$ \AA, \fev $\lambda4227$ \AA, \fevii\ $\lambda\lambda5146,5177$ \AA, along with broad emission lines of H$\beta$, \oiii\ $\lambda\lambda4959,5007$ \AA \ and H$\alpha$ in very low metallicity and dense interstellar medium which are believed to be due to supernovae (SNe) events and/or stellar winds (\citealt{izo07}, and references therein). Whether these low-metallicity BCDs can host an active nucleus is another current area of research. Chandra X-ray observations of star bursting dwarf galaxies, such as NGC 1569 and NGC 3077 \citep{grimes05} seem to indicate that the material is blown out into the halo and consequently even removed from the galaxy leading to enrichment of the intergalactic medium. \

BCDs harbour appreciable amounts of dust \citep{tsm99}, confirmed from the far-IR (FIR) emission at 60 $\mu$m, $90\mu$m and $140\mu$m \citep{hirashita09}. The optical properties of dust are similar to the dust in the Milky Way \citep{hirashita09}.  However the dust in BCDs appears to be warmer. Weak polycyclic aromatic hydrocarbons (PAH) emission in the bands at 6.2, 7.7, 11.2 and $12.8 \mu$m is detected in some BCDs. PAH emission is suppressed in most metal-poor BCDs, believed to be because of a metallicity threshold below which PAHs cease to form (\citealt{wu09}; \citealt{dwek05}). \cite{engel05} and \cite{engel08} found an anticorrelation between the dust temperature and metallicity implying warmer dust at lower metallicities of log(O/H)+12$\sim8$ and temperature continues to fall with further reductions in metallicity. The dependence on metallicity is found out to be $\sim Z^{-2.5}$ down to log(O/H)+12$\sim8$. The change in dust behaviour in terms of PAH emission, FIR colour temperatures and dust/gas mass ratio, all near metallicity log(O/H)+12$=8$ indicate that near this metallicity there is a general modification of the components of the interstellar dust that dominates the infrared emission \citep{engel08}. 

Radio observations which include the 21cm spectral line of \hi and radio continuum emission are useful in estimating the neutral gas content and kinematics, star formation rates and possible signatures of interactions. A sample of BCDs observed in \hi confirms that metal poor systems tend to be gas-rich low-luminosity galaxies \citep{hutchmeier07}. A range of spectral shapes at radio frequencies have been observed for BCDs (\citealt{hunt05}; \citealt{yin03}; \citealt{deeg93}; \citealt{kle91}). The observed radio continuum spectrum is attributed to star formation. The FIR-radio correlation of BCDs is similar to that of normal galaxies \citep{yun01}. 

The initial triggering mechanism, evolution of starburst and evolution of BCDs as a whole is not yet understood. Several mechanisms have been proposed, ranging from internal instabilities to external (especially tidal) triggers. If systems are isolated, star formation can be explained using the stochastic self propagating star formation mechanism (SSPSF), first proposed by \cite{gerola80}. Recent studies of large samples of star-forming dwarf galaxies \citep{noeske01} that look for faint companions support the hypothesis of interaction-induced star formation in BCDs. A lower limit for the fraction of star forming dwarf galaxies found with companions is $\sim30\%$ \citep{noeske01}. Thus, both the mechanisms are plausible and it is difficult to quantify the relative influence of the two mechanisms at a particular epoch. 

The five blue compact dwarf galaxies studied here (Mrk 104, Mrk 108, Mrk 1039, Mrk 1069 and I Zw 97) are selected from a larger sample chosen for an optical study (\citealt{ramya09}; \citealt{ram10}). In this paper, we present the 21cm \hi tracing the neutral atomic gas and radio continuum observations at 240, 325 and 610 MHz tracing the combined distribution of thermal and non-thermal radiation for the five BCDs. This is the first time that many of these galaxies have been detected at frequencies $<1$ GHz. Combined with the higher frequency observations from literature, where available, the radio spectra can be modelled. The distance to these galaxies is between 20 and 40 Mpc. Table \ref{tab:para} lists the general properties of these galaxies.

Mrk 104 belongs to a loose group UZC-CG 94 consisting of 3 members, with the closest member, UGC 4906 an Sa galaxy separated from Mrk 104 in the sky plane by $\sim330$ kpc. The other member is PGC 26253. Mrk 104 has been classified as having a double nucleus in the process of merging \citep{mazarella93}. \cite{ramya09} resolve the two  nuclei and note that both show \hii region like spectra thus ruling out the presence of an AGN in the centre of the galaxy. Mrk 104 is situated at a distance of 31.2 Mpc and at this distance $1\arcsec$ corresponds to $\sim151$ pc.

Mrk 108 classified as an I0 in NED is one of the four members of the group Holmberg 124. NGC 2820 (of Hubble type SBc) is the closest neighbour. There is also a radio bridge connecting NGC 2820 and Mrk 108 to the third member of the group NGC 2814 \citep{nim05} clearly indicating a tidal interaction. Several other signatures of hydrodynamic processes are also observed in the group \citep{nim05}. This galaxy hosted a type IIp supernova, SN 1998bm indicating recent massive star production. Mrk 108 is situated at a distance of 22.2 Mpc and at this distance $1\arcsec$ corresponds to $\sim108$ pc.

Mrk 1039 is a member of the group USGC S087 \citep{ramella02} and LGG 59 (Lyon group of galaxies: \citealt{garcia93}). DDO 023, DDO 020 and Mrk 1042 are the other members of the group. All the companion members are dwarf galaxies. Though located close to the Eridanus supergroup, it is not considered to be part of it \citep{brough06}. A type II supernova, SN 1985S is recorded in Mrk 1039. The galaxy is located at a distance of 28.8 Mpc (galactocentric distance taken from NED) and at this distance $1\arcsec$ corresponds to $\sim139$ pc.

The fourth galaxy in our sample, Mrk 1069 also lies close to the Eridanus supergroup but is  not considered to be member of the group. No group membership is assigned to Mrk 1069. This galaxy is located at a distance of 20.7 Mpc from us and at this distance $1\arcsec$ corresponds to a distance of $\sim100$ pc.

I Zw 97 is an isolated galaxy with no neighbour within about $50\arcmin$ (525 kpc). \cite{thuan81} do not detect this galaxy in \hi and conclude that the atomic gas surface density is $< 2.7\times10^6\, M_\odot\,$Mpc$^{-2}$ and the upper limit on the \hi mass is $7.3\times10^8\, M_\odot$. Type II SN 2008bx was discovered recently in this galaxy, and \cite{atel09} reported the detection of radio continuum emission at 610 MHz from this supernova. The galaxy is at a distance of 36.1 Mpc. At this distance $1\arcsec$ corresponds to 
$\sim175$ pc.\\

This paper is structured as follows. Section \ref{sec:obs_rad} gives an account of the observations and data reduction. Section \ref{sec:notes_rad} gives a note on individual galaxies. Section \ref{sec:discus_rad} is a detailed discussion on our results and section \ref{sec:con_rad} summarizes the study.

\begin{landscape}
\begin{table*}[h!]\footnotesize
\begin{center}
\caption{\footnotesize The general parameters of the five galaxies collected from literature.} 
\begin{tabular}{llllll}
\\
\hline\hline
\textbf{Parameter} & \multicolumn{5}{c}{\textbf{Galaxy}} \\
\hline
--- & \textbf{Mrk 104} & \textbf{Mrk 108} & \textbf{Mrk 1039} & \textbf{Mrk 1069} & \textbf{I Zw 97} \\
\hline \\
\textbf{Hubble type}$^e$ & Pec & I0 pec & Sc, edge-on, \hii & Sa$^*$ & --- \\	
\textbf{Helio vel}$^e$ (\kms) & 2235 & 1534 & 2111 & 1562 $^d$ & 2518 \\	
\textbf{Central vel}$^a$ (\kms) &  2203   &  1574 & 2098$^d$ & 1562$^d$ & 2530   \\	
\textbf{Group} & UZC-CG 94$^b$ & Holm 124 & USGC S087$^c$  & --- & --- \\
\textbf{members} & UGC 4906, & NGC 2820, & DDO 023, & UGCA 052	& --- \\
--- & PGC 26253 & NGC 2814 & DDO 020 & --- & --- \\	
--- & --- & --- & Mrk 1042 & --- & --- \\
\textbf{Single dish \hi mass}$^a$ (M$_\odot$) & $5.0\times10^8$ & $9.4\times10^9$ & $1.2\times10^9 \ ^d$ & $8.5\times10^8 \ ^d$ & {$<7.3\times10^8$}\ \\
\textbf{50\% line width}$^a$ (\kms) & 163 & 324 & 149 $^d$ & 106 $^d$ & --- \\ 
\textbf{Galactocentric Distance}$^{e,h}$ (Mpc) & 31.2 & 22.2 & 28.8 & 20.7 & 36.1 \\
\textbf{linear scale - $1''$}\ $^{e,h}$ (pc) & 151 & 108 & 139 & 100 & 175 \\
\textbf{m$_B$}$^f$ (mag) & $15.1\pm0.4$ & $15.5\pm0.3$ & 13.94$^g$ & 14.55$^g$ & $14.9\pm0.3$ \\
\textbf{M$_B$} (mag) & -17.46 & -16.47 & -18.46 & -17.32 & -16.76 \\
\textbf{L$_B$} ($10^9\times L_\odot$) & 1.5 & 0.6 & 3.8 & 1.3 & 0.79 \\
\textbf{L$_{FIR}$} ($10^9\times L_\odot$) & 0.54 & --- & 1.96 & 0.88 & 0.82 \\
\textbf{SN recorded} & --- & SN1998bm & SN1985S & --- &	SN2008bx \\
\hline 
 \end{tabular}
\end{center}
$^a$ - \cite{thuan81}, $^b$ - \cite{focardi02}, $^c$ - \cite{ramella02}, $^d$ - \cite{thuan99}, $^e$ - NASA 
Extragalactic Database, $^f$ - \cite{rc391}, $^g$ - \cite{doy05}, $^h$ - Assuming $H_0$ = 73 km~s$^{-1}$~Mpc$^{-1}$, * - Hyperleda.
\label{tab:para}
\end{table*}
\end{landscape}

\section{\textsc{Observations and data reduction}}
\label{sec:obs_rad}
\hi and radio continuum observations of Mrk 104, Mrk 108, Mrk 1039, Mrk 1069 and I Zw 97 were obtained using the Giant 
Meterwave Radio Telescope (GMRT: \citealt{swa91}). GMRT consists of 30 antennas of 45m diameter distributed along a Y.  
14 antennas are situated in the central 1 km region whereas the rest are spread over a 25 km region. GMRT operates at 
five frequency bands namely 150 MHz, 240 MHz, 325 MHz, 610 MHz and 1420 MHz. We observed all the galaxies except Mrk 108 
in the dual frequency mode in which data in two frequency bands namely 240 MHz and 610 MHz are simultaneously observed. We
 obtained the observations of Mrk 104, Mrk 108, Mrk 1039 and Mrk 1069 at 325 MHz bands also. All the galaxies except 
I Zw 97 were observed in the 21cm line of \hi in the 1420 MHz band. We used a bandwidth of 4 MHz for all the galaxies and 
with 128 spectral channels this resulted in a channel width of 31.2 kHz ($\sim 6.6$ \kms). Appropriate phase and 
amplitude (which also doubled as bandpass) calibrators were selected for the observations. Data reduction was carried 
out using the standard tasks in AIPS\footnote{The National Radio Astronomy Observatory is a facility of the National 
Science Foundation operated under co-operative agreement by Associated Universities, Inc.}.

To summarise our data reduction procedure, we imported the FITS file of the raw data to AIPS and selected a spectral
 channel after examining the data on the amplitude calibrator which was relatively free of radio frequency interference. 
 This spectral channel of the amplitude calibrator was then gain calibrated and bad data were edited. Once the gain 
calibration was satisfactory, the data were used to obtain the bandpass calibration tables. In the next step, the 
visibility data on all sources were bandpass calibrated and then every 10 channels of the central 100 channels were 
averaged to avoid the effects of bandwidth smearing in the outer parts of the primary beam. The data on both the
 amplitude and phase calibrator of one of the 10 resultant channels were then used to obtain the complex antenna gains.  
After an iterative procedure involving calibrating the calibrator data and removing bad data, the final gain tables were
 generated and the target source data were calibrated.

The data were then imaged by using 9 facets across the primary beam with cellsize of $1\arcsec$ at 1420 MHz, 25 facets 
with cellsize of $2\arcsec$ at 610 MHz and 49 facets with cellsize of $4\arcsec$, $3\arcsec$ at 240 MHz and 325 MHz
 respectively. The analysed data on Mrk 108 were obtained from the authors of \cite{nim05}. The continuum data were then
 self-calibrated. At  least three iterations of phase self-calibration were required. After the first self-calibration 
run, it was noticed that the 610, 325 and 240 MHz images improved considerably. A final round of amplitude and phase
 self-calibration was done for each of the data sets. The rms noise in the final 610 MHz maps is in the range  65--150 
$\mu$Jy whereas it is $1.2-1.9$ mJy for the 240 MHz images. The rms noise in 325 MHz maps varies from 300--900~$\mu$Jy. 
The rms noise in the 1420 MHz continuum maps range from 50~$\mu$Jy for VLA Archive image to 150~$\mu$Jy for the maps 
obtained from GMRT. Self-calibration at 1.4 GHz did not improve the image quality and hence the iteration procedure was 
stopped after two rounds of phase self calibration. The line-free channels were used to obtain the continuum images at 
1.4 GHz.

The spectral line cube was made after removing the continuum emission using the task {\sc uvsub}. We also constructed 
image cubes of angular resolutions varying from around $40\arcsec$ to $10\arcsec$ to examine both large scale and finer
 features in the \hi distribution. The moment maps of these galaxies were created using the {\sc momnt} task in {\sc aips}.
 The galaxy I Zw 97 was not observed in \hi as this galaxy was not detected in single dish observations \citep{thuan81}. 
 More details of these observations and the results are presented in the Tables \ref{tab:obs_cont} and \ref{tab:obs_hi}.
 We also made images of Mrk 1039 using archival data from the Very Large Array (VLA) at 1.4 GHz, 4.8 GHz and 14.9 GHz.

While all the observed galaxies were detected at 325 MHz, 610 MHz and 1420 MHz, only Mrk 104 and Mrk 1069 were detected at 240 MHz. 
All the observed galaxies were also detected in the 21cm spectral line of hydrogen. The contours of the continuum images and \hi column density, velocity maps of these galaxies overplotted on the optical images from DSS are shown in Figures \ref{fig:mrk104_r}--\ref{fig:izw97_r}. Figure \ref{fig:mrk104_r} shows the radio \hi line and continuum data maps of Mrk 104, Figure \ref{fig:mrk108_r}
 displays the radio maps of Mrk 108 at \hi line and radio continuum bands. Figures \ref{fig:mrk1039_r},
 \ref{fig:mrk1069_r} and \ref{fig:izw97_r} show the radio continuum and \hi line maps of Mrk 1039, Mrk 1069 and I Zw 97 
respectively. Details of the flux densities along with rms noise of the maps are presented in Table \ref{tab:obs_cont}. 
Table \ref{tab:obs_hi} shows the \hi properties of these galaxies and Table \ref{tab:fir_rad} gives the radio-FIR 
properties of these BCDs.

\section{\textsc{Results: Notes on individual sources}}
\label{sec:notes_rad}

\begin{enumerate}
\item{Mrk 104 : Figure \ref{fig:mrk104_r} (a) and (b) show the \hi column density and velocity field of Mrk 104. The \hi 
extent ($D_{\textrm{H{\sc i}}}$, diameter of \hi upto a column density of $5\times10^{19}$ cm$^{-2}$) is about $\sim2.7$ 
times the optical size ($D_{25}$, diameter at 25 mag~arcsec$^{-2}$) of the galaxy. A small offset is noticed in the 
kinematic centre with respect to the optical centre. The \hi mass is estimated to be $\sim 2.2\times10^8\, M_\odot$ 
(refer Table \ref{tab:obs_hi}) which is within a factor of 2 of the mass estimated from single dish measurements. Figure
 \ref{fig:mrk104_r}(b) shows the kinematics of the \hi gas. We note that Mrk 104 shows a distorted velocity field. The 
velocity is redshifted outwards from the centre of the galaxy (see Figure \ref{fig:mrk104_r}b) reaching velocities 
$\sim2260$ \kms\ in the northern parts and velocities of $\sim2250$ \kms\ in the southern parts.  We detect a distinct 
cloud near the northern edge of the galaxy at velocities $\sim2211$ \kms \ (see Figure \ref{fig:mrk104_r}b). This cloud has
 no optical counterpart.

Figure \ref{fig:mrk104_r}(c) shows the \hi column density contours overlaid on the H$\alpha$ image of the galaxy taken
 from \cite{ramya09}. The \hi peak is located in between the two \hii regions as seen in the H$\alpha$ image. Figure
 \ref{fig:mrk104_r}(d) shows the position-velocity curve drawn along the major axis of the galaxy. Deviations from simple 
rotation are clearly seen in this diagram. As seen in the lower part of the figure, the edges of the galaxy show high 
velocity whereas the central parts show lower velocities.  The northern cloud seen near declination
 $53\arcdeg27\arcmin10\arcsec$ shows a very different velocity compared to the edge of the
 galaxy indicating it is not a part of the galaxy.

\begin{landscape}
\begin{table}[h!]\footnotesize
\begin{center}
\caption{\footnotesize The radio continuum flux density and RMS noise on the GMRT images of the five galaxies.}
\label{tab:obs_cont}
\begin{tabular}{lllllllll}
\\
\hline\hline 
 \textbf{Galaxy} & \textbf{RA} & \textbf{Dec} & \textbf{Date of obs} & \textbf{Bands} & \textbf{Bandwidth} & \textbf{on-source time} & \textbf{Flux density} & \textbf{RMS noise} \\
 ---& \textit{hh:mm:ss} & \textit{dd:mm:ss} &  & \textit{MHz} & \textit{MHz} & hrs & \textit{mJy} & \textit{mJy/beam} \\
\hline
Mrk 104 & 09:16:45.5  & +53:26:35  & 16/06/08 & 240 & 6 & 5 & $7.33\pm2.6$ & 1.66 \\
 --- & --- & ---      & 12/06/10   & 325 & 16 & 2.5 & $3.95\pm0.86$ & 0.22 \\
 --- & --- & ---      & 16/06/08   & 610 & 16 & 5 & $2.40\pm0.3$ & 0.12 \\
 --- & --- & ---      & 21/06/08   & 1420$^b$ & 1.6 & 7.5 & $1.66\pm0.5$ & 0.18 \\
Mrk 108 & 09:21:30.1  & +64:14:19  & 20/08/02 & 325 & 16 & 5 & $16.52\pm3.80$ & 0.77 \\
 --- & --- & ---      & 06/09/02   & 610 & 16 & 5 & $9.03\pm3.6$ & 0.31 \\
 --- & --- & ---      & 16/07/02   & 1280 & 16 & 7.5 & $1.95\pm0.6$ & 0.75 \\
Mrk 1039 & 02:27:32.8 & -10:09:56  & 16/06/08 & 240 & 6 & 5 & $<5.00\pm1.7$ & 1.66 \\
 --- & --- & ---      & 12/06/10   & 325 & 16 & 2.5 & $5.8\pm1.1$ & 0.90 \\
 --- & --- & ---      & 16/06/08   & 610 & 16 & 5 & $7.85\pm0.4$ & 0.17 \\
 --- & --- & ---      & 13/07/09   & 1420$^a$ & 50 & 4 & $3.00\pm0.2$ & 0.05 \\
 --- & --- & ---      & 14/06/00   & 4800$^a$ & 50 & 4 & $1.45\pm0.1$ & 0.05 \\
 --- & --- & ---      & 16/07/09   & 14900$^a$ & 50 & 4 & $1.26\pm0.4$ & 0.15 \\
Mrk 1069 & 03:08:19.0 & -13:54:11  & 09/06/09 & 240 & 6 & 5 & $10.33\pm1.3$ & 1.99 \\
 --- & --- & ---      & 12/06/10   & 325 & 16 & 2.5 & $12.86\pm2.73$ & 1.30 \\
 --- & --- & ---      & 09/06/09   & 610 & 16 & 5 & $13.92\pm0.4$ & 0.06 \\
 --- & --- & ---      & 31/05/09   & 1420$^b$ & 1.6 & 7.5 & $9.57\pm1.9$ & 0.15 \\
I Zw 97 & 14:54:39.0  & +42:01:25  & 04/06/09 & 240 & 6 & 5 & $<4.50\pm1.50$ & 1.50 \\
 --- &  --- &  ---    & 04/06/09   & 610    & 16 & 5 & $1.14\pm0.9$ & 0.06 \\
\hline\hline
\end{tabular}
 \end{center}
$^a$ Higher frequency data of Mrk 1039 is obtained from the VLA Archives.\\
$^b$ This data is obtained after collapsing the line free channels of \hi data observed from the GMRT to obtain
 the continuum map. RMS noise here corresponds to the noise in the 50 channels that are collapsed.  
\end{table}
\end{landscape}

\begin{figure}[h!]
\begin{center}\footnotesize
\subfloat[][]{\includegraphics[width=5cm,height=5cm]{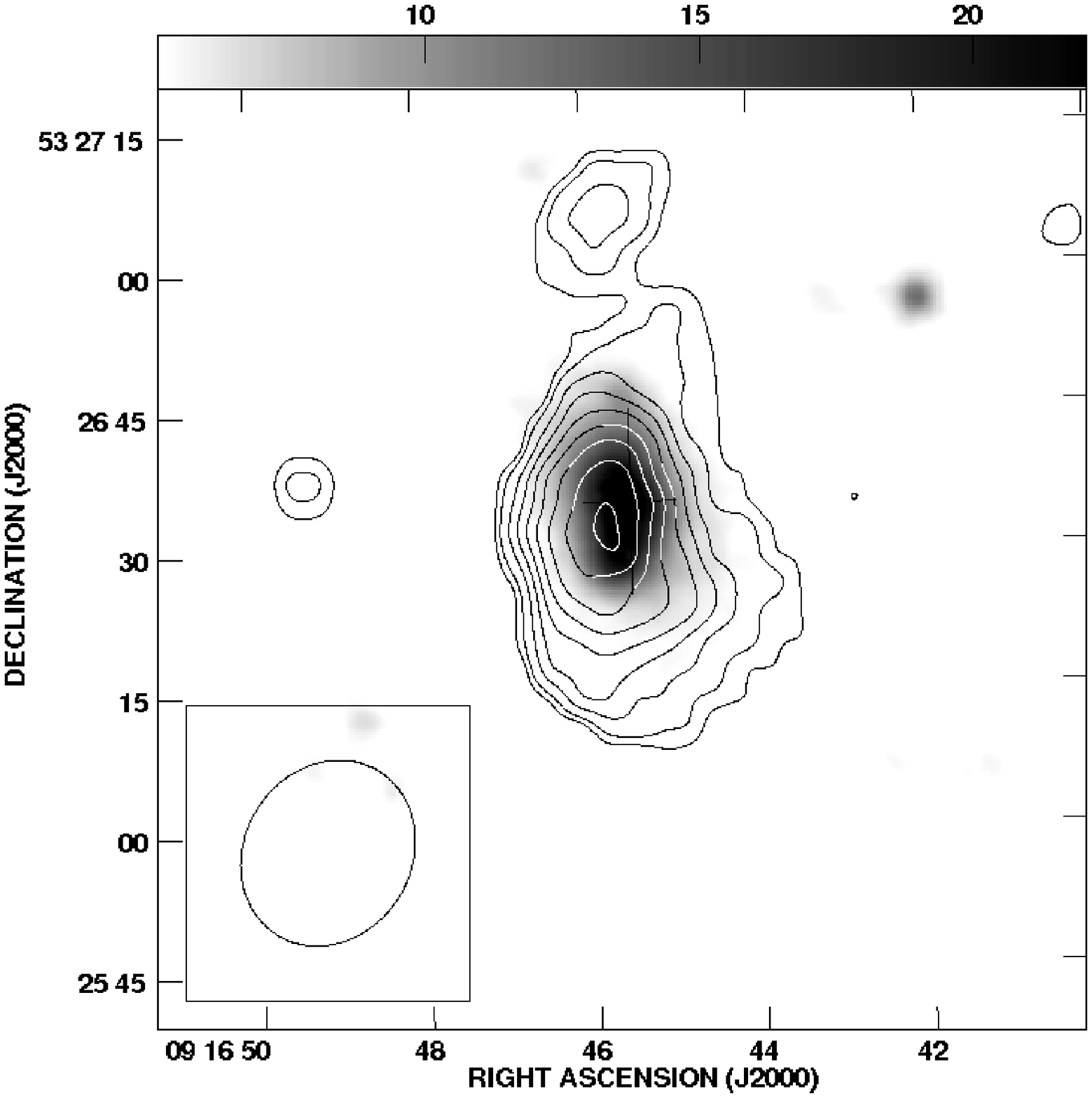}}
\subfloat[][]{\includegraphics[width=5cm,height=5cm]{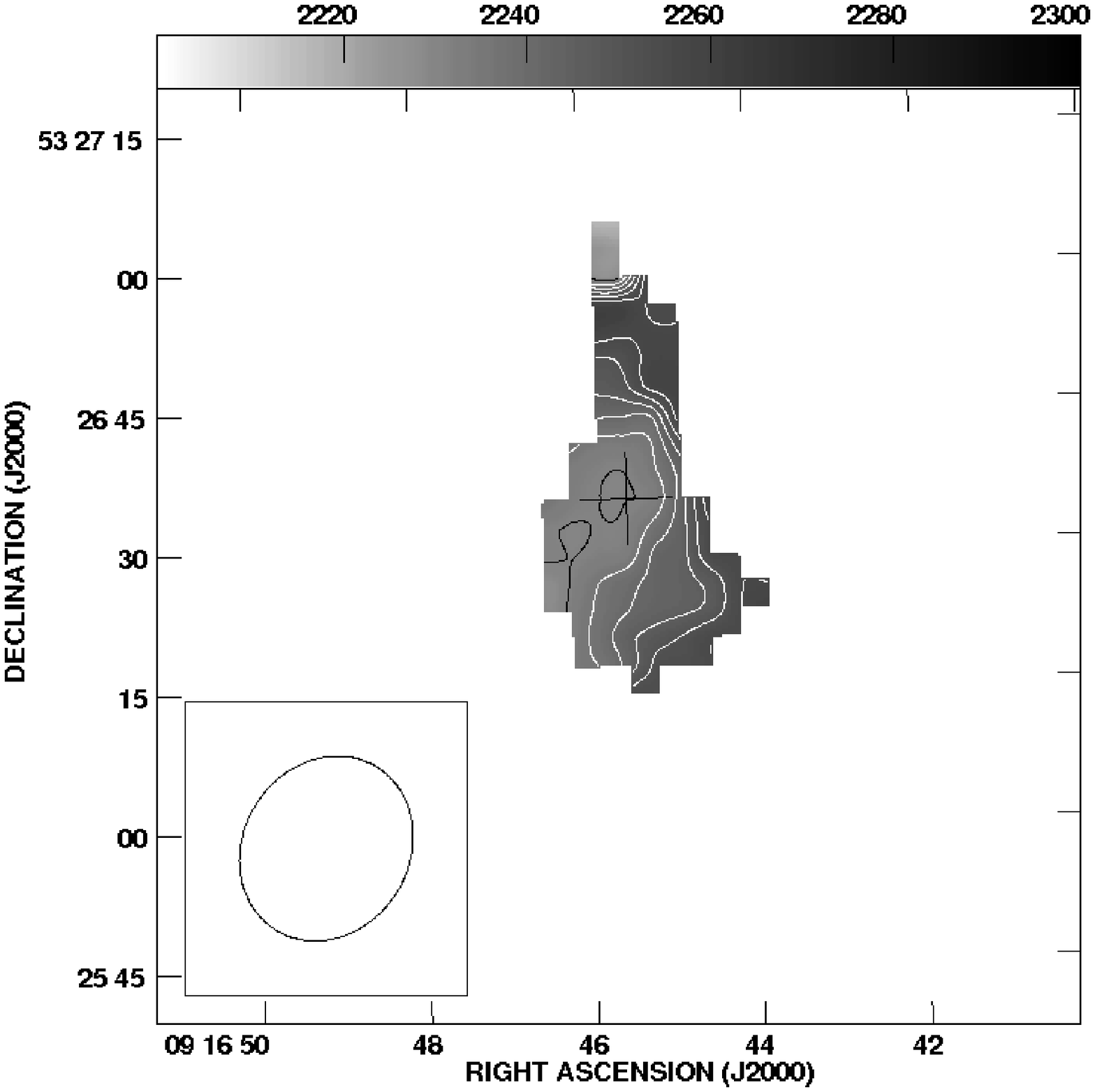}} 
\subfloat[][]{\includegraphics[width=5cm,height=5cm]{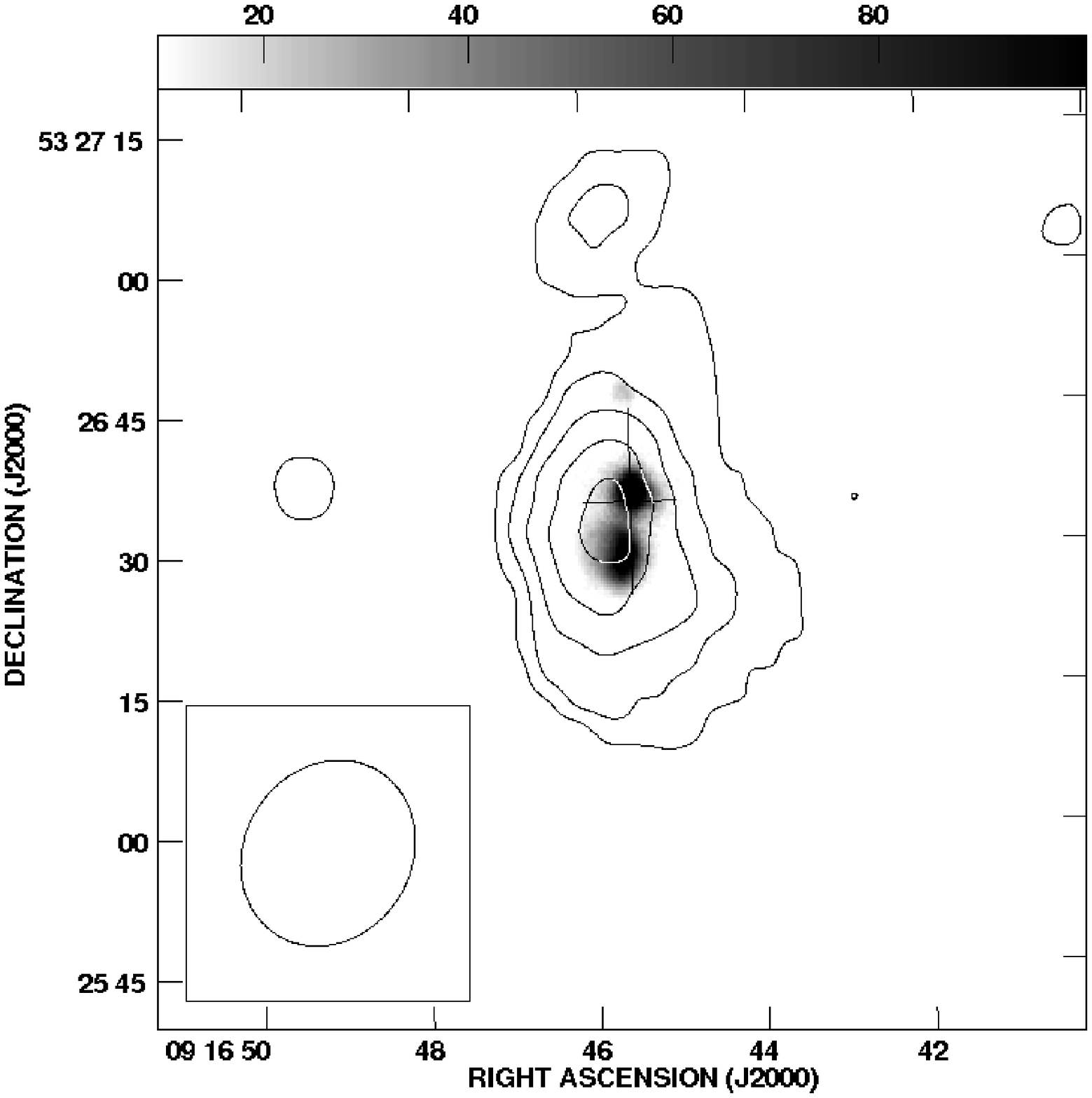}}
\qquad
\subfloat[][]{\includegraphics[width=5cm,height=5cm]{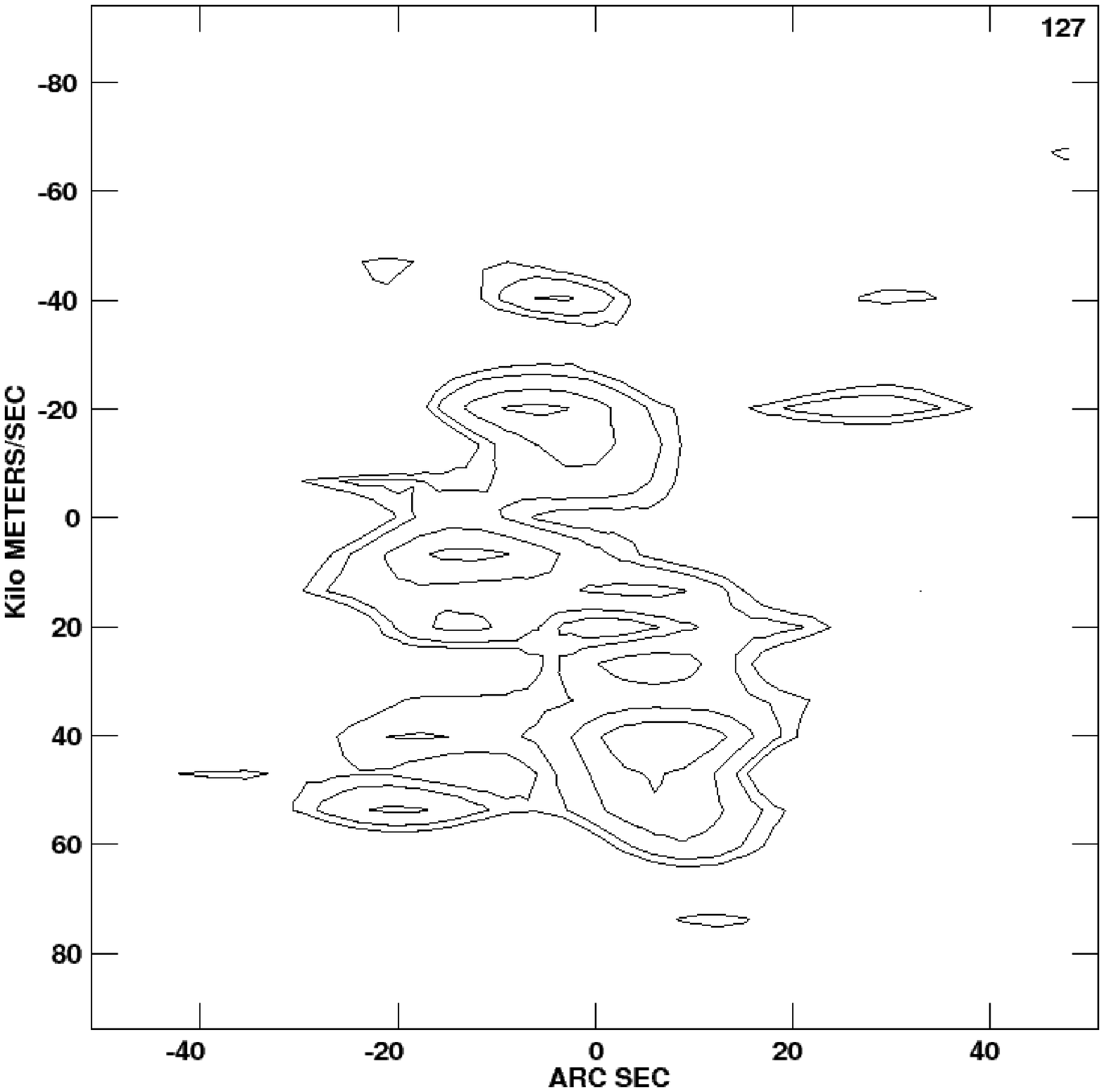}}
\caption{\footnotesize {\bf Mrk 104 :} {\bf (a)} Column density, $N$(\hi) contours at (0.5, 1, 2, 3, 4, 5, 6, 7, 7.8) 
$\times 10^{20}$ cm$^{-2}$ overlaid on the optical $B$ band DSS image. {\bf (b)} The velocity map(\textit{MOMENT 1} 
map) with velocity contours drawn from 2231, 2236, 2241, 2246, 2251, 2256 \kms\ for this galaxy. The central contour in 
black is 2231 \kms. {\bf (c)} The \hi column density contours ((0.5, 2, 4, 6, 7.5) $\times 10^{20}$ cm$^{-2}$) overlaid 
on the continuum subtracted H$\alpha$ image \citep{ramya09}. The peak contour shows that the peak \hi emission lies
 between the two \hii regions. {\bf (d)} Position-velocity curve along the major axis ($5\arcdeg$ east of north) of the
 galaxy. Note that the major axis coordinates are plotted along the x-axis. The contours are plotted at 
(2.5, 3, 4, 5)$\times1.3$ mJy/beam.  Note the similar velocities observed in the north and south of the galaxy and the 
cloud showing a distinct identity in the PV diagram. It could be an infalling cloud. The angular resolution of all the
 maps is $21\arcsec\times18\arcsec$.  }
\label{fig:mrk104_r}
\end{center}
\end{figure}

\begin{figure}[h!]\ContinuedFloat
\begin{center}\footnotesize
\subfloat[][]{\includegraphics[width=5.5cm,height=5.5cm]{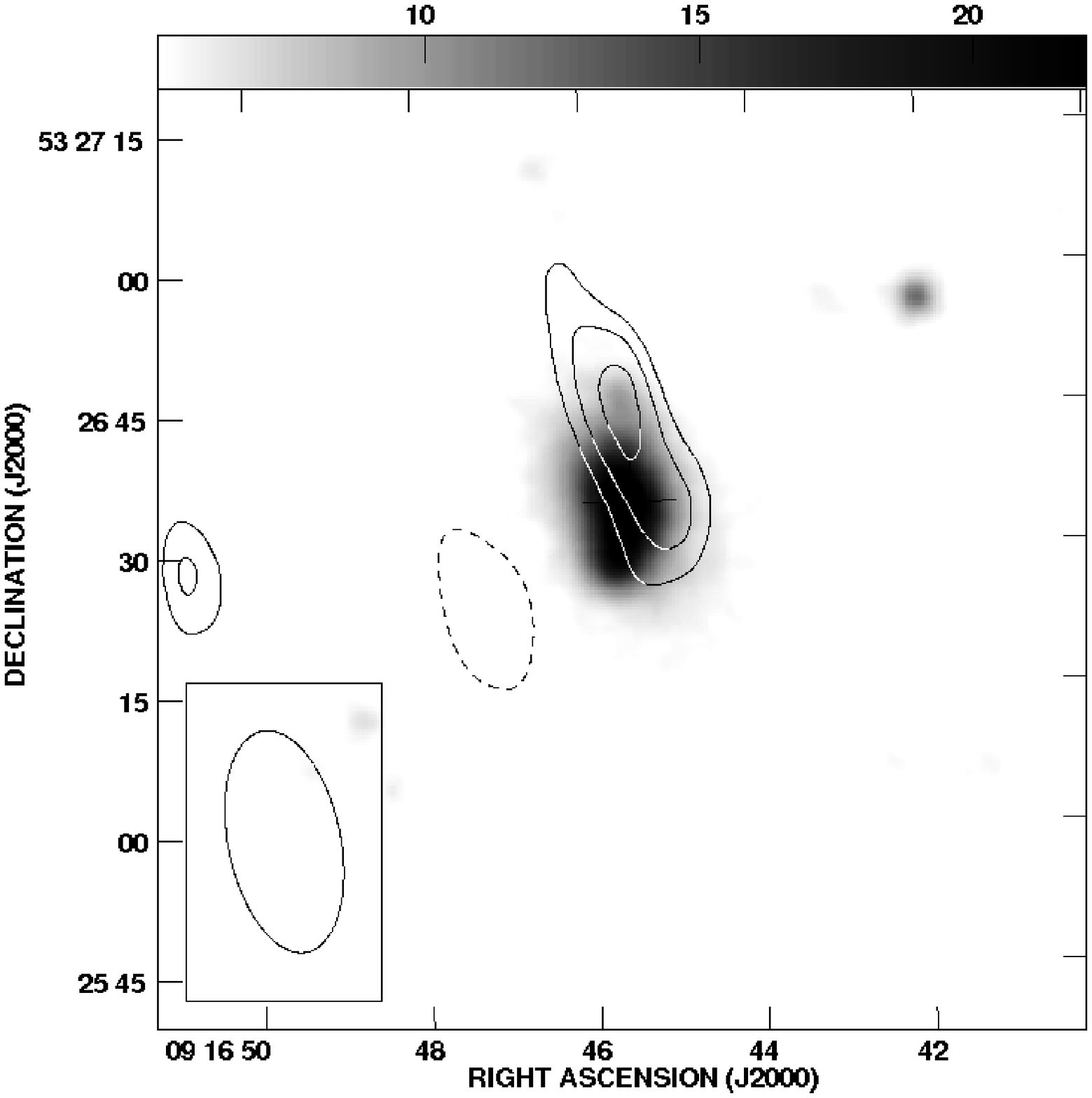}}
\subfloat[][]{\includegraphics[width=5.5cm,height=5.5cm]{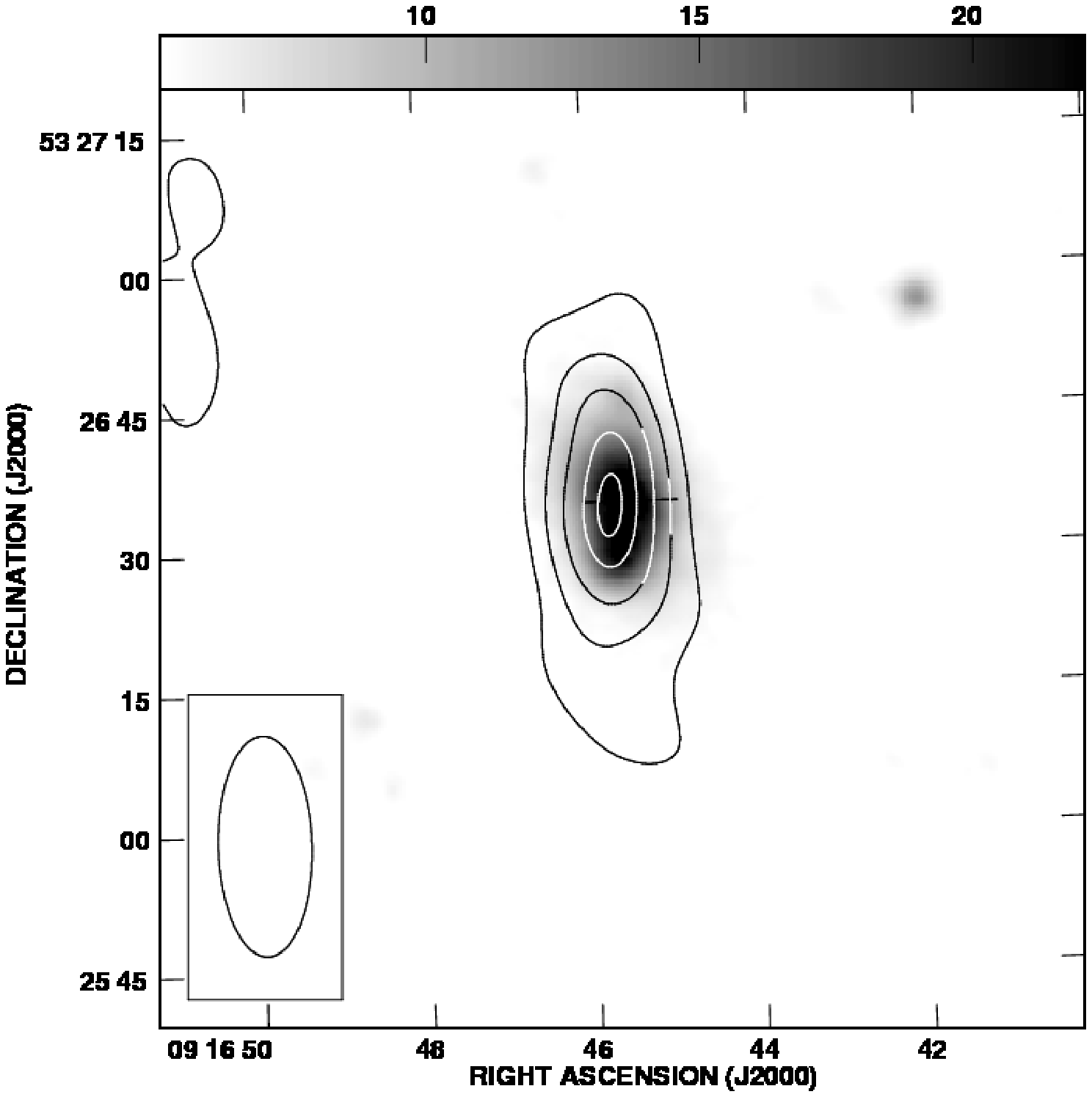}}
\subfloat[][]{\includegraphics[width=5.5cm,height=5.5cm]{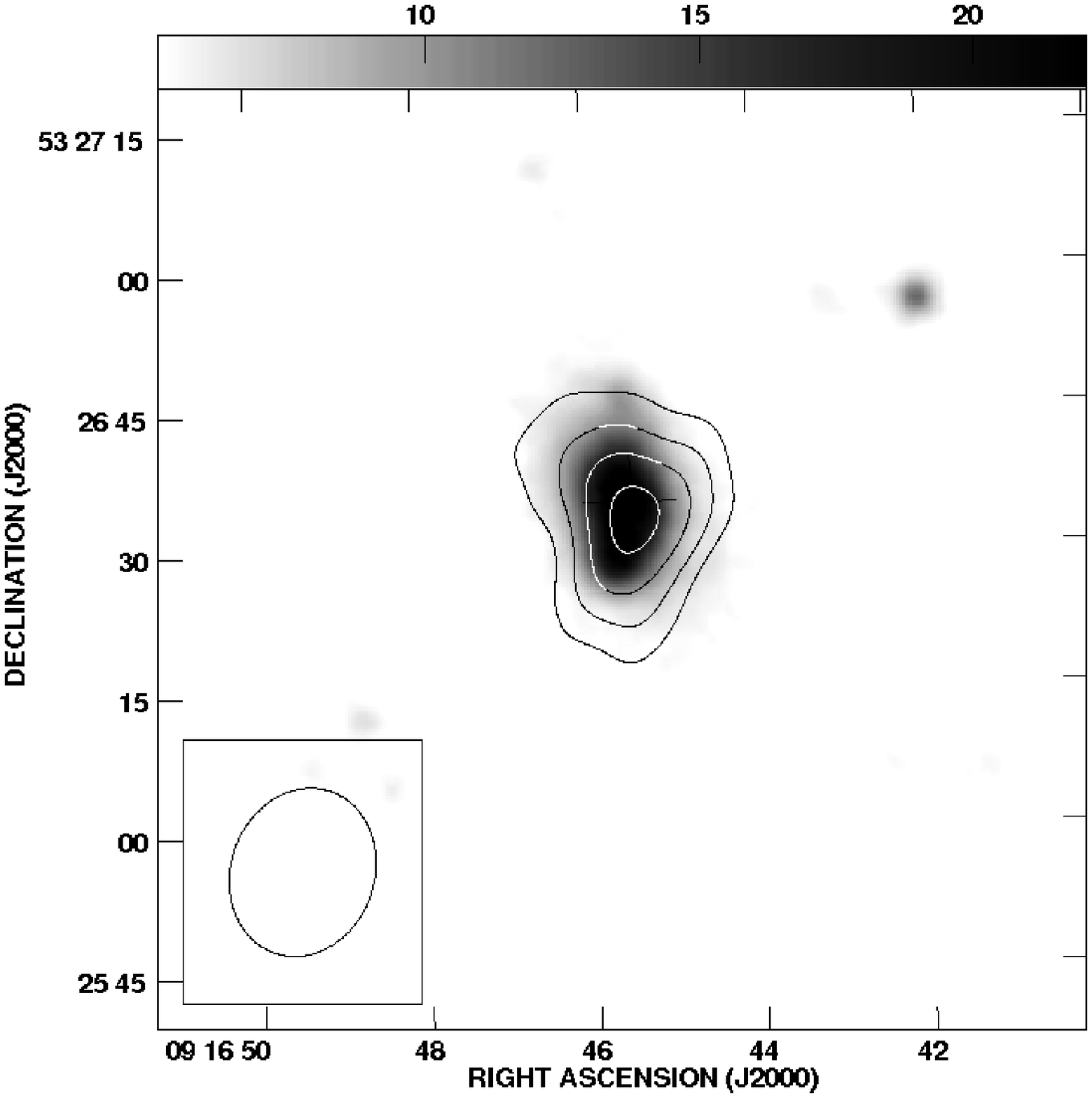}} 
\qquad
\subfloat[][]{\includegraphics[width=5.5cm,height=5.5cm]{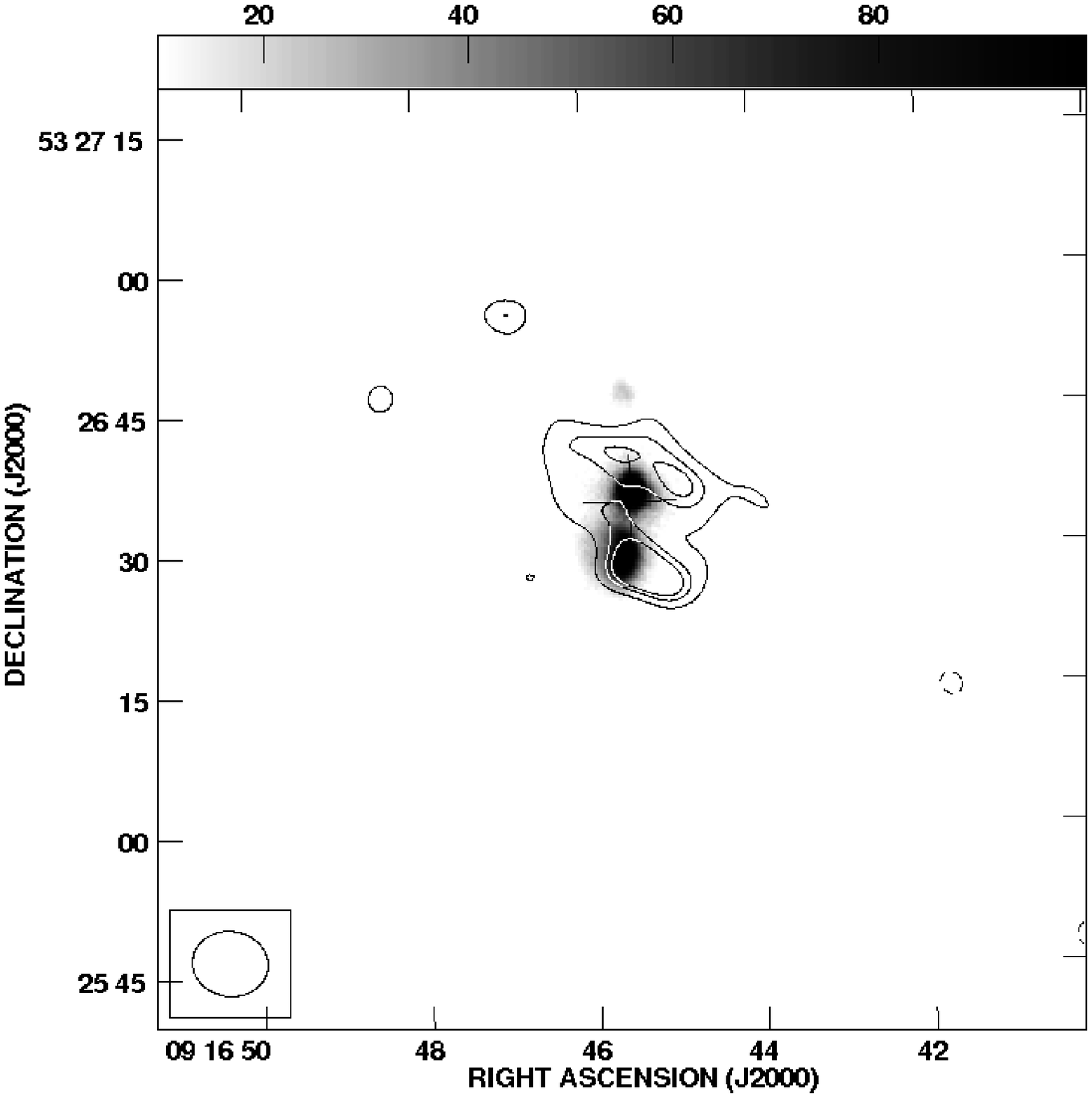}} 
\subfloat[][]{\includegraphics[width=5.5cm,height=5.5cm]{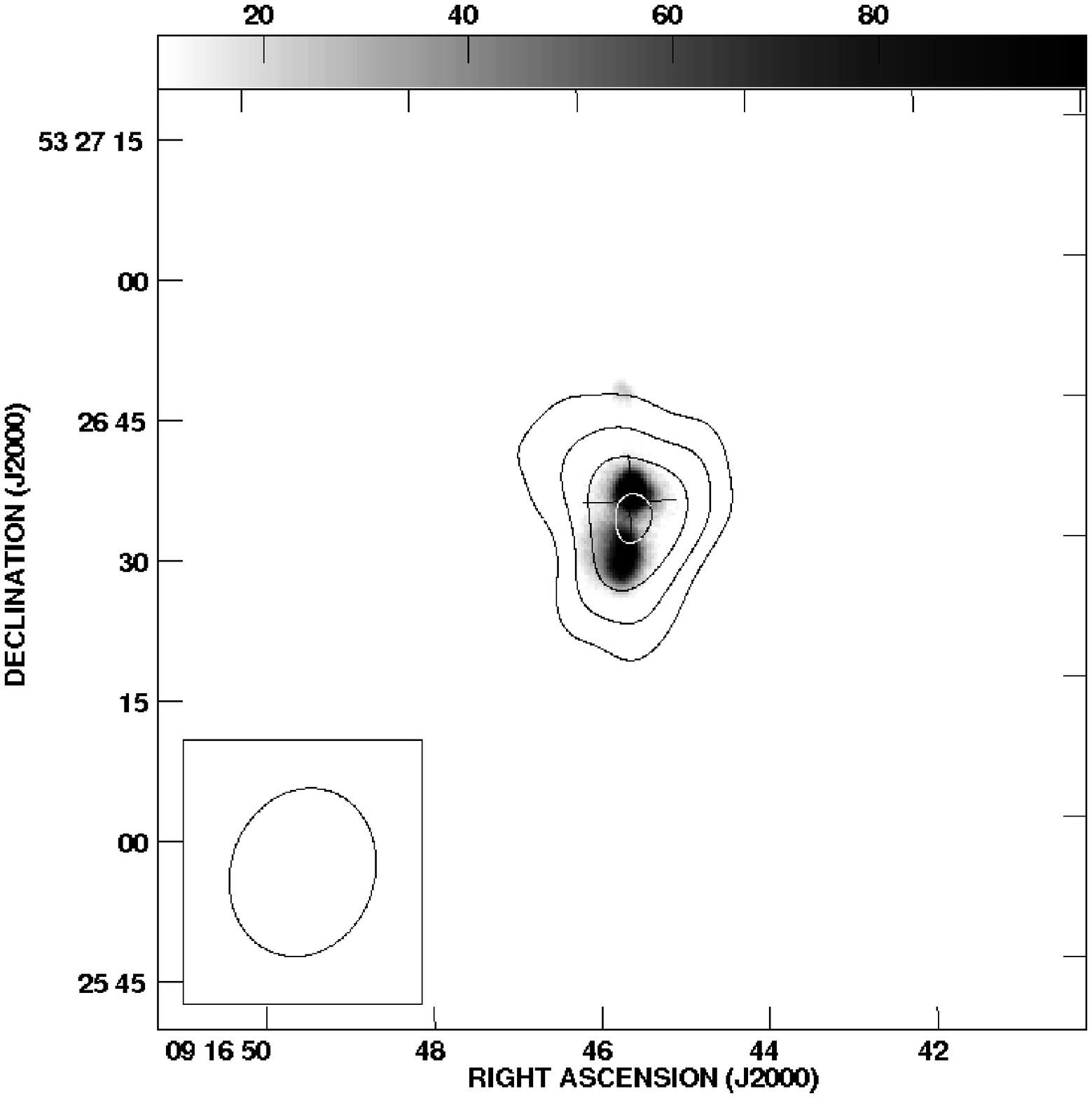}}
\caption{\footnotesize {\bf Mrk 104 {\it Continued:}} {\bf (e)} contours ($1.3 \ \times$ (-6, -4, 4, 5, 6) mJy/beam) show 240 MHz and the grey 
scale represents the $B$ band optical image. The resolution of the map is $24\arcsec\times12\arcsec$. {\bf (f)} contours 
($300 \ \times$ (-6, -4, 4, 6, 8, 10, 11) $\mu$Jy/beam) show 325 MHz overlaid on the grey scale DSS $B$ band optical 
image. Beam size is $24\arcsec\times10\arcsec$. {\bf (g)} contours ($140 \ \times$ (-6, -4, 4, 6, 9, 10) $\mu$Jy/beam) 
show 610 MHz and the grey scale represents the $B$ band optical image. Beam size is $18\arcsec\times15\arcsec$. {\bf (h)} 
contours ($145 \ \times$ (-5, -4, 4, 5, 5.5) $\mu$Jy/beam) show 1.4 GHz continuum map and the grey scale represents the 
H$\alpha$ narrow band optical image taken from \cite{ramya09}. The 1.4 GHz continuum emission shows emission peaks
 coincident with the bright H~{\sc~ii} regions. Beam size is $8\arcsec\times7\arcsec$. {\bf (i)} 610 MHz continuum map
 (natural weighted) is overlaid on the H$\alpha$ image taken from \cite{ramya09}. The contours levels drawn are
 $144\times$(-6, -4, 4, 6, 8, 10)$\mu$Jy/beam. In this figure and all the figures to follow, North is up and East is 
towards left.}
\end{center}
\end{figure}

The galaxy is detected in the radio continuum in all the four bands --- 240, 325, 610 MHz and 1.4 GHz
(Figure \ref{fig:mrk104_r}e,f,g,h). The continuum emission encompasses the entire optical galaxy and appears to consist of 
two radio peaks almost coincident with the \ha peaks. The global spectral index, $\alpha(325,610)=-0.8$ 
(where $S_{\nu}\propto\nu^{\alpha}$) and $\alpha(610,1420)=-0.4$ imply 
dominance of non-thermal emission at the lower frequencies and probable contribution from thermal processes at the
 higher frequencies. Figure \ref{fig:mrk104_r}(i) shows the 610 MHz map overlayed on the H$\alpha$ images taken 
from \cite{ramya09}. The radio peak is coincident with the two bright \hii regions seen in Figure 
\ref{fig:mrk104_r}(c) as well.}

\item{Mrk 108 : This galaxy is in a group, namely Holmberg 124 and group interaction is clearly noticeable (refer
 Figure \ref{fig:mrk108_r}(a)). Figure \ref{fig:mrk108_r}(a) shows a larger region in 325 MHz radio continuum in which
 the group members NGC 2820 (eastern edge), NGC 2814 (western edge) are also seen. Figure \ref{fig:mrk108_r}(b) shows the 
\hi intensity map. Figure \ref{fig:mrk108_r}(c) shows the velocity field of Mrk 108 which is distinct from that of NGC 
2820. The systemic velocity of the galaxy given in NED is 1562 \kms. We estimate a \hi mass of $1.6\times10^8\, M_\odot$
 with an error of $\sim30\%$ due to the large \hi disk of its neighbour NGC 2820. 

The galaxy is detected at 325 MHz, 610 MHz and 1280 MHz (see Figure \ref{fig:mrk108_r} d,e,f).

\begin{figure}[h!]\footnotesize
\centering
\subfloat[][]{\includegraphics[width=5.5cm,height=5.5cm]{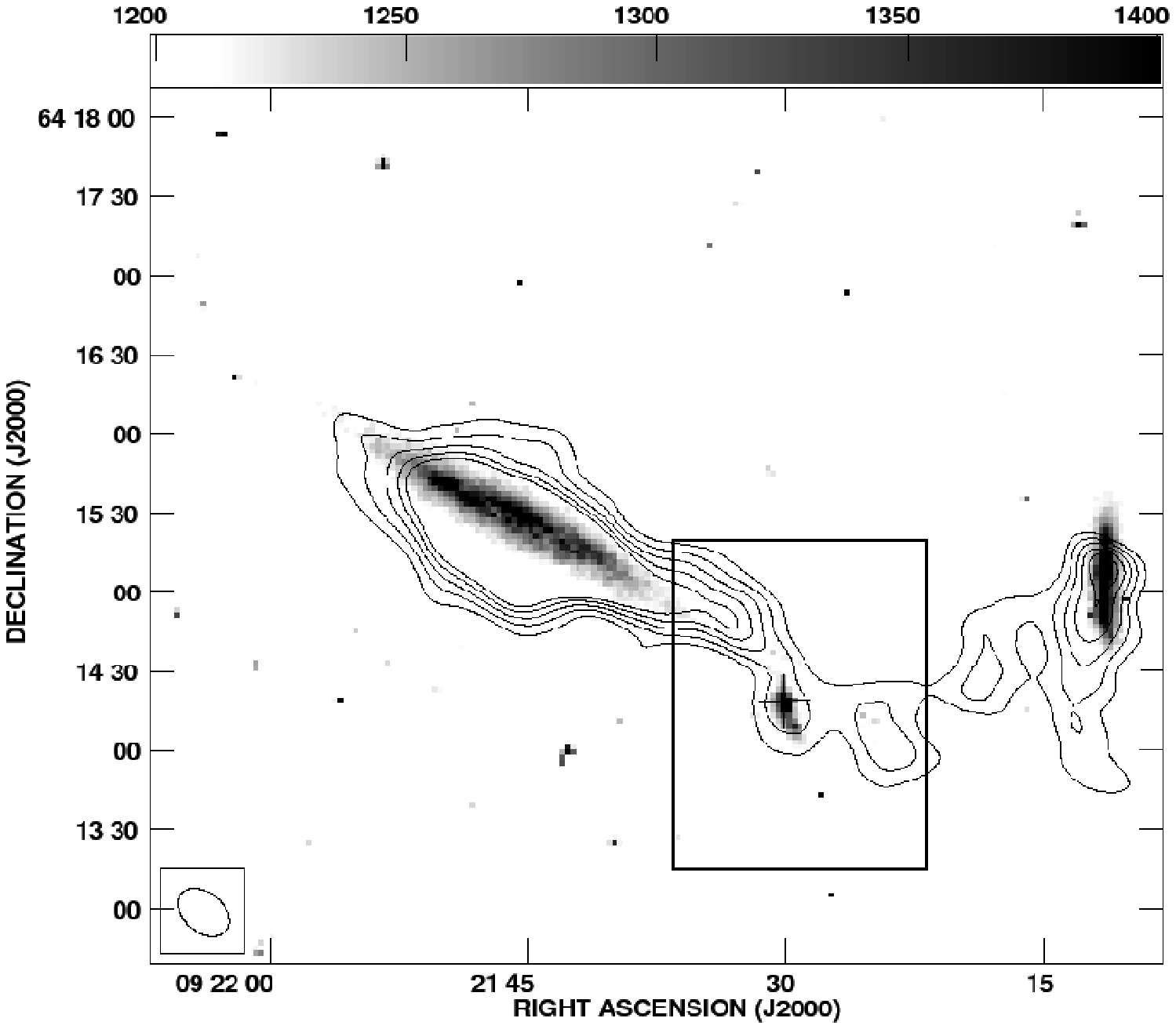}}
\subfloat[][]{\includegraphics[width=5.5cm,height=5.5cm]{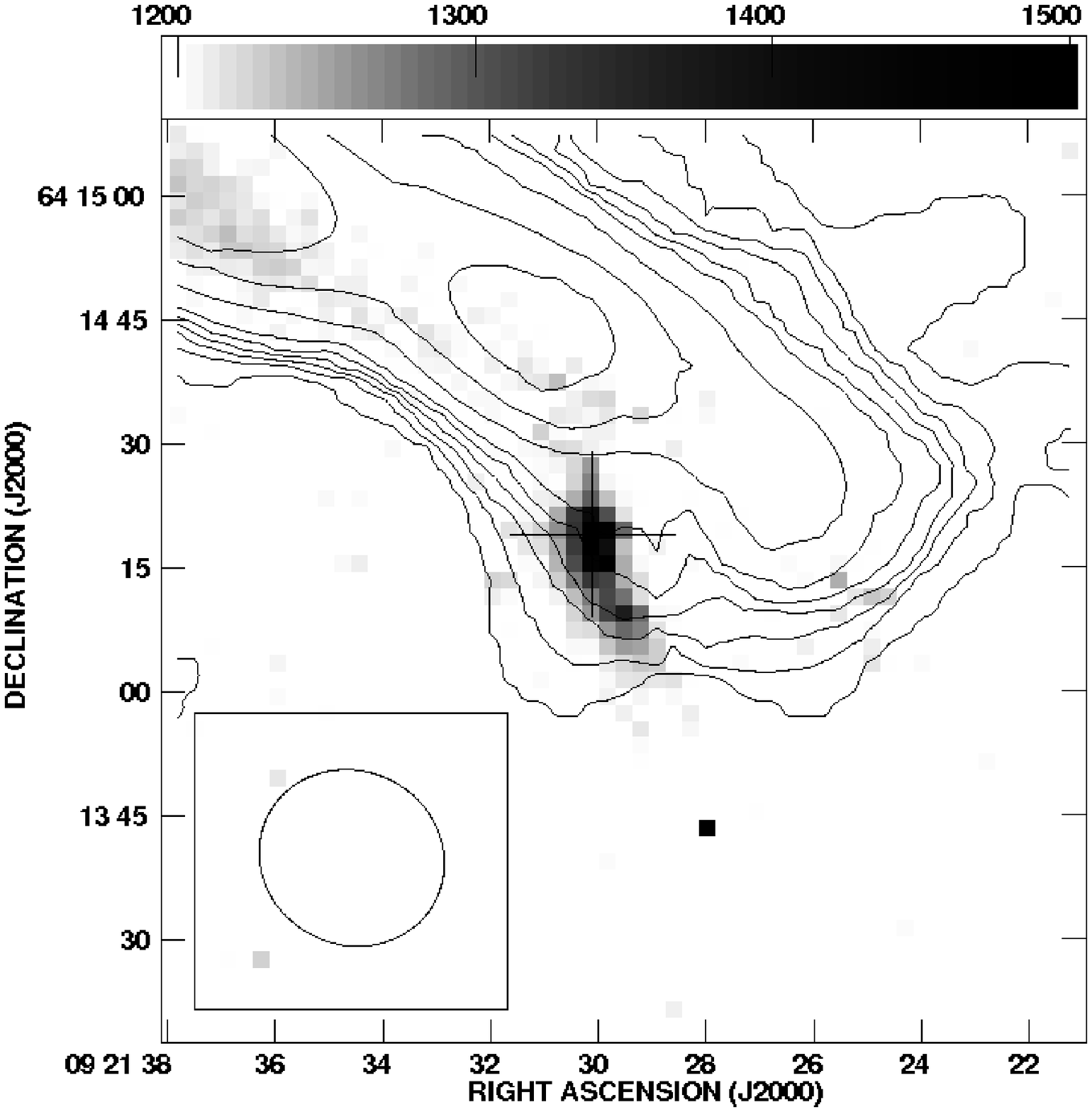}} 
\subfloat[][]{\includegraphics[width=5.5cm,height=5.5cm]{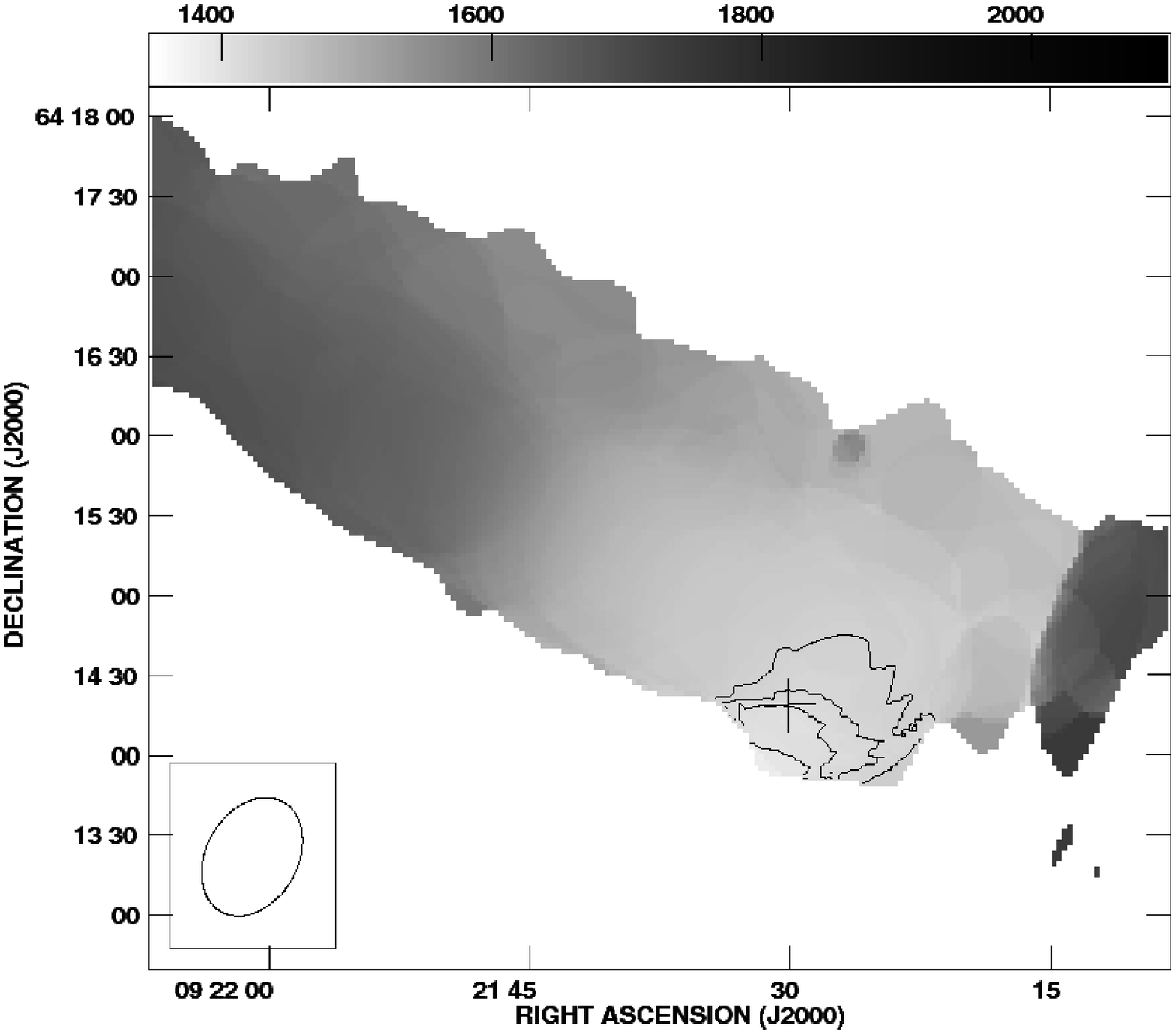}}
\qquad
\subfloat[][]{\includegraphics[width=5.5cm,height=5.5cm]{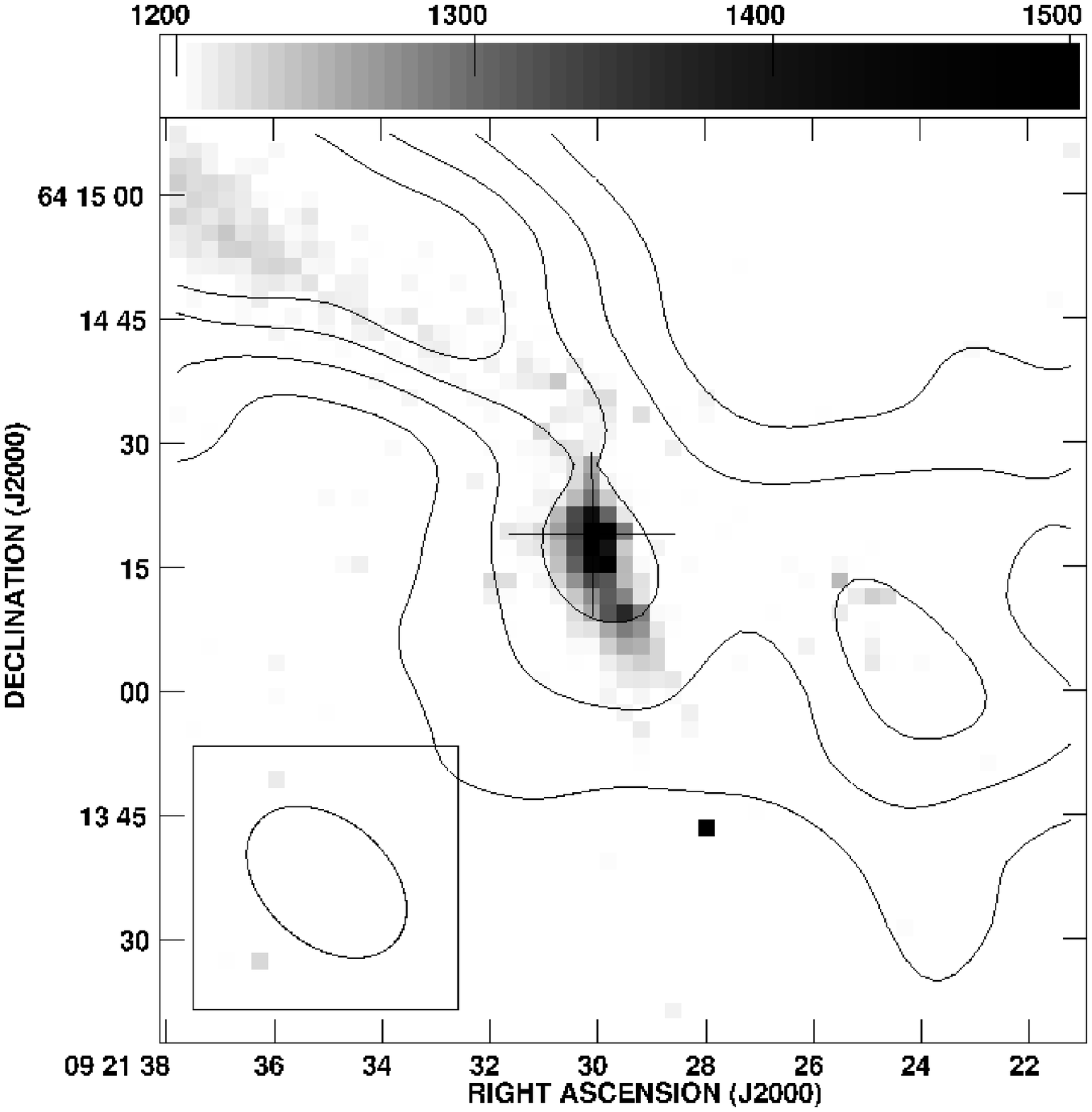}}
\subfloat[][]{\includegraphics[width=5.5cm,height=5.5cm]{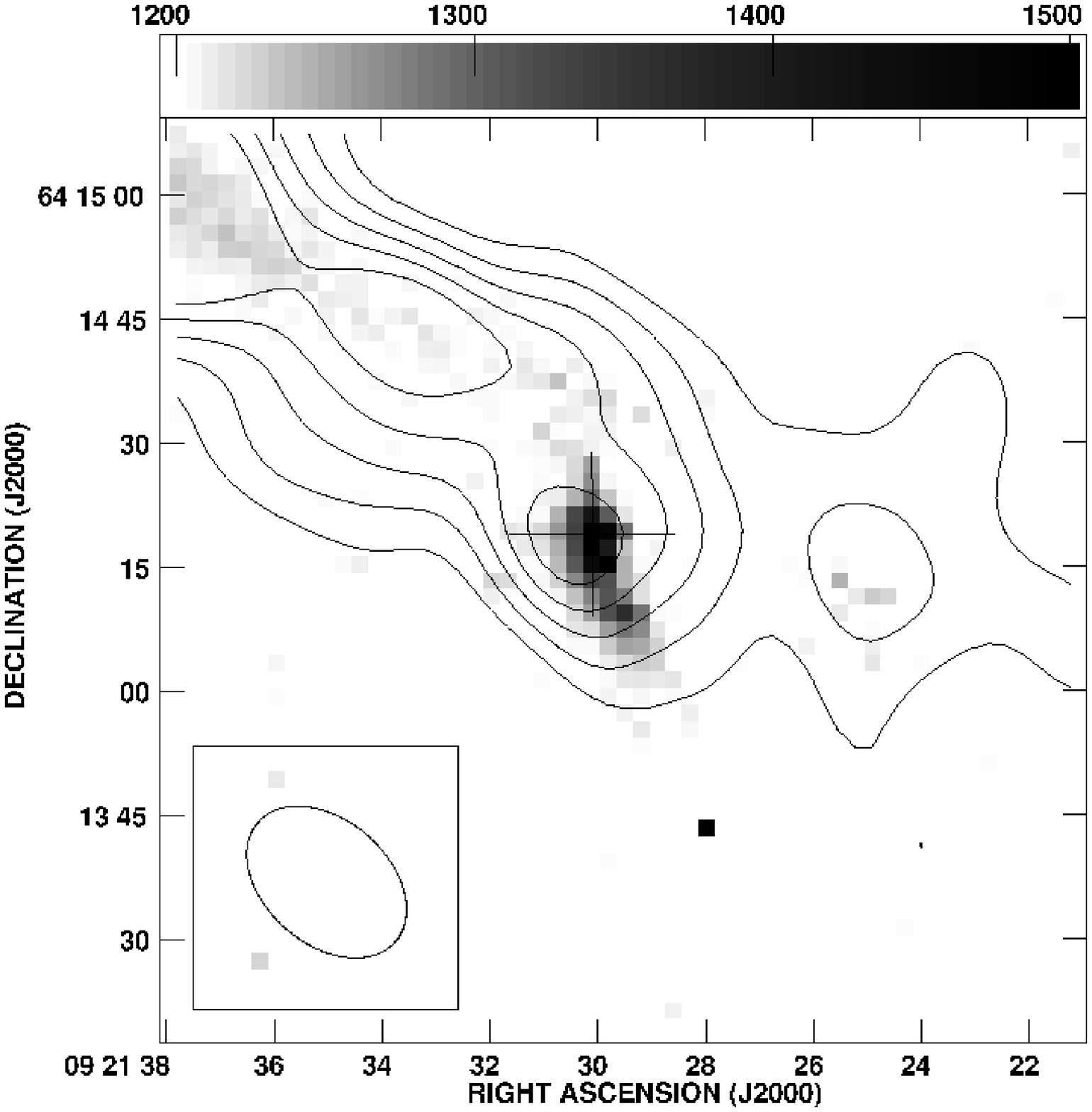}}
\subfloat[][]{\includegraphics[width=5.5cm,height=5.5cm]{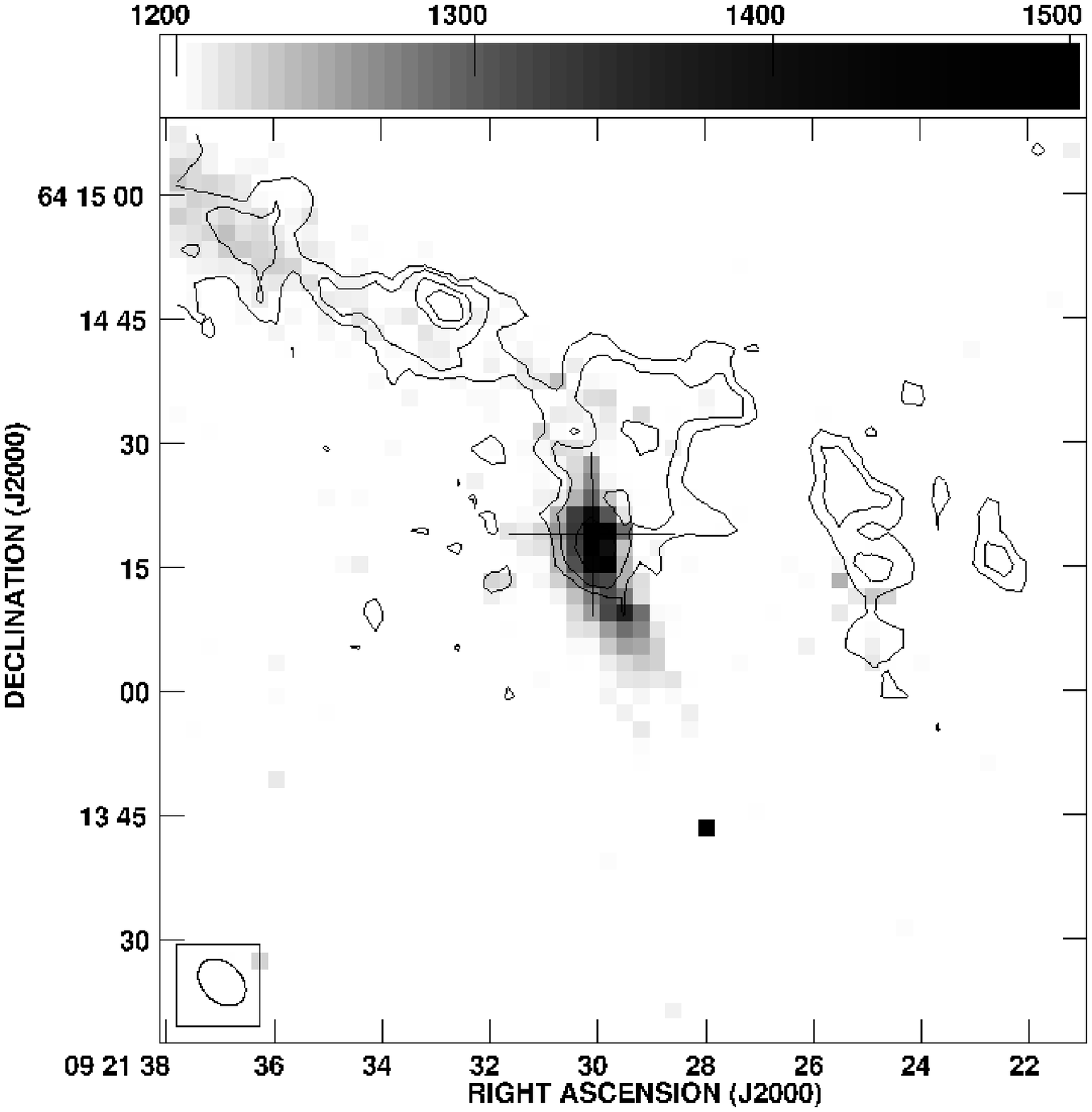}}
\caption[\footnotesize \hi and radio continuum images of Mrk 108]{\footnotesize {\bf Mrk 108 :} Figure {\bf (a)} shows the triplet in the group Holmberg 124. Contours are 325 MHz contours overlaid on $B$ band image. The cross marks the position of Mrk 108. The box drawn in the image represents the sizes of the image in Figures (b), (d), (e), (f). Figure {\bf (b)} represents \hi column density contours of Mrk 108 overlaid on optical DSS $B$ band  image. The edge on galaxy located on the eastern side is NGC 2820 (refer Figure (a)). Contours are plotted at (3, 5, 7, 9, 11) $\times10^{20}$ cm$^{-2}$. The angular resolution of the map is  $23\arcsec\times21\arcsec$. Figure {\bf (c)} shows the {\sc{moment 1}} map including the velocity fields of NGC 2820 (eastern edge) and NGC 2814 (western edge). Contours are plotted at 1390, 1400, 1410, 1420, 1430 \kms.  The kinematically distinct nature of Mrk 108 compared to NGC 2820 is visible.  This map is taken from \cite{nim05} and has an angular resolution of $47\arcsec\times34\arcsec$. {\bf (d)} shows the 325 MHz radio image of Mrk 108 zoomed in and overlaid on the optical $B$ band image.  Contour levels are $656\times$(-4, -3, 3, 4, 6, 8, 10) $\mu$Jy/beam. The resolution of the map is $22\arcsec\times15\arcsec$. {\bf (e)} shows the 610 MHz contours of Mrk 108 overlaid on the optical $B$ band image.  Contour levels are $245 \ \times$ (-6, -4, 4, 6, 8, 10, 12) $\mu$Jy/beam. The resolution of the map is $22\arcsec\times15\arcsec$. {\bf (f)} shows the 1280 MHz contours overlaid on optical DSS image. Levels are $90 \ \times$ (-6, -3, 3, 4, 6, 8) $\mu$Jy/beam. Resolution of the maps is $6\arcsec\times5\arcsec$. Maps are taken from \cite{nim05}.} 
\label{fig:mrk108_r}
\end{figure}


}

\begin{landscape}
\begin{table*}[h!]\footnotesize
\begin{center}
\caption{\footnotesize The observation and results of \hi for the four galaxies observed using GMRT.}
\label{tab:obs_hi}
\begin{tabular}{lllll}
 \\
\hline\hline
 \textbf{parameter} & \multicolumn{4}{c}{\textbf{Galaxy}} \\
\hline
--- & \textbf{Mrk 104} & \textbf{Mrk 108} & \textbf{Mrk 1039} & \textbf{Mrk 1069} \\
\hline \\ 
\textbf{Date of obs} & 21/06/2008 & 28/10/2002 & 13/07/2009 & 31/05/2009 \\
\textbf{Bandwidth} (MHz/\kms) & 4/858 & 4/858 & 4/858 & 4/858 \\
\textbf{Channelwidth} (KHz/\kms) & 31.25/6.7 & 31.25/6.7 & 31.25/6.7 & 31.25/6.7 \\
\textbf{on-source time} (hr) &	7.5 & 7.5 & 7.5 & 7.5 \\
\textbf{\hi highest resolution}$^a$ (arcsec)& $21''\times18''$ & $11''\times10''$ & $13''\times10''$ & $21''\times13''$ \\
\textbf{RMS in the highest resolution map} (mJy) & 1.1 & 1.2 & 1.9 & 2.1 \\
\textbf{line width at 50\% peak} & $68.5\pm1.6$ & $51.6\pm1.9$ & $124.0\pm1.1$ & $61.5\pm0.1$ \\
\textbf{line width at 20\% peak} & 152.6 & 190.4 & 197.5 & 104.8 \\
\textbf{Total flux} (Jy \kms) & $0.97\pm0.05$ & $1.42\pm0.41$ & $6.53\pm0.10$ & $2.75\pm0.20$ \\
\textbf{$D_{\rm \hi}/D_{25}$} $^b$  & 2.66 & 2.13 & 1.1 & 5.43 \\
\textbf{M(\hi)} ($M_\odot$) & $2.17\times10^8$ & $1.6\times10^8 \ ^d$ & $1.25\times10^9$ & $2.71\times10^8$ \\
\textbf{M(*)} ($M_\odot$) $^c$ & $1.4\times10^9$ & $3.1\times10^8$ & $2.2\times10^9$ & $2.13\times10^9$ \\
\textbf{M(dyn)} ($M_\odot$)$^e$ & $1.17\times10^9$ & $9.4\times10^8$ & $4.7\times10^9$ & $5.3\times10^9$ \\
\textbf{M(\hi)/L$_B$} ($M_\odot/L_\odot$) & 0.15 & 0.27 & 0.30 & 0.20 \\
\textbf{M(\hi)/L$_K$} ($M_\odot/L_\odot$) & 0.12 & 0.41 & 0.40 & 0.10 \\
\textbf{M(dyn)/L$_K$} ($M_\odot/L_\odot$) & 0.83 & 2.86 & 1.47 & 1.99 \\
\hline\hline
\end{tabular}
\end{center}
$^a$ - gives the details of highest resolution map that we have used. \\
$^b$ - Except for the galaxy Mrk 1069, the optical diameter ($D_{25}$) is measured at 25 mag arcsec$^{-2}$ from \cite{rc391}. 
For Mrk 1069, the optical diameter is taken from $K$ band image as given in NED.\\
$^c$ - We adopt the relation $M_*/L_K\sim0.8$ ($M/L_{K}$)$_\odot$ given by \cite{kir08} for galaxies with $B -R\sim2$ \citep{vad07}, 
the average colour of the dwarfs. The $K$ band luminosity is estimated from 2 MASS magnitudes.\\
$^d$ - This is lower by an order of magnitude compared to \cite{thuan81} who                      used single dish observations and 
hence included contribution from the massive neighbouring spiral NGC 2820. Our interferometric estimate is a 
better estimate of the mass of this galaxy. \\
$^e$ - M(dyn) is estimated using the simple assumption of circular orbits i.e. using $M = {\frac{v^2~R}{G}}$. \\
\end{table*}
\end{landscape}

\item{Mrk 1039 : The \hi intensity and kinematics are shown in Figure \ref{fig:mrk1039_r}(a),(b).  A warped gas disk is visible in Figure \ref{fig:mrk1039_r}(a) whereas the stellar disk does not show any signature of the warp.  The velocity field (Figure \ref{fig:mrk1039_r}(b)) shows the rotating gas disk with  small distortions in the outer parts especially in the western half. Such small kinks and bends were also noticed by \cite{thuan04}. The velocity dispersion of Mrk 1039 varies from $\sim15$ \kms \ at the edges of the galaxy to 45 \kms \ in the central regions. The \hi mass is estimated to be $1.25\times10^9\, M_\odot$ (Table \ref{tab:obs_hi}) which is in good agreement with the mass estimated from single dish measurements. The \hi disk is about $\sim1.1$ times the ($D_{25}$) optical size of the galaxy. Figure \ref{fig:mrk1039_r}(c) shows the \hi column density map at a resolution of $13\arcsec\times10\arcsec$. The smooth \hi distribution seen in the lower resolution image is resolved into several high column density clumps, noticeably the clouds on the eastern and western edges. Figure \ref{fig:mrk1039_r}(d) shows the position-velocity diagram along the major axis ($90\arcdeg$ east of north). The disk seems to execute solid body rotation till the outskirts of the galaxy. Figure \ref{fig:mrk1039_r}(e) shows the \hi \ column density (high resolution) contours overlaid on the H$\alpha$ images taken from \cite{ram10}. It is seen that the high column density \hi gas overlaps with the strong \hii regions in the eastern side of the galaxy, from where radio continuum emission is also detected.

\begin{figure}[h!]\footnotesize
\centering
\subfloat[][]{\includegraphics[width=5.5cm,height=5.5cm]{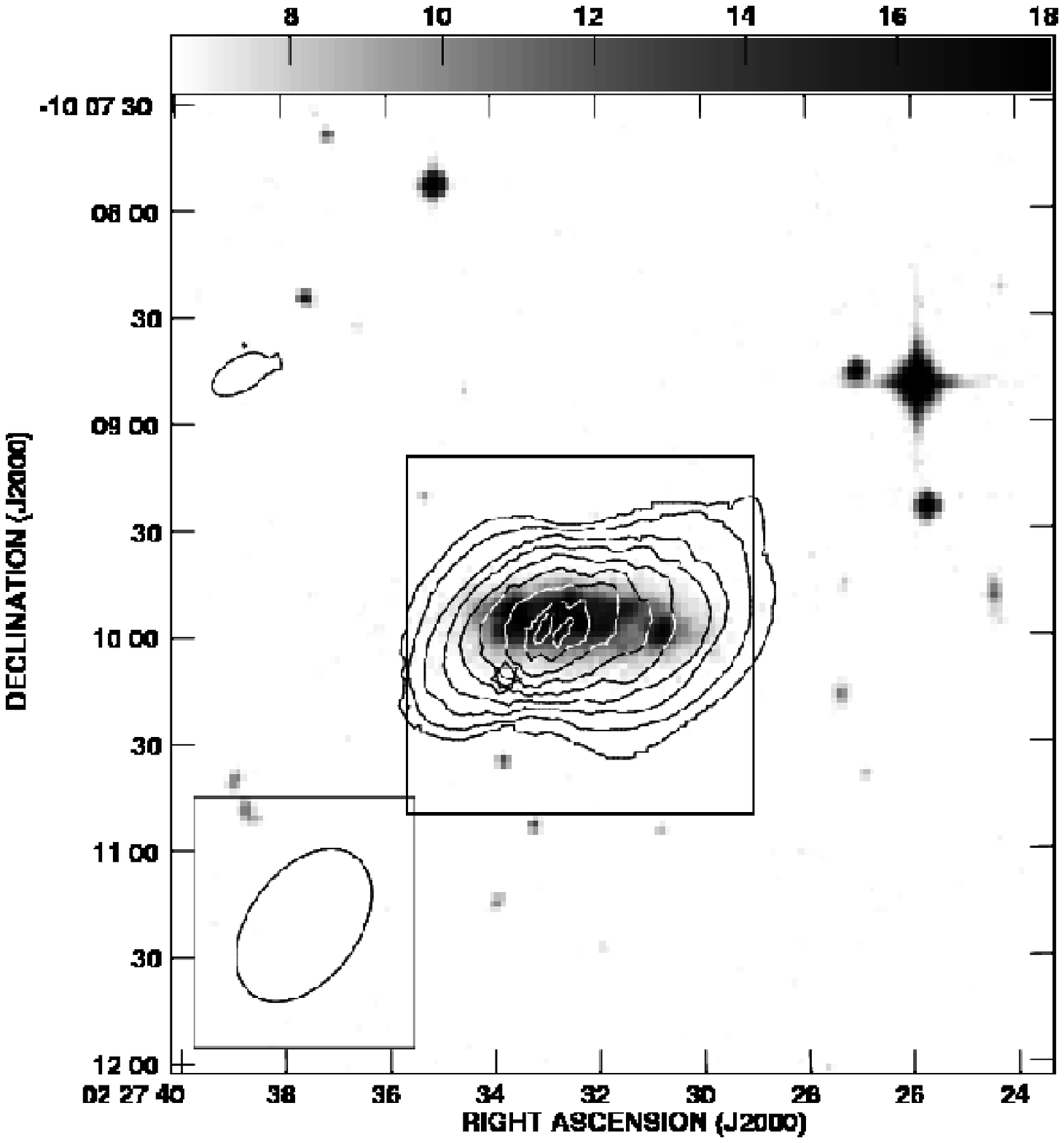}}
\subfloat[][]{\includegraphics[width=5.5cm,height=5.5cm]{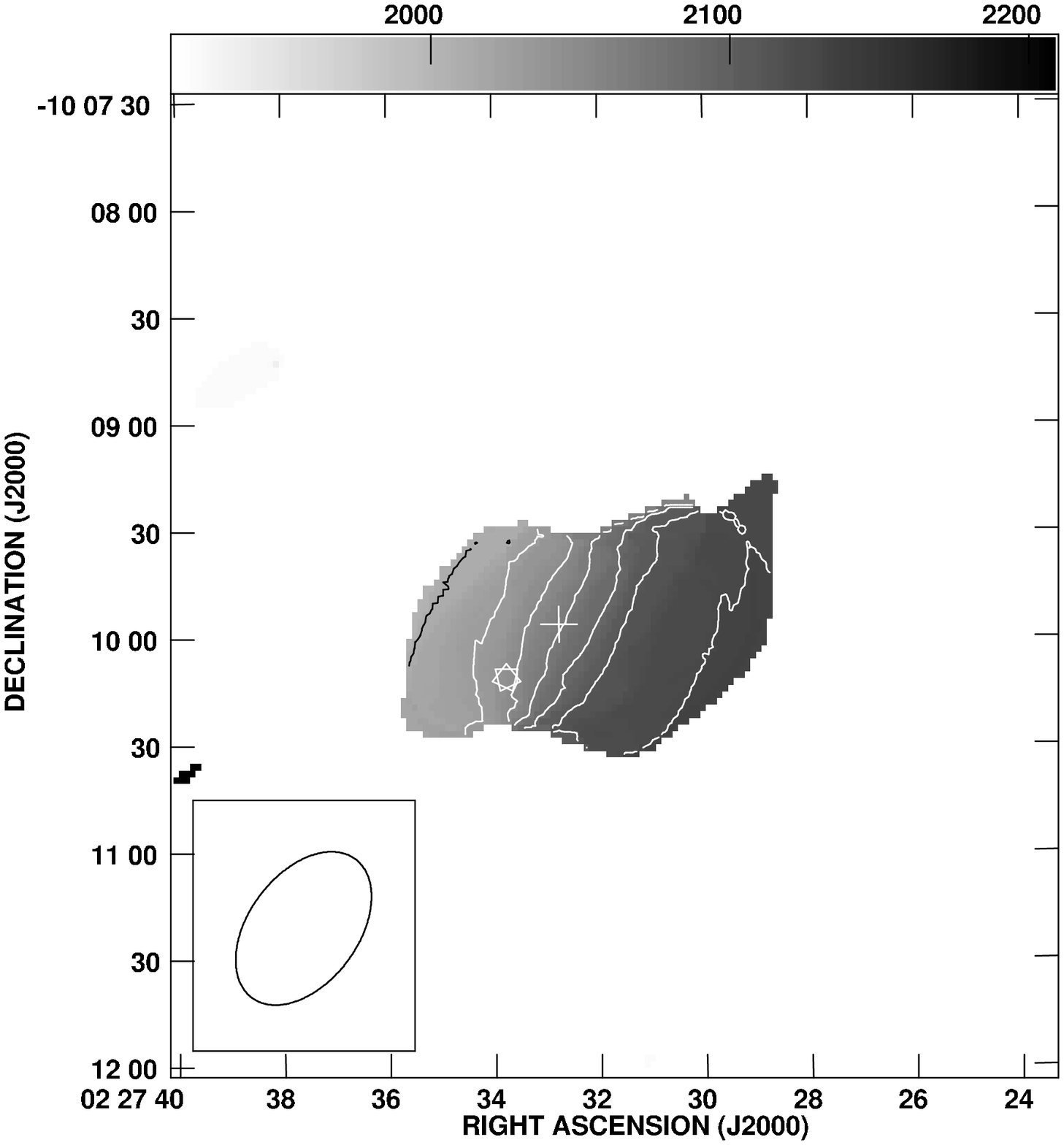}}
\subfloat[][]{\includegraphics[width=5.5cm,height=5.5cm]{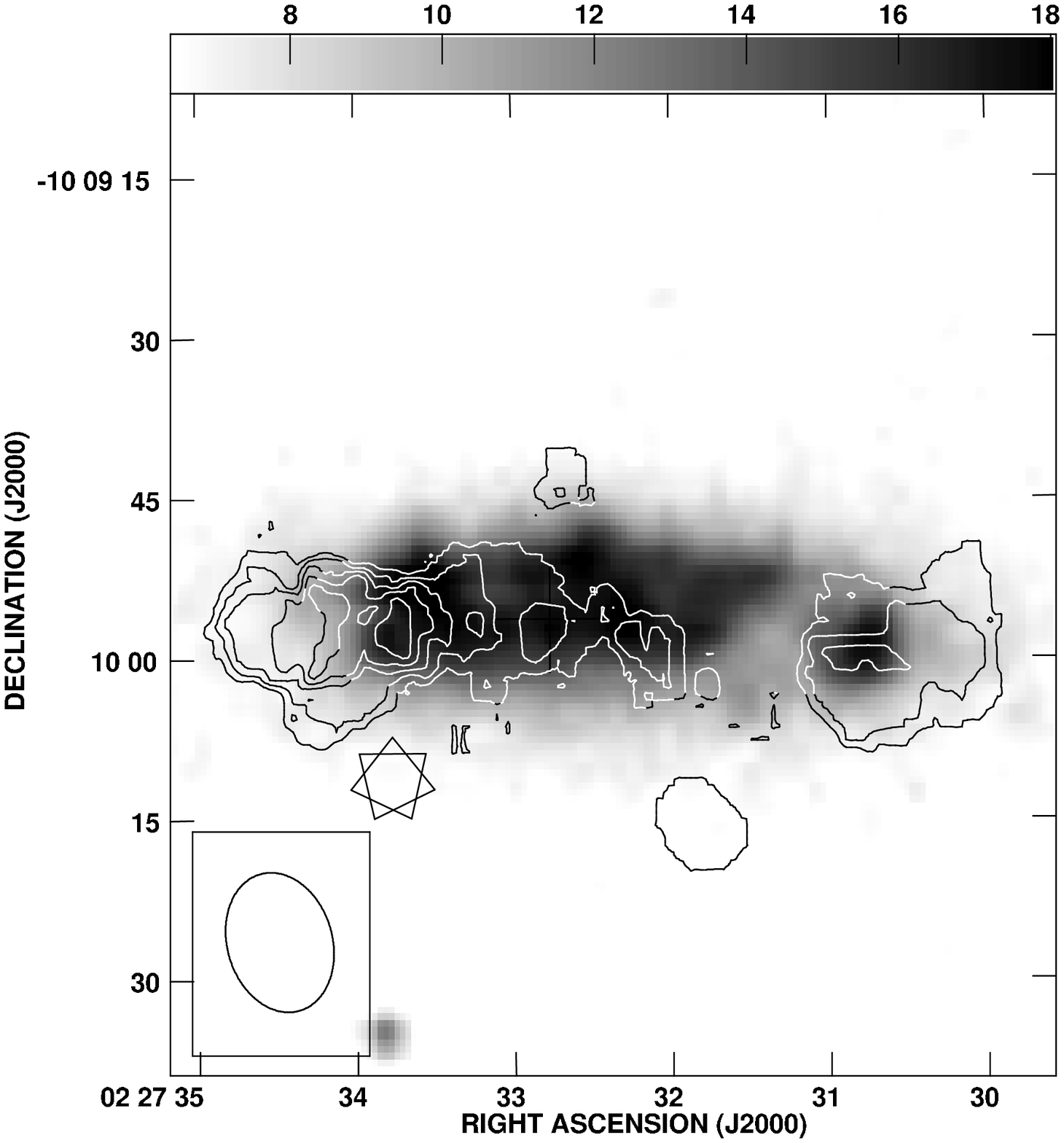}}
\qquad
\subfloat[][]{\includegraphics[width=5.5cm,height=5.5cm]{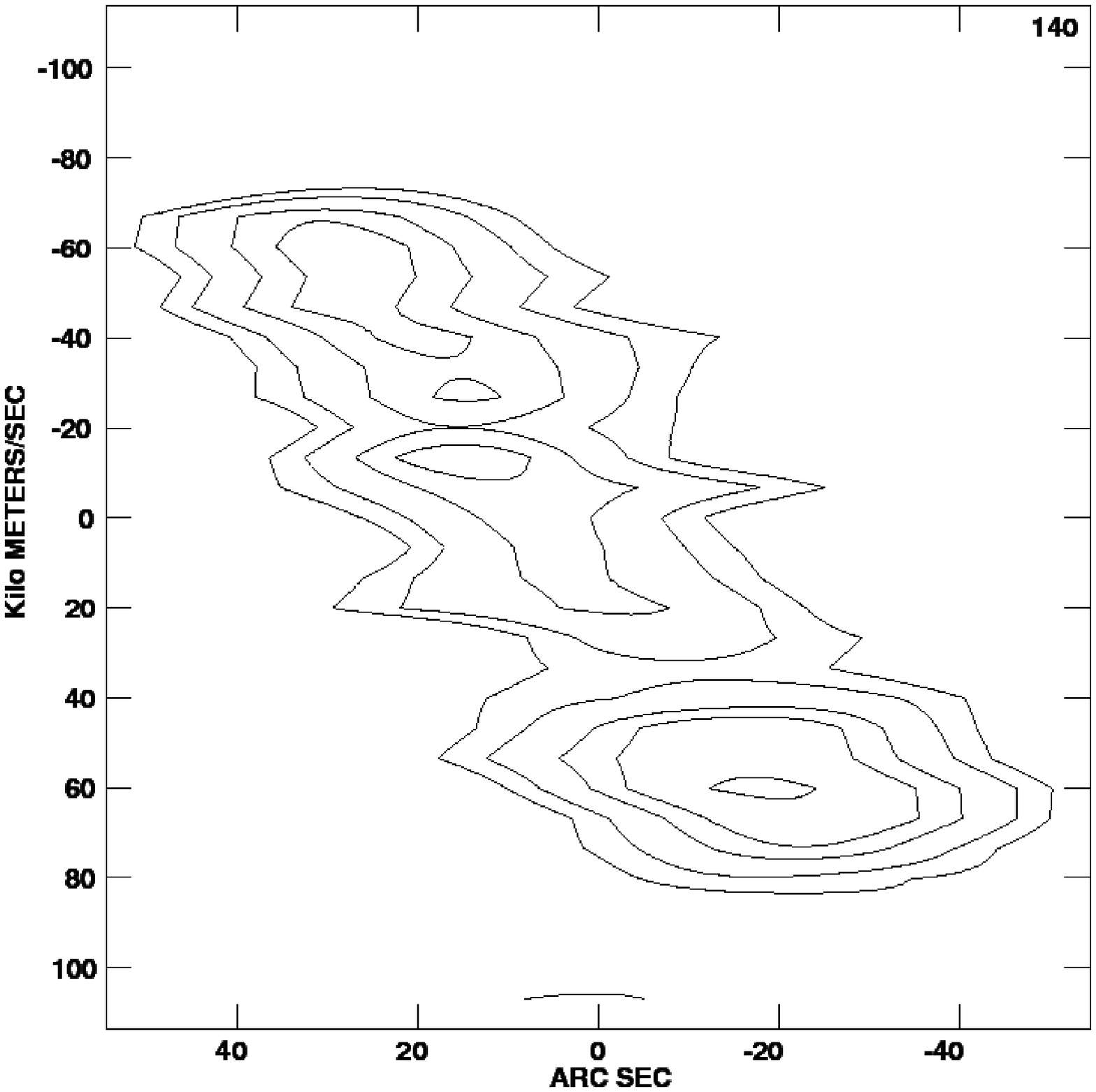}}
\subfloat[][]{\includegraphics[width=5.5cm,height=5.5cm]{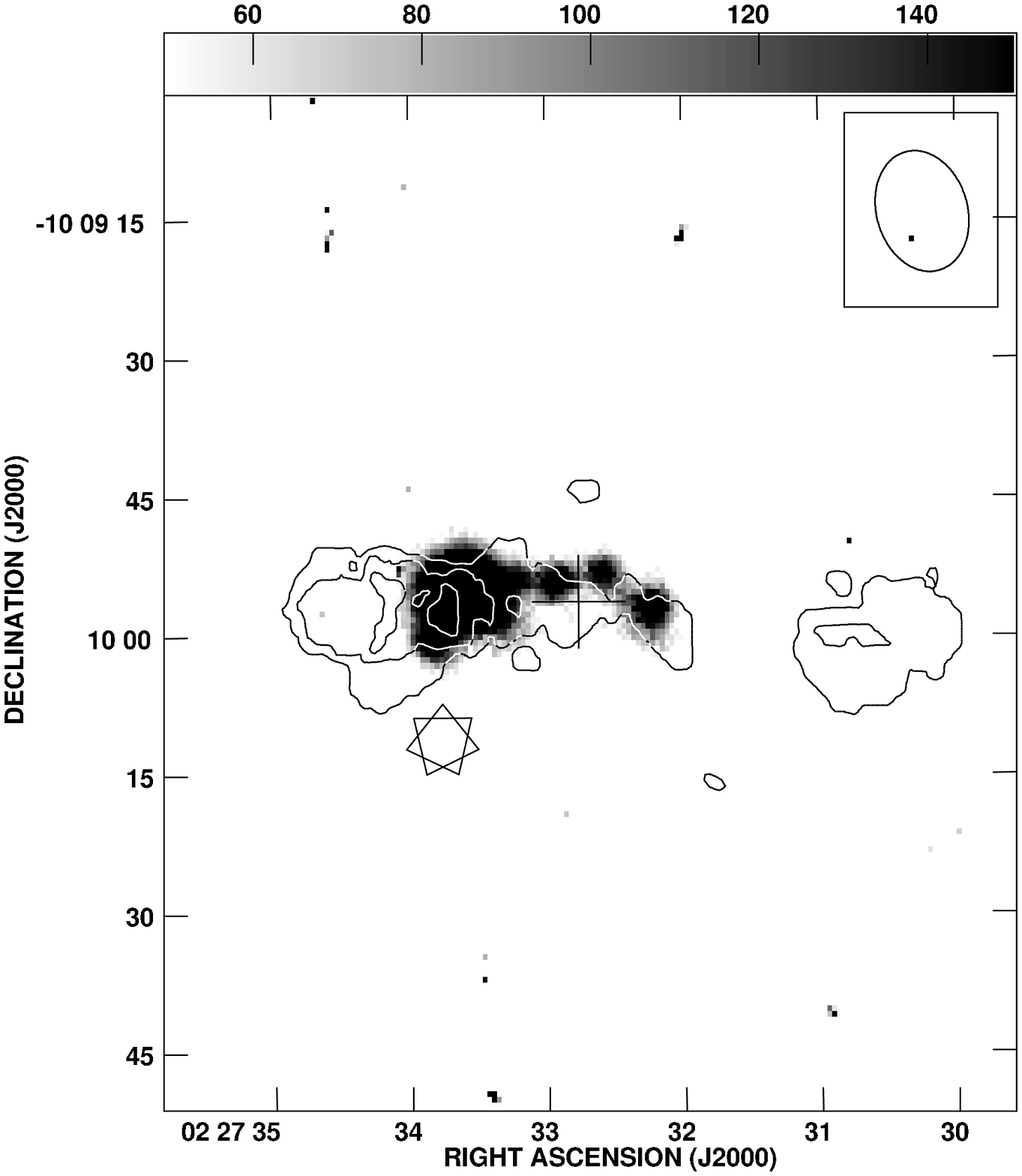}}
\caption{ \footnotesize {\bf Mrk 1039 :} {\bf (a)} The \hi column density contours at (2, 4, 6, 8, 10, 12, 14, 16) 
$\times 10^{20}$ cm$^{-2}$ are overlaid on DSS $B$ band image. Angular resolution of the map is 
$48\arcsec\times31\arcsec$. The \hi disk is warped. The \hi emission ($D_{\textrm{H\sc~i}}$) is $\sim1.1$ 
times the optical size ($D_{25}$) of the galaxy. The box drawn in the picture represents the size of the images in 
Figures (c)-(k). {\bf (b)} The velocity field of Mrk 1039 is shown in this figure. Contours are plotted at 2020, 2040, 2060, 2080, 2100, 2120 \kms.  
{\bf (c)} The high resolution ($13\arcsec\times10\arcsec$) \hi column density map of Mrk 1039.  The \hi gas resolves into
 several clouds at the higher resolution. Contours are drawn at 4, 8, 12, 16, 20 $\times10^{20}$ cm$^{-2}$.
{\bf (d)} Position-velocity curve along the major axis ($90\arcdeg$ east of north) of the galaxy. The contours are 
plotted at $3.1\times$(4, 5.5, 6.5, 8, 9) mJy/beam. This P-V diagram is created using the lowest resolution
 ($48\arcsec\times31\arcsec$) \hi cube and shows the solid body rotation in the galaxy.
{\bf (e)} The high resolution \hi column density contours are overlaid on the H$\alpha$ image taken from
 \cite{ram10}. The levels are plotted at (5, 12, 20) $\times10^{20}$ cm$^{-2}$. The highest column density 
(2 $\times10^{21}$ cm$^{-2}$) \hi is coincident with the strong H$\alpha$ emitting \hii region in the east.}
\label{fig:mrk1039_r}
\end{figure}

\begin{figure}[h!]\footnotesize\ContinuedFloat
\centering
\subfloat[][]{\includegraphics[width=5.5cm,height=5.5cm]{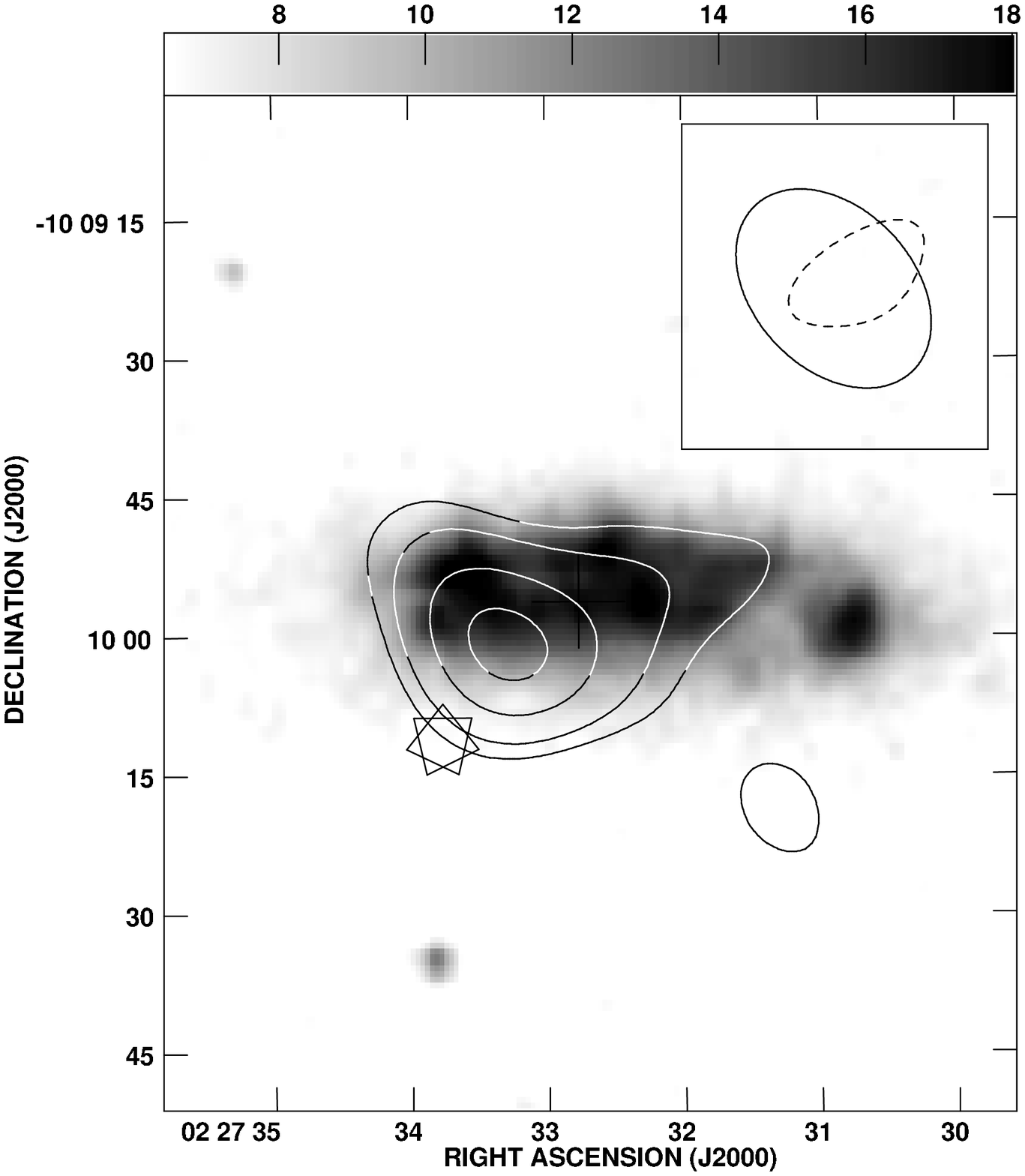}}
\subfloat[][]{\includegraphics[width=5.5cm,height=5.5cm]{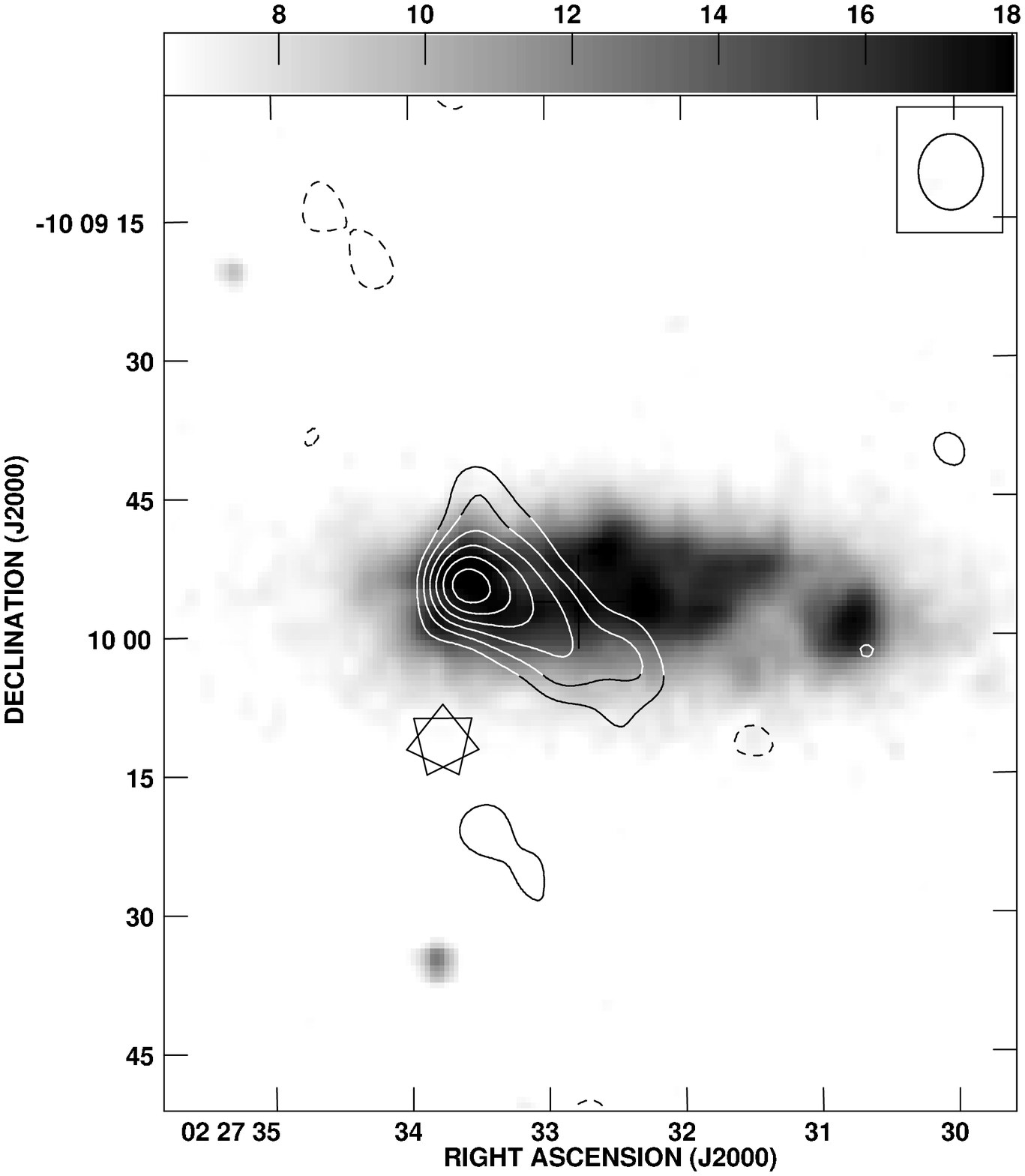}}
\subfloat[][]{\includegraphics[width=5.5cm,height=5.5cm]{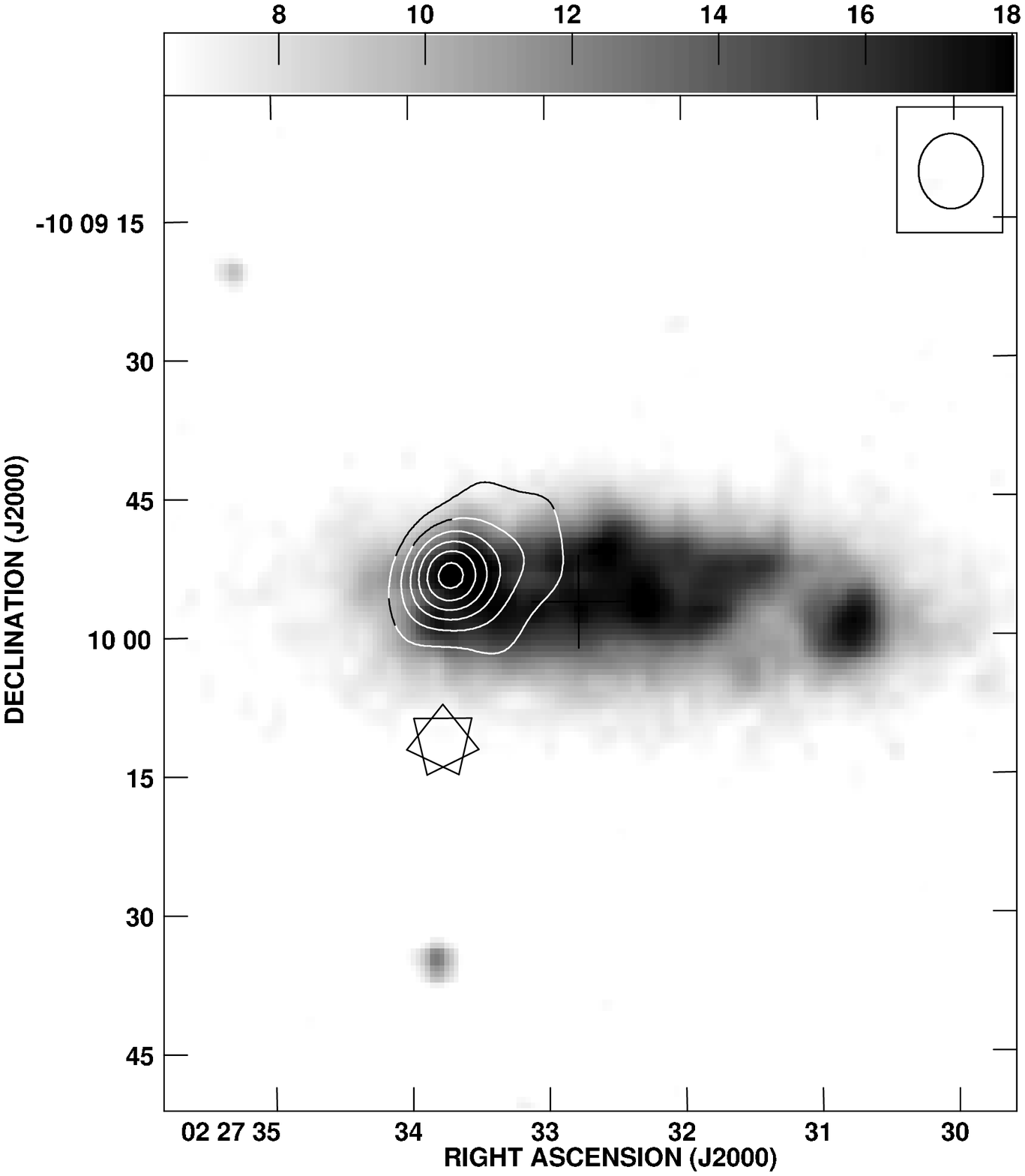}}
\qquad
\subfloat[][]{\includegraphics[width=5.5cm,height=5.5cm]{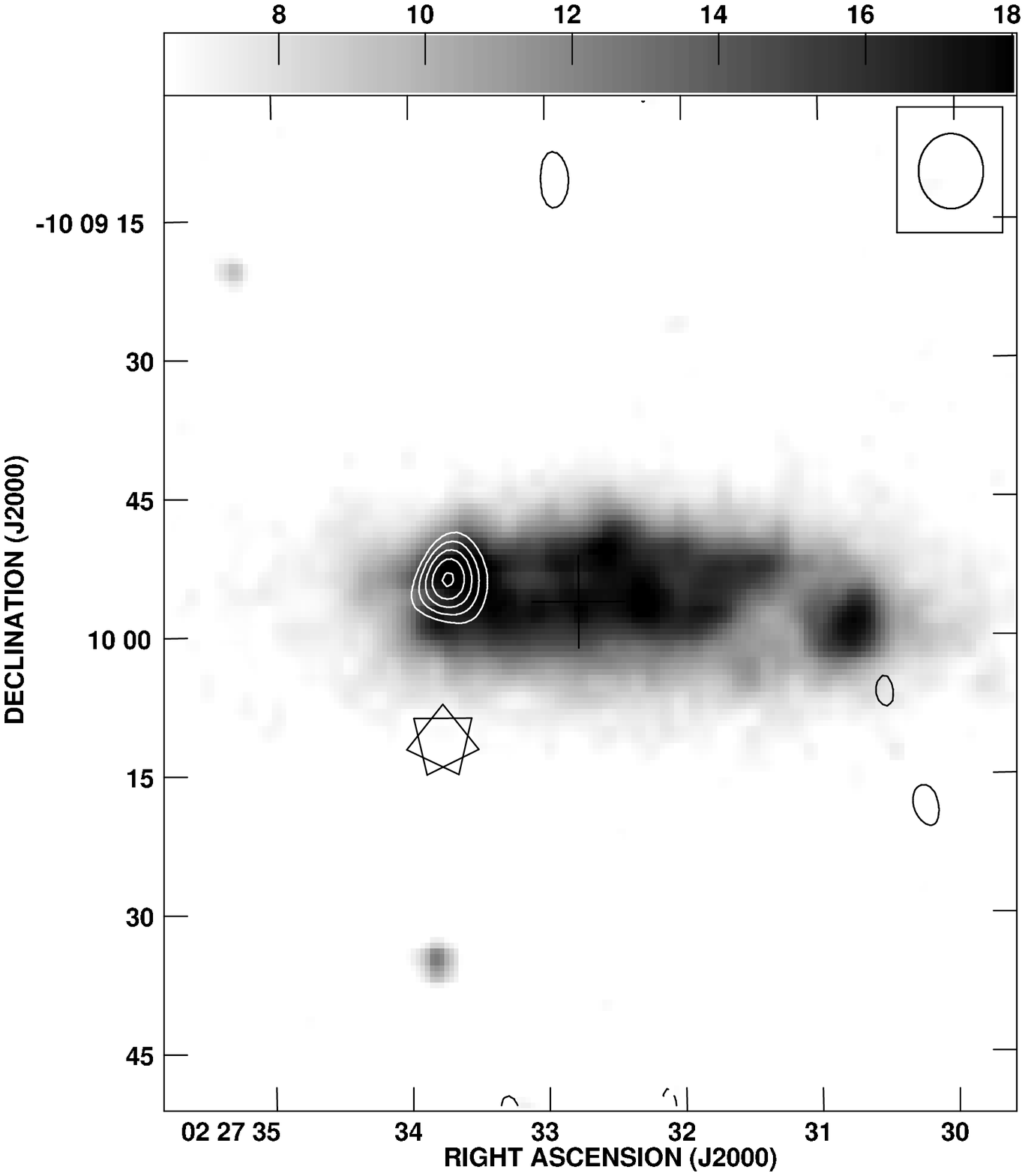}}
\subfloat[][]{\includegraphics[width=5.5cm,height=5.5cm]{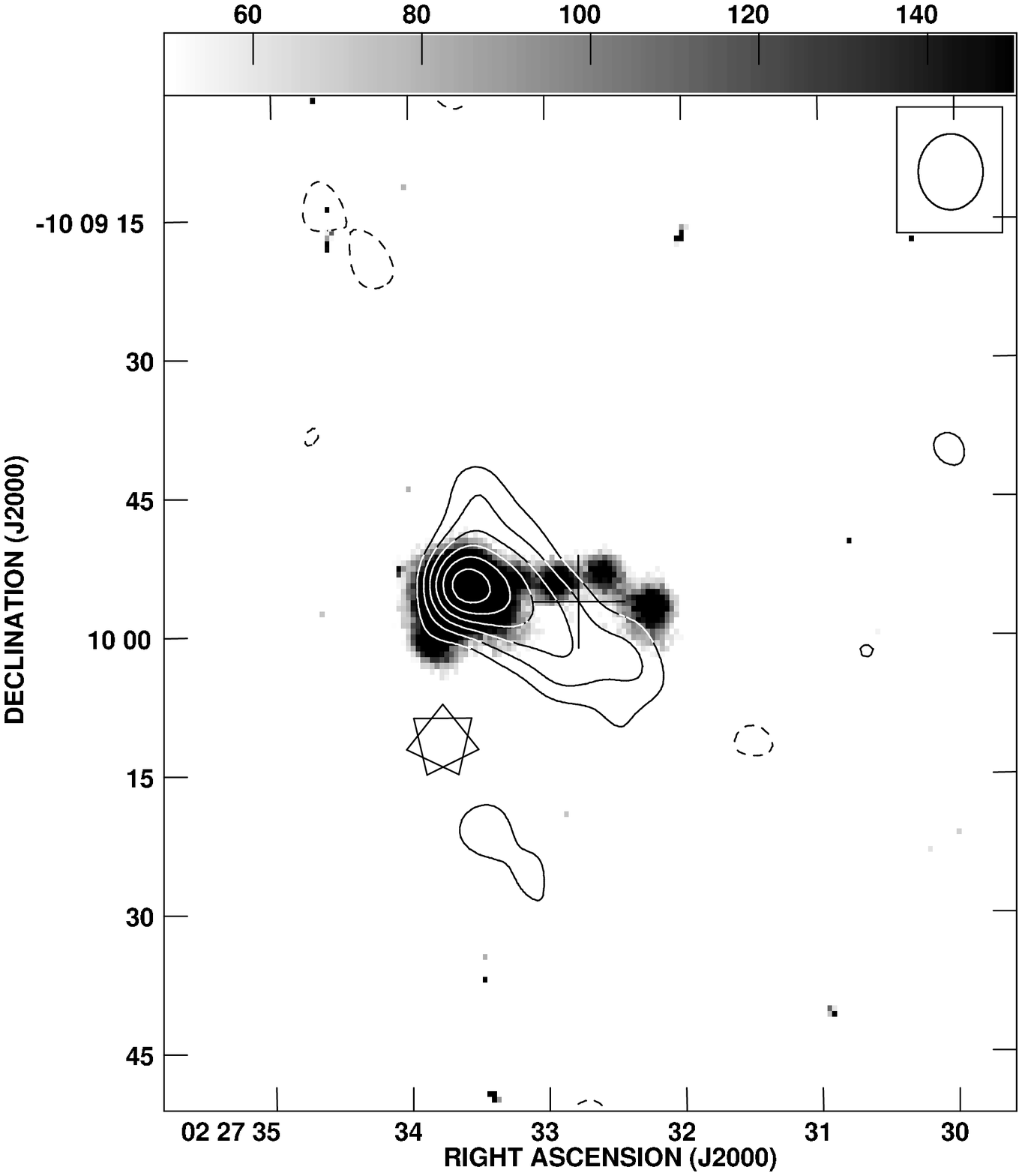}}
\subfloat[][]{\includegraphics[width=5.5cm,height=5.5cm]{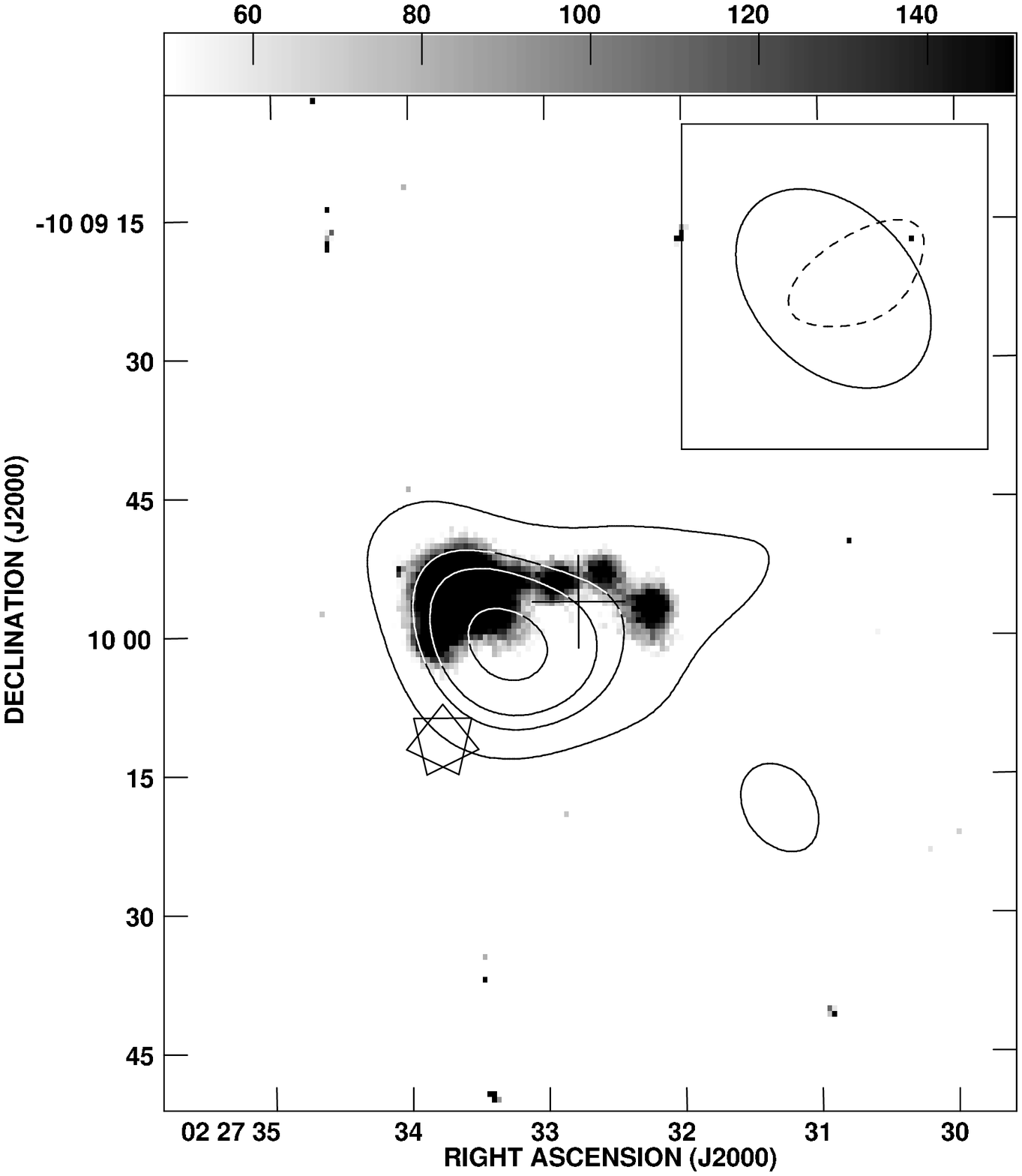}}\\
\caption{\footnotesize {\bf Mrk 1039 {\it Continued:}}  Figure {\bf (f)} shows 325 MHz contours overlaid on the optical DSS image.  Contour levels are 
$1.2 \ \times$ (-4, -3, 3, 4, 6, 8) mJy/beam. The resolution of the map is $25\arcsec\times17\arcsec$.
{\bf (g)} shows 610 MHz contours overlaid on the optical DSS image.  Contour levels are $165 \ \times$ (-6, -4, 4, 6, 8,
 10, 12, 14, 18) $\mu$Jy/beam. The resolution of the map is $8\arcsec\times6\arcsec$.
{\bf (h)} shows the 1.4 GHz contours obtained from VLA archival data overlaid on the optical DSS image. Levels are 
$84.0 \ \times$ (-8, -4, 4, 8, 12, 16, 20, 24) $\mu$Jy/beam. The resolution of the map is $8\arcsec\times7\arcsec$. 
{\bf (i)} shows the 14.9 GHz contours obtained from VLA archival data overlaid on the 
optical DSS image. Levels are $207.0 \ \times$ (-4, -3, 3, 4, 5) $\mu$Jy/beam. The emission is localised being
 coupled to a star-forming region in the east.  Note that a \hi cloud is also seen to be coincident with this 
region. Angular resolution of this map is $8\arcsec\times7\arcsec$.
 {\bf (j)} Shows the 610 MHz contours overlaid on H$\alpha$ image taken from \cite{ram10}. The contour levels are 
$165 \ \times$ (-6, -4, 4, 6, 8, 10, 12) $\mu$Jy/beam. 
{\bf (k)} Shows the 325 MHz contours overlaid on H$\alpha$ image taken from \cite{ram10}. The 
contour levels are $1.2 \ \times$ (-4, -3, 3, 5, 6, 8) mJy/beam. The radio emission encompasses all the \hii 
regions seen in the H$\alpha$ image. The star marked in all the images towards south-west direction reperesents 
the location of the SN 1985S.}
\end{figure}

The continuum maps at 325, 610 MHz (GMRT data), 1.4 and 14.9 GHz (from VLA archival data) are shown in 
the Figures \ref{fig:mrk1039_r}f--i. The radio emission appears to be coupled to the bright \hii region near the 
eastern edge of the galaxy. Note that the emission at 610 MHz is more extended compared to that at 14.9 GHz 
indicating the presence of non-thermal emission. 
Figures \ref{fig:mrk1039_r}(j) and (k) show 325 and 610 MHz contours overlaid on the H$\alpha$ image shown in 
\cite{ram10}. This galaxy hosted a type II supernova SN 1985S close to the region from where we detect radio 
continuum emission (marked by a star in Figure \ref{fig:mrk1039_r}). We do not detect the galaxy at 240 MHz
 with a $3\sigma$ limit of 5 mJy. The integrated spectral index between 610 MHz and
 1.4 GHz is $\alpha\sim-0.8$ and $\alpha(325,610)=+0.48$. }

\item{Mrk 1069 : The \hi moment maps are shown in Figure \ref{fig:mrk1069_r}(a),(b).  This galaxy, classified as a 
Sa type by Hyperleda\footnote{http://leda.univ-lyon1.fr} \citep{pat03} shows a large \hi disk (see Figure 
\ref{fig:mrk1069_r}(a)) about 6 times (Table \ref{tab:obs_hi}) the optical size of the galaxy. The \hi disk shows
 a slight warp in the northern parts.  The southern part of the disk appears to be abruptly truncated. The galaxy 
shows rotation (Figure \ref{fig:mrk1069_r}(b)). A slight offset between the optical centre and the kinematic 
centre is noticed in this galaxy similar to Mrk 104. We estimate a \hi mass of $2.7\times10^8\, M_\odot$ (Table
 \ref{tab:obs_hi}) which is lower by a factor of about 3 compared to the single dish estimate. Higher resolution 
($21\arcsec\times13\arcsec$) column density distribution is shown in Figure \ref{fig:mrk1069_r}(c). Figure
 \ref{fig:mrk1069_r}(d) shows the higher resolution column density map overlaid on the H$\alpha$ images of 
\cite{ram10}. There appears to be an offset between the higher column density 
\hi peaks and the H$\alpha$ peaks. The position-velocity curve (refer Figure \ref{fig:mrk1069_r}(e)) along the 
major axis of the galaxy shows rotation.

\begin{figure}[h!]\footnotesize
\begin{center}
\subfloat[][]{\includegraphics[width=5.5cm,height=5.5cm]{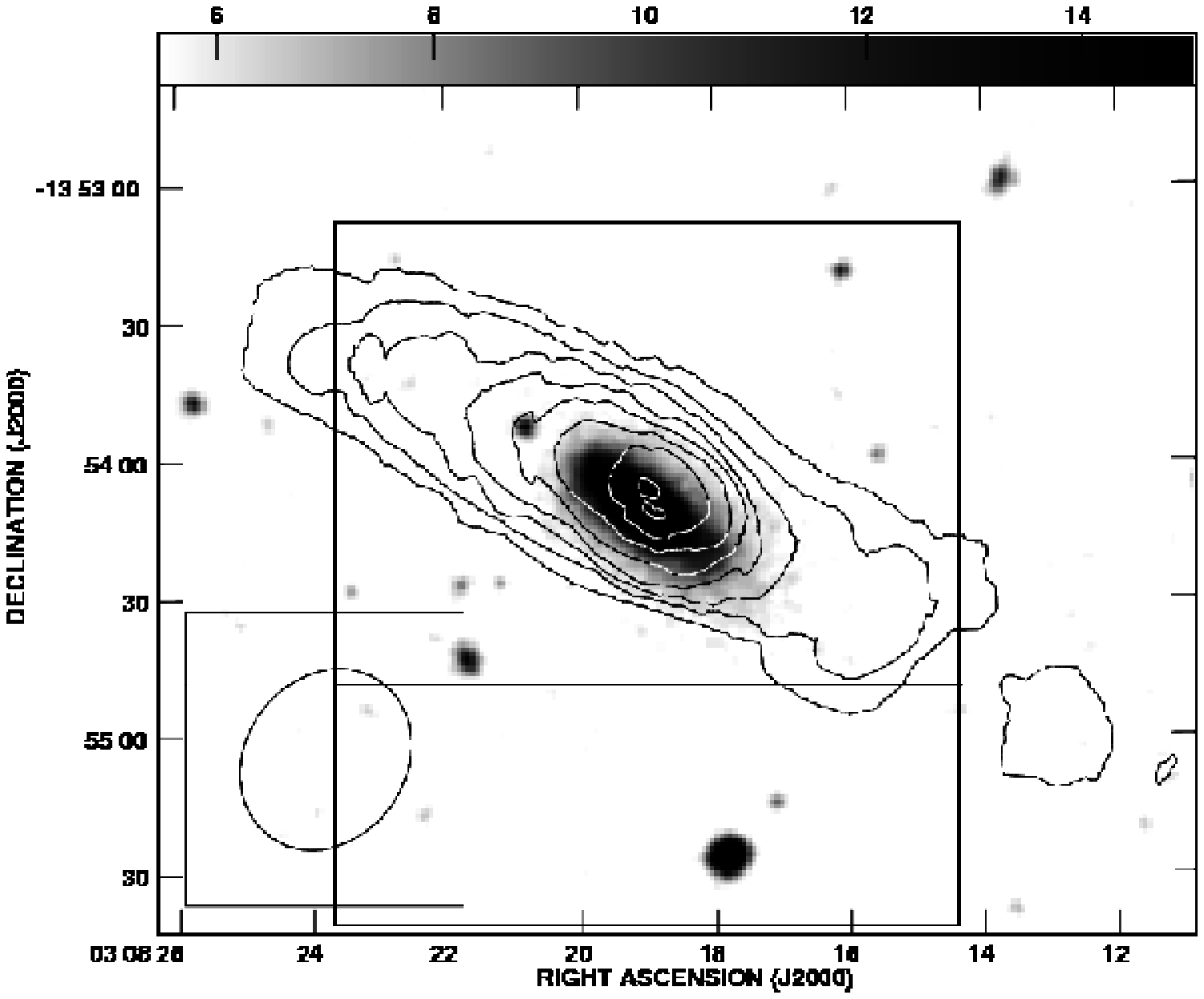}}
\subfloat[][]{\includegraphics[width=5.5cm,height=5.5cm]{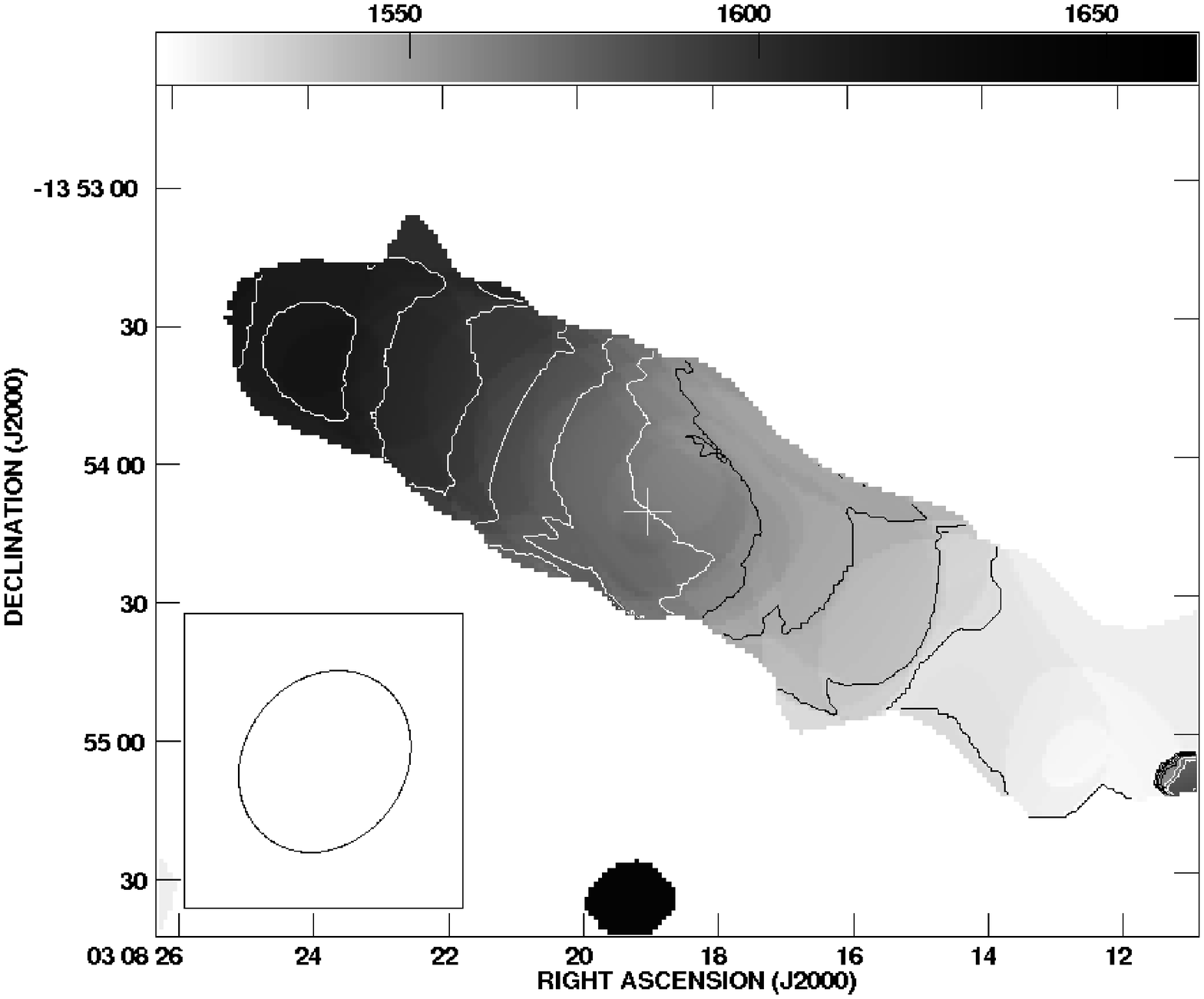}}
\subfloat[][]{\includegraphics[width=5.5cm,height=5.5cm]{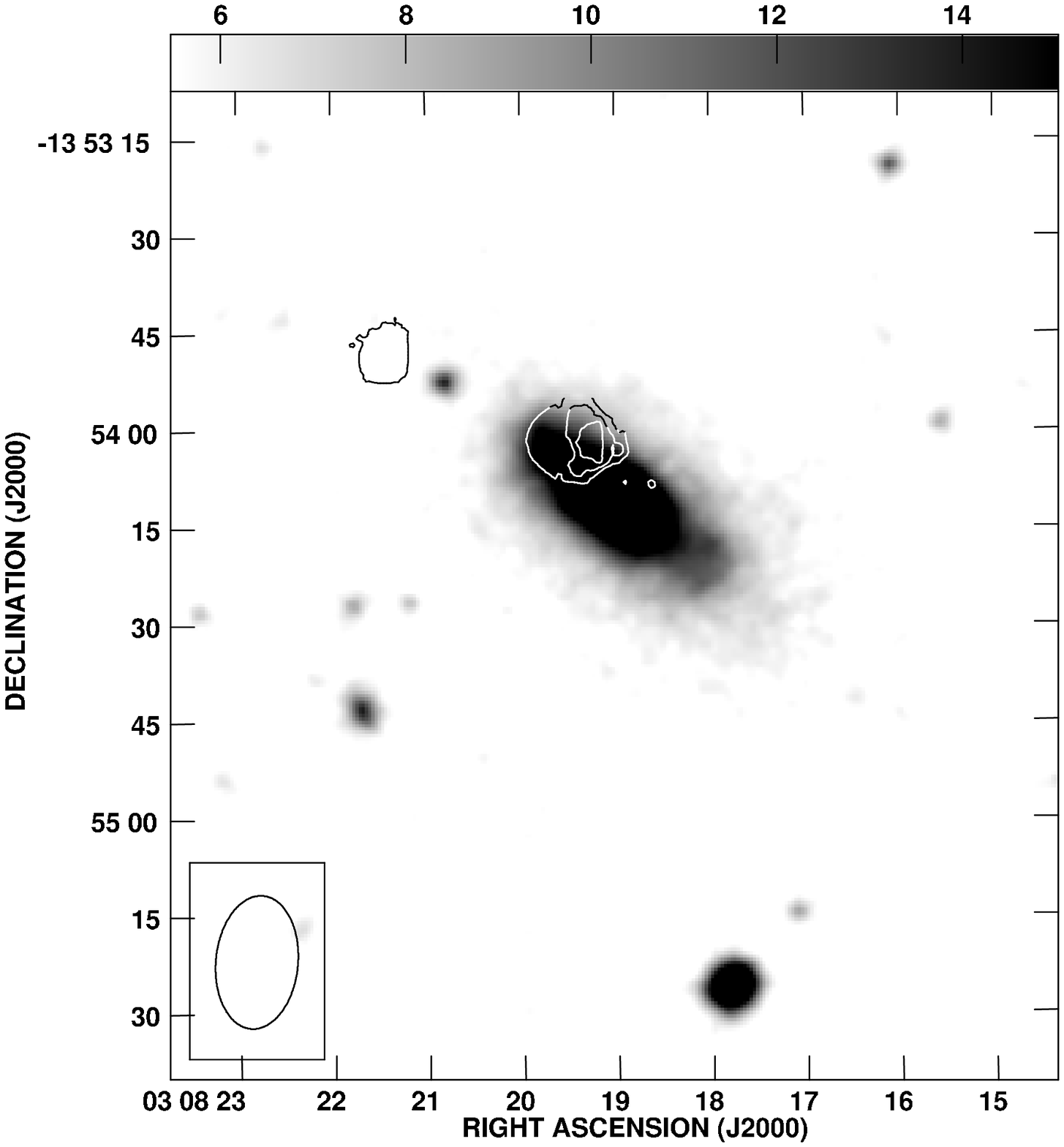}}
\qquad
\subfloat[][]{\includegraphics[width=5.5cm,height=5.5cm]{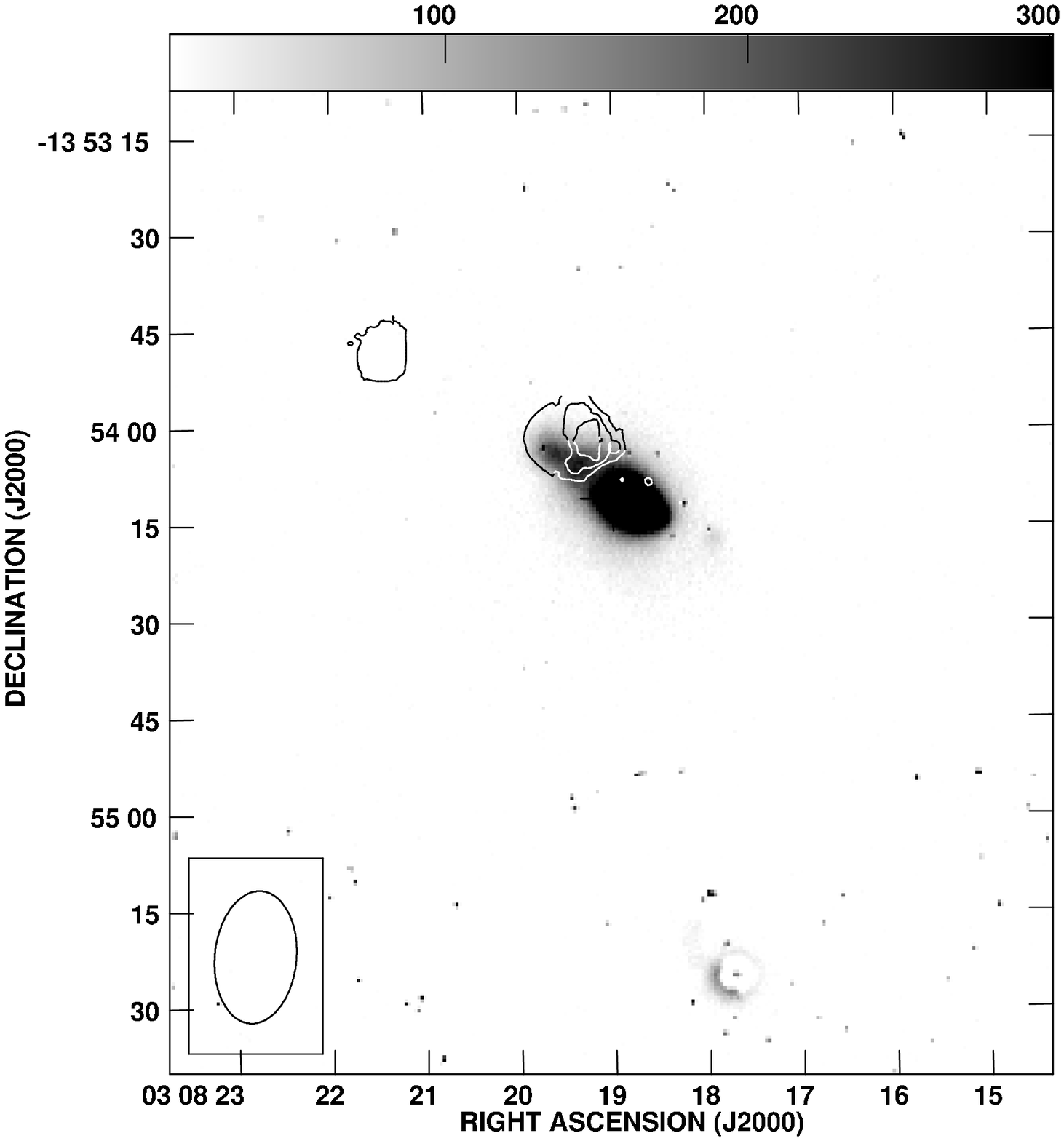}}
\subfloat[][]{\includegraphics[width=5.5cm,height=5.5cm]{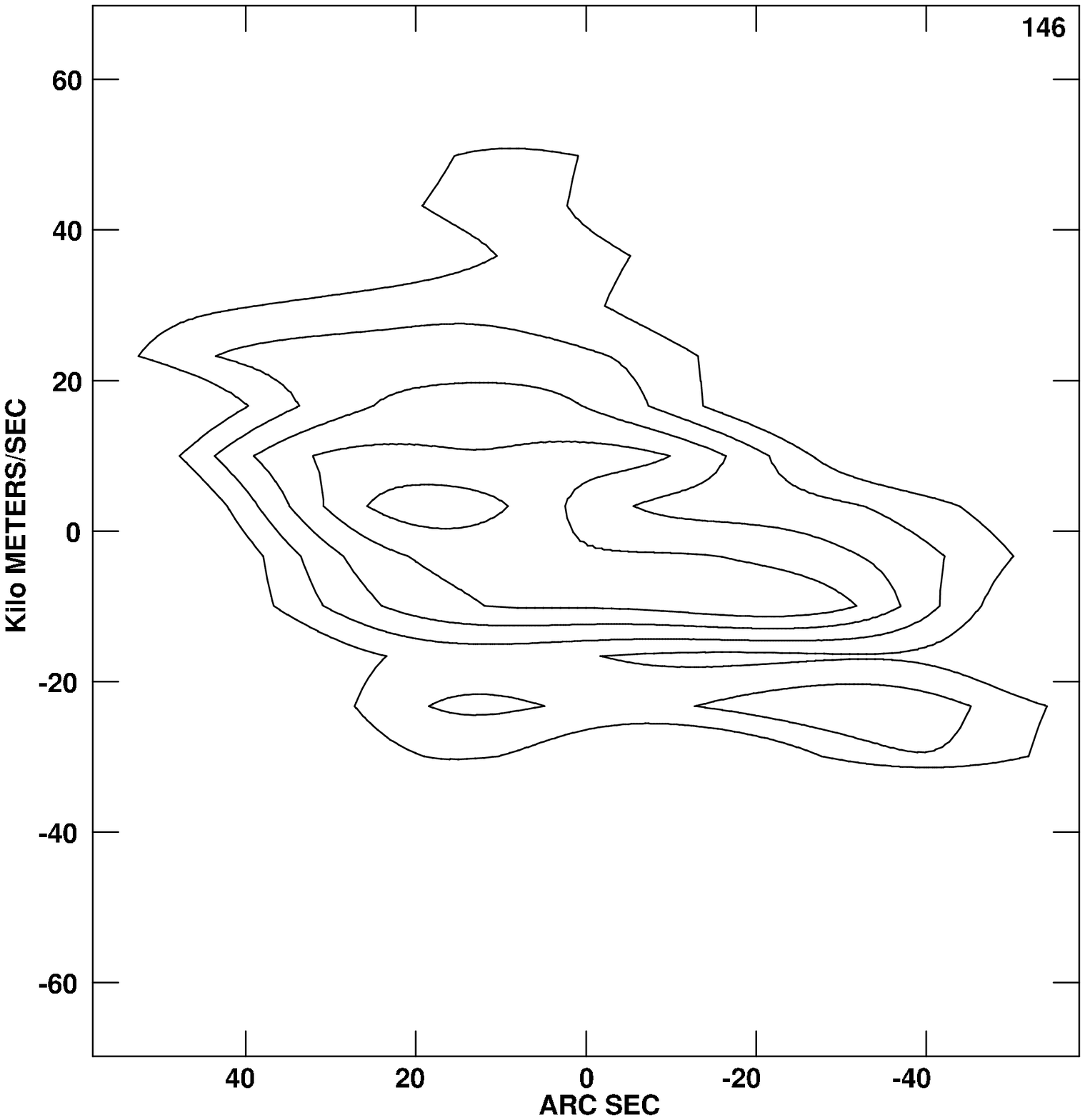}}
\caption{\footnotesize {\bf Mrk 1069: } Figure {\bf (a)} shows \hi column density contours (1, 2, 3, 4, 5, 6, 7, 8$\times \ 10^{20}$ cm$^{-2}$)
overlaid on the $B$ band DSS image. Angular resolution of the map is $41\arcsec\times35\arcsec$.
Note the truncation of the \hi disk in the southern part of the galaxy and the large \hi disk compared
to the optical disk. The box represents the size of the images in Figures (c)-(j).
{\bf (b)} \hi iso-velocity map (\textit{MOMNT 1}). Contours are plotted between
1515 and 1615 \kms \ in steps of 10 \kms. The velocity field shows a disturbed rotation field.
{\bf (c)} Higher resolution ($21\arcsec\times13\arcsec$) \hi column density map for this galaxy is shown here. Contours
 are plotted at 6, 8, 10 $\times10^{20}$ cm$^{-2}$.  Only the central $1\arcmin.3\times1\arcmin.3$ of the galaxy shown in (a) 
is zoomed-in and shown in this map.  Note the fragmentation of the \hi disk into smaller clouds. 
{\bf (d)} The higher resolution \hi column density map is overlaid on the H$\alpha$ image taken from 
\cite{ram10}. Contour levels are same as in (c). It is noticed that the highest column density contour is off-centered 
from the H$\alpha$ image. {\bf (e)} Position-velocity curve along the major axis ($60\arcdeg$ east of north) of the galaxy. 
The contours are plotted at $2.3\times$(3, 4, 5, 6, 7) mJy/beam. 
This P-V diagram is created using the lowest resolution ($41\arcsec\times35\arcsec$) \hi cube.\label{fig:mrk1069_r}}
\label{fig:mrk1069_r}
\end{center}
\end{figure}

\begin{figure}[h!]\footnotesize\ContinuedFloat
\begin{center}
\subfloat[][]{\includegraphics[width=5.5cm,height=5.5cm]{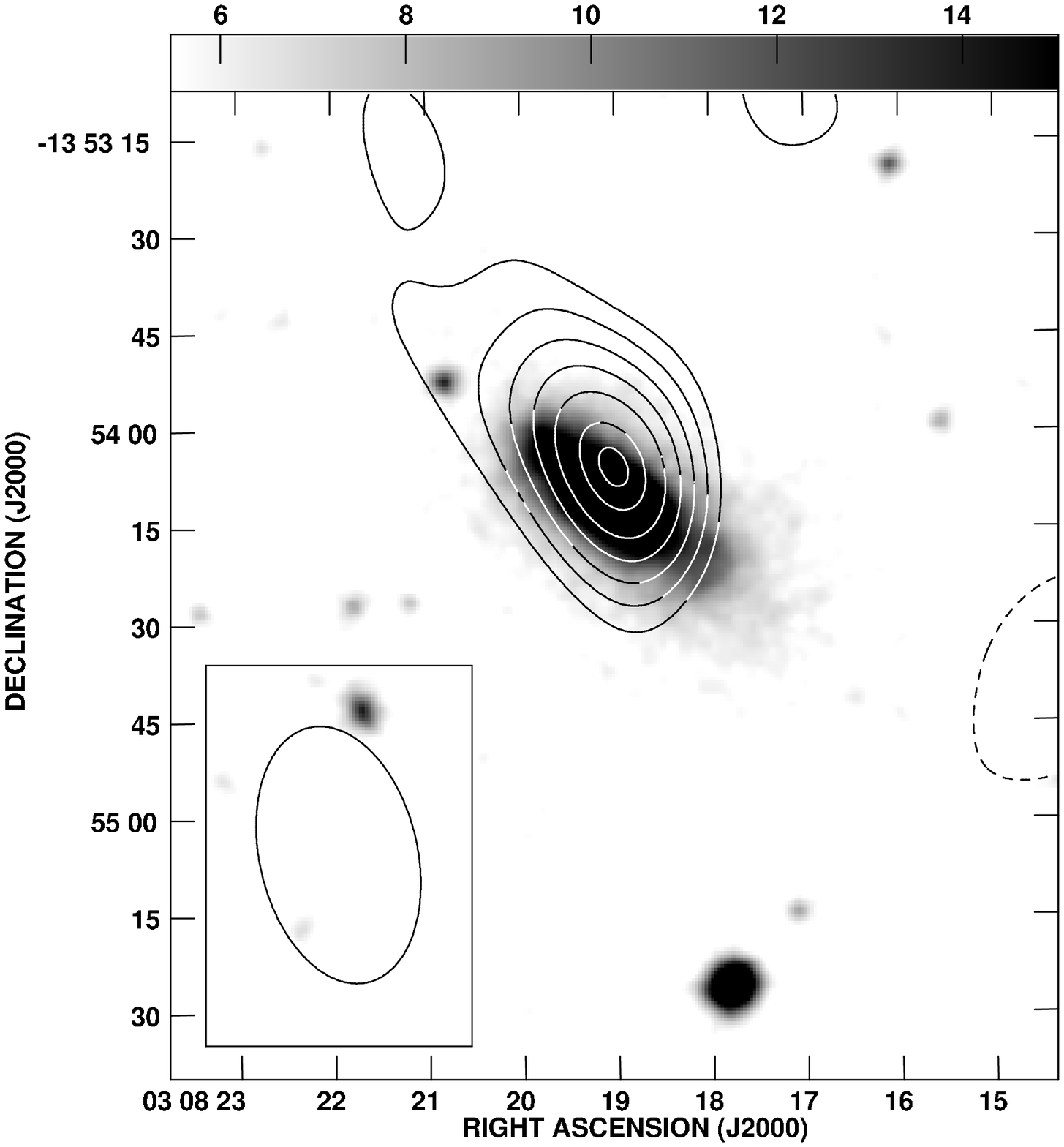}}
\subfloat[][]{\includegraphics[width=5.5cm,height=5.5cm]{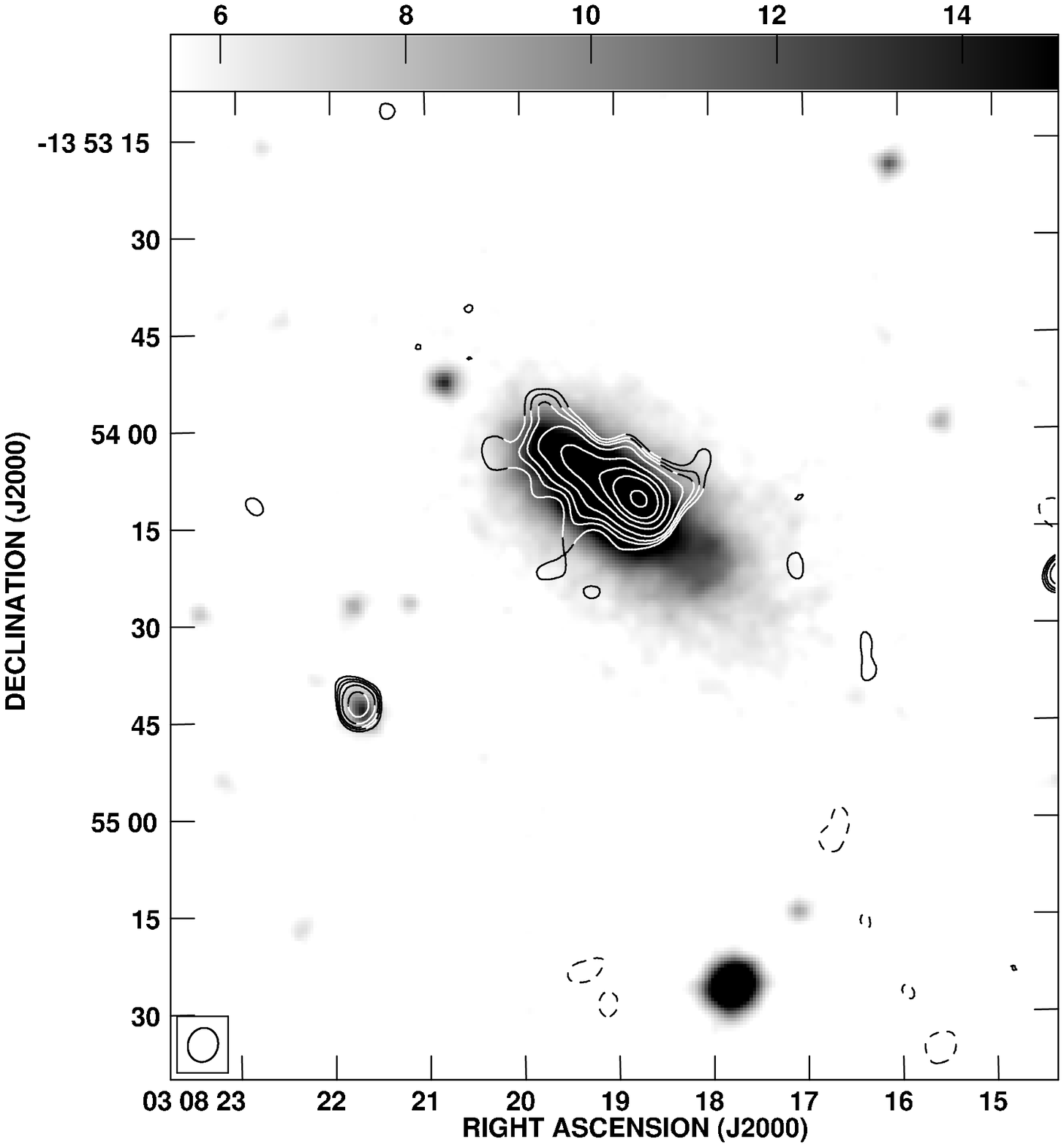}}
\subfloat[][]{\includegraphics[width=5.5cm,height=5.5cm]{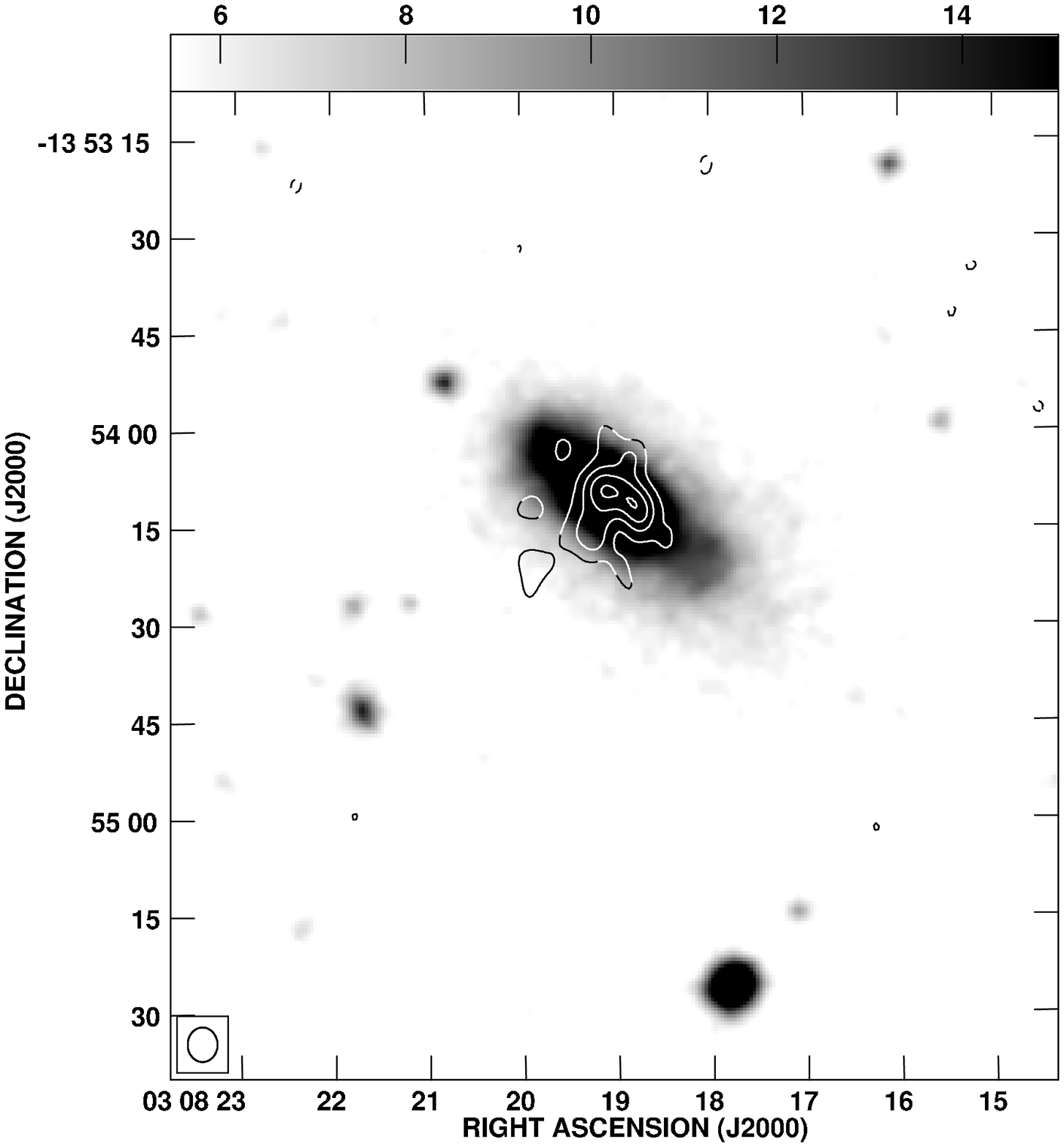}}
\qquad
\subfloat[][]{\includegraphics[width=5.5cm,height=5.5cm]{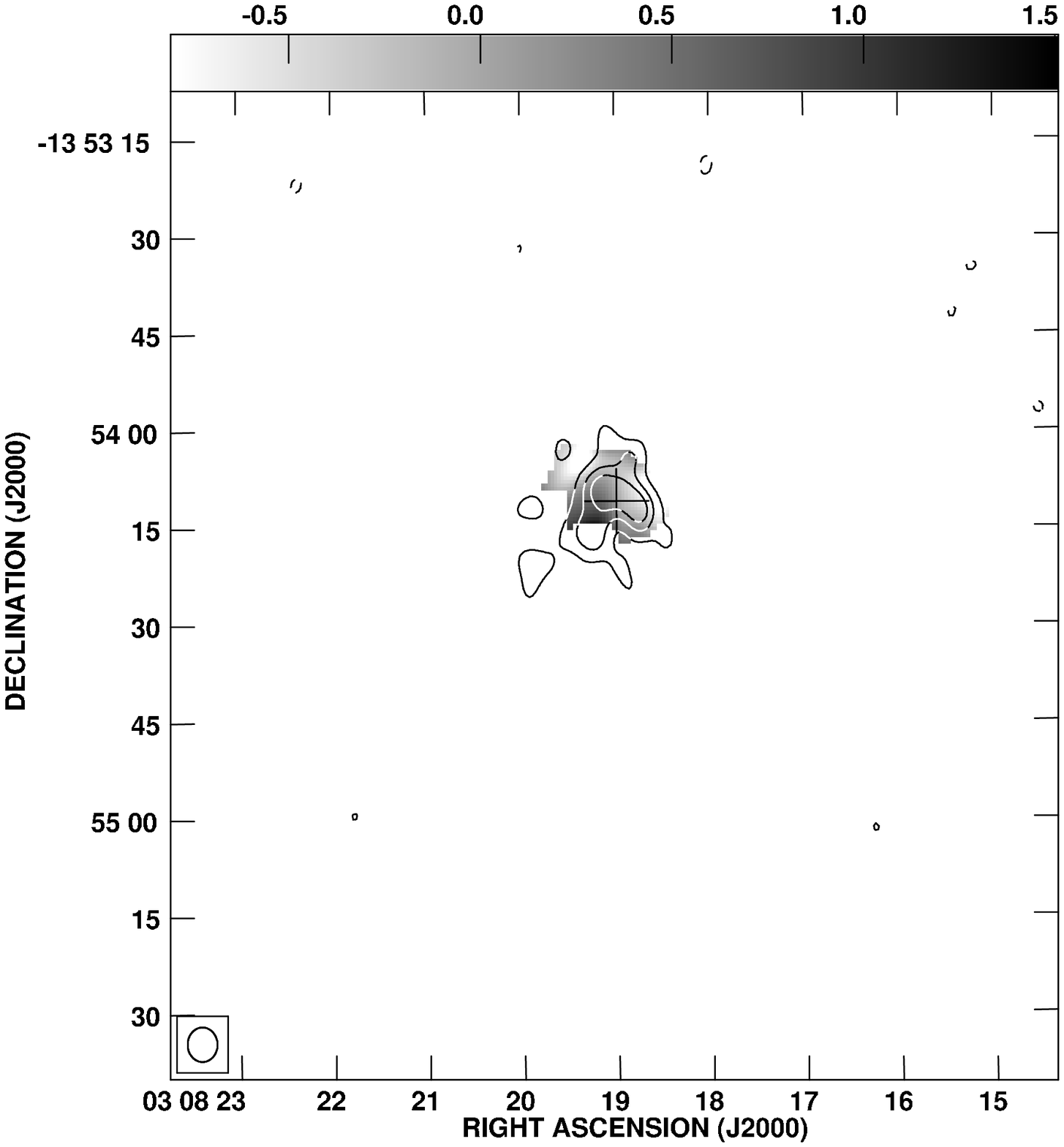}} 
\subfloat[][]{\includegraphics[width=5.5cm,height=5.5cm]{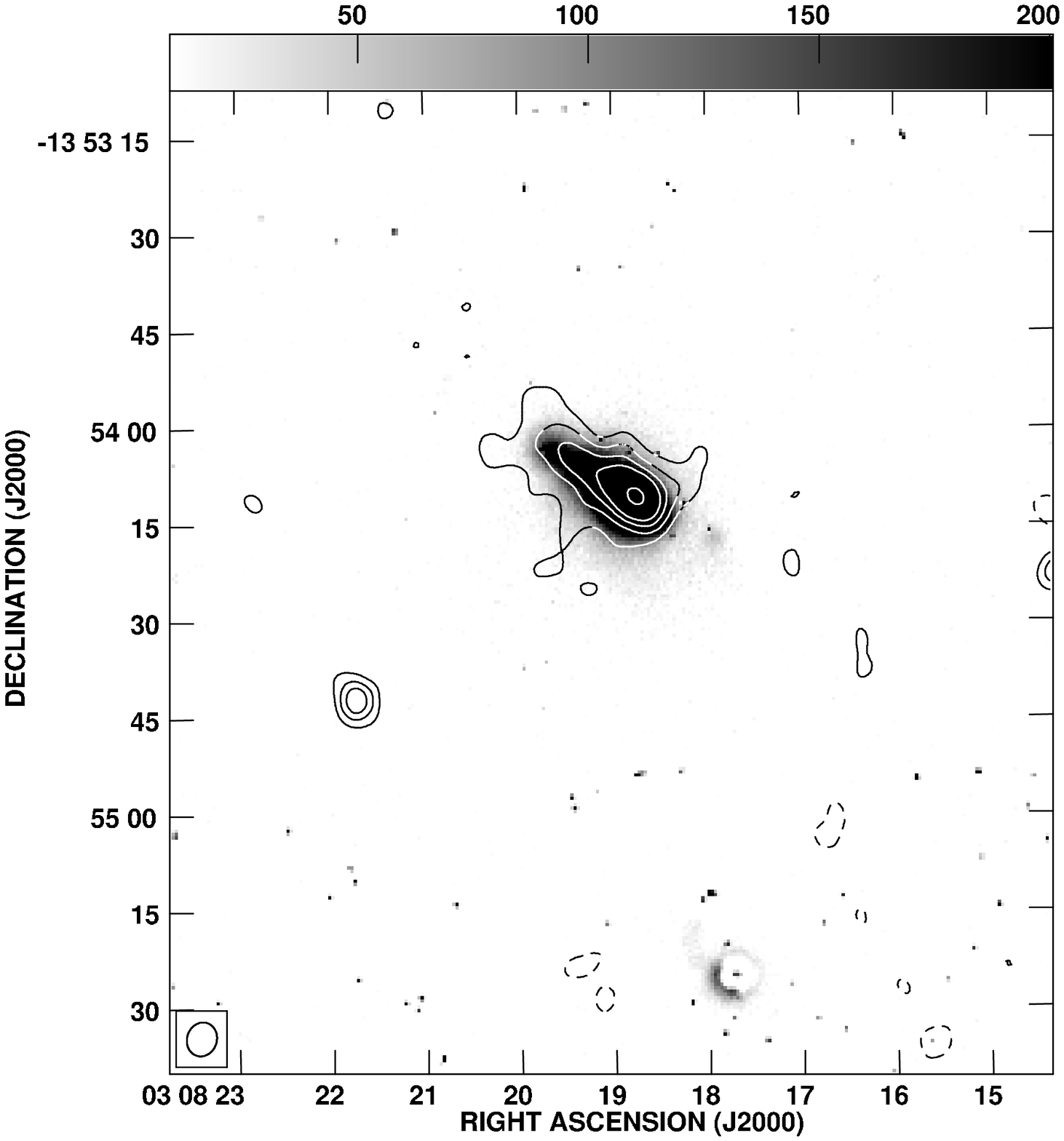}}
\caption{\footnotesize {\bf Mrk 1069 {\it Continued :}}  Figure {\bf (f)} shows 325 MHz contours (levels = $1.7 \ \times$ 
(-5, -4, 4, 5, 6, 7, 8, 9) mJy/beam) overlaid on DSS $B$ band optical image. The resolution of the map is 
$25\arcsec\times16\arcsec$. {\bf (g)} shows 610 MHz contours (levels = $70.0 \ \times$ (-6, -4, 4, 5, 6, 8, 
12, 16, 20, 28) 
$\mu$Jy/beam) overlaid on DSS $B$ band optical image. The resolution of the map is $5\arcsec\times5\arcsec$. 
{\bf (h)} shows the 1.4 GHz continuum map obtained from GMRT overlaid on the DSS optical image. Contour levels are 
$140.0 \ \times$ (-6, -4, 4, 6, 8, 10) $\mu$Jy/beam. The resolution of the map is $5\arcsec\times5\arcsec$. Note the
 bipolar feature extending along the minor axis of the galaxy. {\bf (i)} shows the spectral index map in gray scale 
plotted with 1.4 GHz contours at (-6, -4, 4, 6, 8) 
$\times140\mu$Jy/beam.  Spectral index varies from $-0.8$ to +1.5. Angular resolution of the maps is 
$5\arcsec\times5\arcsec$.
{\bf (j)} shows the 610 MHz contours overlaid on the H$\alpha$ image taken from \cite{ram10}. The contour levels 
are $70.0 \ \times$ (-6, -4, 4, 8, 12, 16, 28) $\mu$Jy/beam). The resolution of the radio map is 
$5\arcsec\times5\arcsec$. The radio continuum emission is coincident with the H$\alpha$ emission and has a similar
 extent.}
\end{center}
\end{figure}

Radio continuum emission from Mrk 1069 is detected at 325 MHz, 610 MHz and 1420 MHz (Figure \ref{fig:mrk1069_r}f--h 
respectively). The radio continuum emission arises in the central parts of the galaxy and extends to the north. The 
global spectral index between 610 MHz and 1.4 GHz is $\sim-0.4$ and between 325 and 610 MHz is $\sim+0.13$. 

The 610 MHz and the 1.4 GHz images show a bipolar outflow-like feature extending along the minor axis. 
Interestingly, this feature shows a slightly flatter spectrum as compared to the surrounding emission as seen in 
the spectral index map between 610 and 1420 MHz (Figure \ref{fig:mrk1069_r}(i)). We note that a similar feature 
has been seen in the galaxy F08208+2816 \citep{yin03}.
 Figure \ref{fig:mrk1069_r}(j) shows 610 MHz contours overlaid on the H$\alpha$ image of 
\cite{ram10}, it is noticed that the extent of emission in 610 MHz
 and in the H$\alpha$ are the same. However, no obvious bi-polar like feature is visible in the H$\alpha$
 map. Higher sensitivity and higher resolution data are required to confirm the existence and nature of this radio
 feature. 

An extended source situated at $03^{\textrm h}08^{\textrm m}21\arcsec.7$  $-13\arcdeg54\arcmin42\arcsec.1$ 
(refer Figure \ref{fig:mrk1069_r}(g)) east of Mrk 1069 is detected at 610 MHz with a flux density of $1.43\pm0.18$ mJy. 
This background source is identified with 2MASX J03082177-1354425. }

\item{I Zw 97 : We do not detect the galaxy at 240 MHz upto a $5~\sigma$ limit of 7.5 mJy 
on the flux density of the galaxy. We detect radio emission from this galaxy at 610 MHz (see Figure \ref{fig:izw97_r}(a)) 
with an integrated flux density of about 1.14 mJy.  
This is the first detection of this galaxy in radio bands to the best of our knowledge. 
We note that it has not been detected in the NVSS \citep{condon98} and FIRST \citep{beck95} surveys at 1.4 GHz. 
The emission at 610 MHz arises in two discrete regions of the galaxy. We also note that a type II supernova SN 2008bx was 
discovered in this galaxy on 2008 April 22 \citep{puc08}. 

\begin{figure}[h!]\footnotesize
\begin{center}
\subfloat[][]{\includegraphics[width=5.5cm,height=5.5cm]{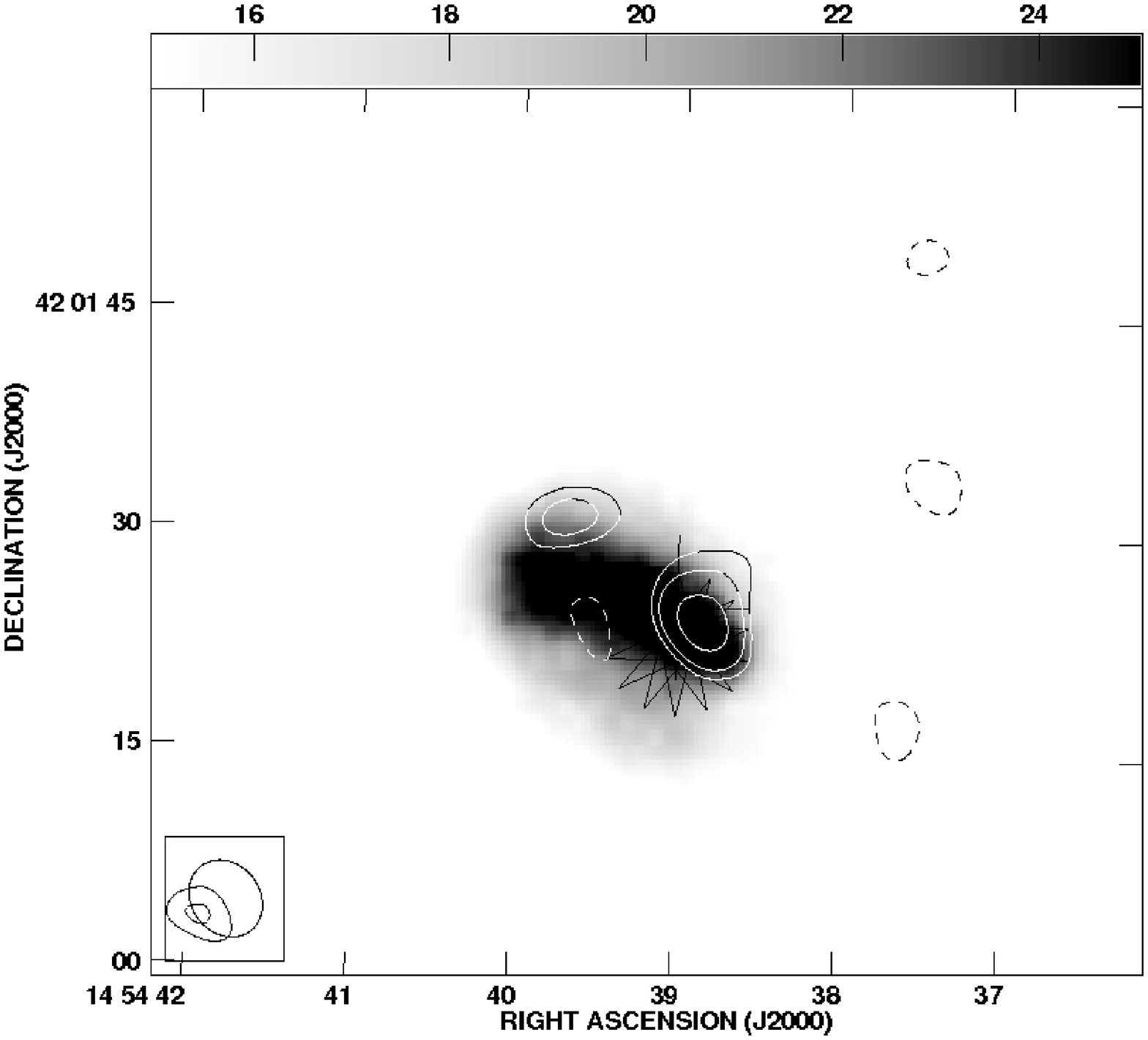}}
\subfloat[][]{\includegraphics[width=5.5cm,height=5.5cm]{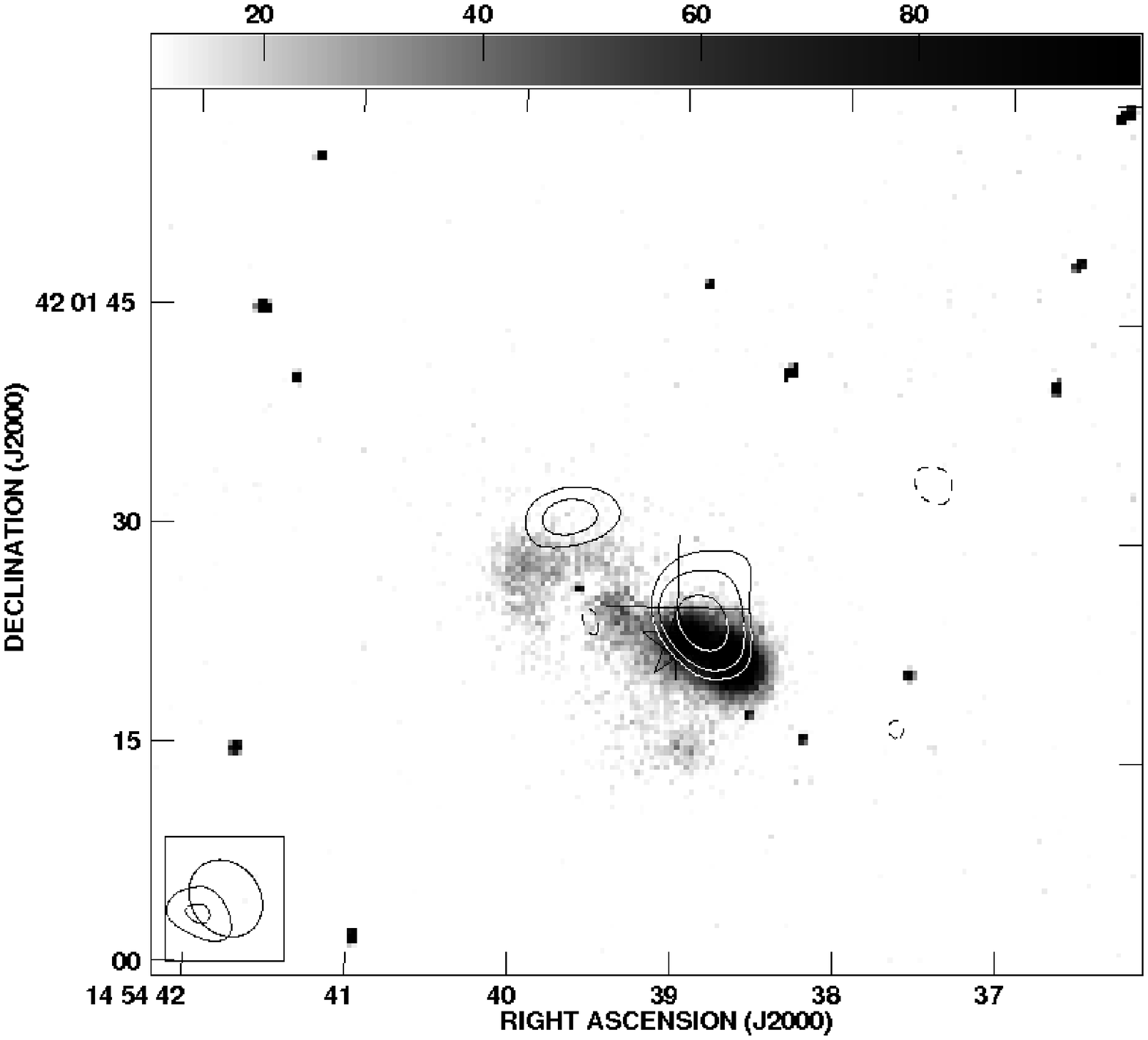}} 
\caption{\footnotesize {\bf I Zw 97 :} 
{\bf (a)} 610 MHz contours superposed on the DSS $B$ band optical image. Contours are plotted at 
$65.0 \ \times$ (-3, 3, 4, 6) $\mu$Jy/beam. The resolution of the map is $6\arcsec\times5\arcsec$. {\bf (b)} 
610 MHz contours overlaid on H$\alpha$ image taken from \cite{ramya09}. The star marks the position of the supernova SN 2008bx. This 
galaxy is not detected at 1.4 GHz either in the NVSS \citep{condon98} or the FIRST \citep{beck95} surveys. }
\label{fig:izw97_r}
\end{center}
\end{figure}

The southern component seen in our 610 MHz image is emission associated with the supernova SN 2008bx 
\citep{atel09}. The radio continuum emission coincides fairly well with the H$\alpha$ emission (see Figure 
\ref{fig:izw97_r}(b), \ha image taken from \cite{ramya09}). The galaxy is not detected in single dish
 \hi observations \citep{thuan81}. }

\end{enumerate}

\section{\textsc{Discussion}}
\label{sec:discus_rad}

\subsection{\textsc{Radio continuum spectra}}
\label{subsec:radio_spec}

\begin{figure}
\includegraphics[width=15cm]{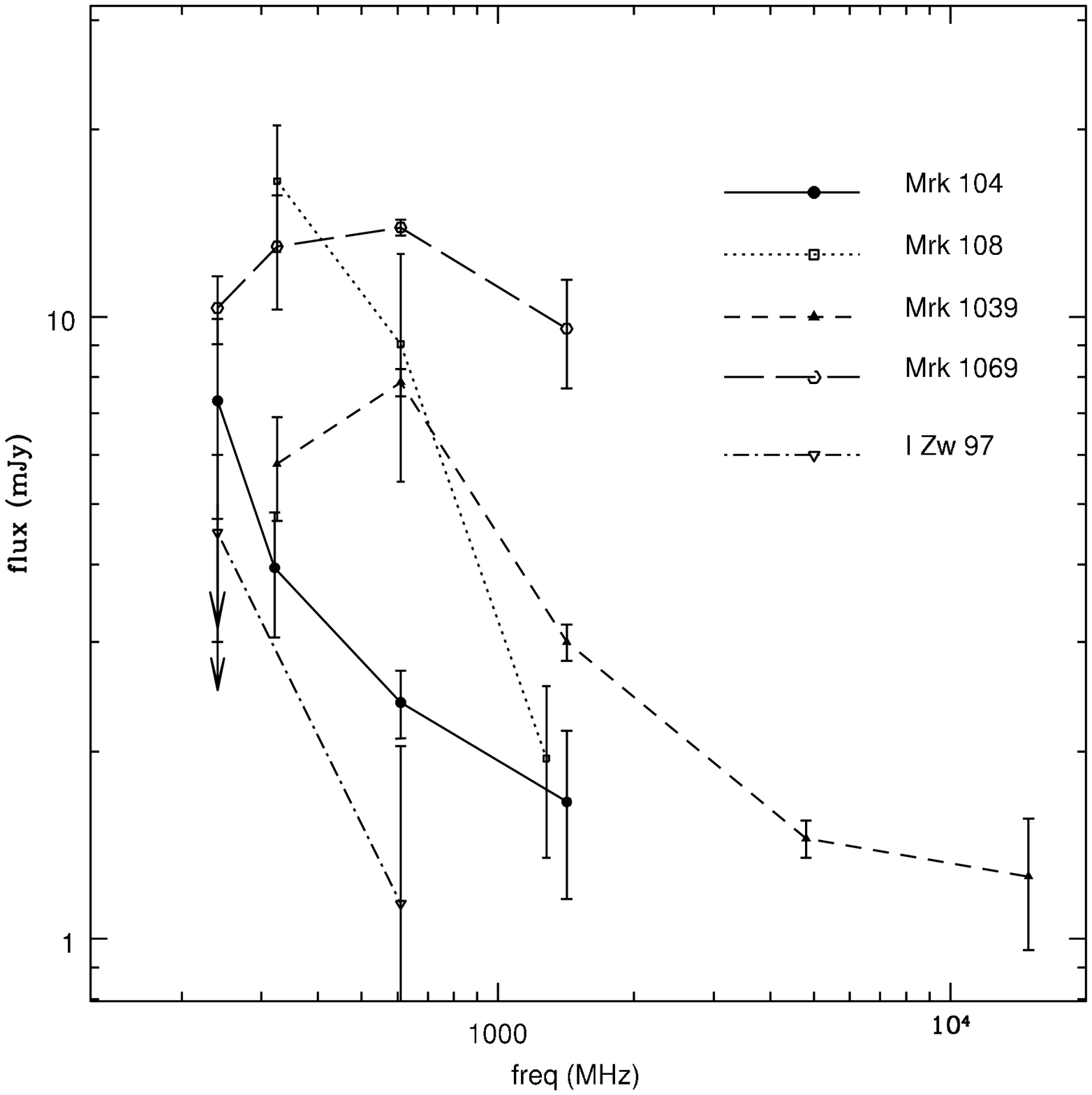}
\caption[\footnotesize The observed radio spectra of the five galaxies]{\footnotesize The observed spectra of the five 
galaxies studied here. The points are the observed data and we have connected these by a line to help the reader
 distinguish the spectra of different galaxies. Solid line represents the data for Mrk 104, dotted line Mrk 108, short
 dashed line Mrk 1039, long dashed line Mrk 1069 and the dot dashed line I Zw 97. The upper limits at 240 MHz for Mrk 
1039 and I Zw 97 are shown as arrows. In the plot, I Zw 97 is shifted by $-125$ MHz for clear viewing. Note the 
differences in shapes of spectra of different galaxies.}
\label{fig:freqflx}
\end{figure}

Figure \ref{fig:freqflx} shows the observed radio spectra of the five galaxies. The spectra of Mrk 1069 and Mrk 1039 are 
seen to turn over at frequencies below 610 MHz, while, the spectra of Mrk 104 and Mrk 108 show a power law upto the lowest 
GMRT frequencies. The observed radio continuum emission for Mrk 1039 and Mrk 1069 is mostly confined to the star forming
 regions traced by H$\alpha$. \citet{kle91} noted that the break frequency is below 1 GHz in their sample of BCDs. We note
 that Mrk 108 (dotted line in Figure \ref{fig:freqflx}) is projected close to NGC 2820 and hence it is difficult to 
separate the emission of NGC 2820 from Mrk 108 at the low frequencies where 
the beamsize is larger. There is a possibility that the lower frequency emission (i.e. frequencies less than 1 GHz) 
includes contribution from the spiral companion.

We have combined VLA archival data at higher frequencies with our low frequency GMRT data for Mrk 1039. The spectrum 
(Figure \ref{fig:freqflx}) flattens for $\nu > 1.4$ GHz and the emission appears to be dominated by thermal emission.
 Since the higher frequency data is not available for other galaxies in our sample, we cannot comment on their behaviour 
at these frequencies. \cite{kle91} have noted flat spectrum at higher frequencies with non-thermal spectral index ranging
 from $-0.7$ to $-2$ for some galaxies in their sample. However the low frequency observations (325 MHz) by \cite{deeg93}
 suggest that the average spectral index of the nonthermal emission is $\sim -0.7$. Moreover, the thermal fraction at 1
 GHz is seen to vary from about 5\% to 72\% \citep{kle91}. Our sample is a subset of a larger optically selected sample
 of BCDs and hence distinct from the sample of \cite{kle91} and \cite{deeg93}. We have been able to model the observed
 integrated radio continuum emission using four models: (1) the observed emission is non-thermal synchrotron emission;
 (2) the observed emission is a combination of thermal free-free and non-thermal synchtrotron emission along the line of 
sight; (3) the non-thermal emission at low frequencies is free-free absorbed by the thermal region in the same volume;
 and (4) the observed emission is a combination of both thermal and non-thermal, and the non-thermal emission is free-free
 absorbed by the thermal gas of optical depth $\tau_1$. 

Since these cases well represent what we understand about radio emission from galaxies and they explain the observed low 
frequency spectrum of the BCD galaxies, we have not explored the case of synchrotron self-absorption which we believe will
 not be important in these galaxies.  \citet{deeg93} have shown that both free-free absorption at lower radio frequencies
 and time-dependent electron injection models explain the flattening of the spectrum at lower radio frequencies. However
 since we detect a clear low frequency turnover in only two galaxies, we have not tried to distinguish between the two
 models. Figure \ref{fig:rad_spec} shows the best fits using one of the four models given below to the observed points.

The four equations which were fitted were:

(1) The observed emission is non-thermal synchrotron emission which is fitted with power law index $\alpha$:
\begin{equation}
 S(\nu) = c_1\nu^{\alpha}
\end{equation} 

(2) The observed emission is a combination of non-thermal (S$_{nth}$) and thermal free-free emission (S$_{th}$): 
\begin{equation}
 S(\nu) = S_{nth} (\propto \nu^{\alpha}) + S_{th} (\propto \nu^{-0.1})
\end{equation} 

(3) The low frequency emission is free-free absorbed by the thermal material intermixed with the emitting region with
optical depth $\tau_1$ (Equation taken from \cite{deeg93}):
\begin{equation}
  S(\nu) = c_1\nu^{(\alpha+2.1)}~[1-\exp(-\tau_1\nu^{-2.1})]
\end{equation} 

(4) The observed emission is a combination of both non-thermal and thermal emission. The non-thermal component at 
lower frequencies is free-free absorbed by the thermal gas intermixed with the emitting region 
with optical depth $\tau_1$ (Equation taken from \cite{deeg93} and modified to include thermal emission):
\begin{equation}
 S(\nu) = c_1\nu^{(\alpha+2.1)}~[1-\exp(-\tau_1\nu^{-2.1})] + c_2 ~ \nu^{-0.1}
\end{equation}

In the above equations, $c_1$ is a constant, while $c_2$ is the thermal emission at frequency $\nu$. While a power law i.e. model 
(1) is the best fit to the spectrum of Mrk 108, the observed spectrum of Mrk 104 is best fit by including contributions from thermal and non-thermal emission i.e. model (2).  The observed spectrum of Mrk 1069 was best fit by model (3) i.e. free-free
absorption of synchrotron emission giving rise to a low frequency turnover. 
Mrk 1039 is best fit using model (4). These fits are shown in Figure \ref{fig:rad_spec} and the parameters are listed in Table \ref{tab:rad_spec}. Since the number of input points  are small, the observed points were first spline interpolated. The interpolated points were then given as input data points to the programme which used the Levenberg-Marquardt algorithm as given by \cite{pres93} to find the best fit. The fitted parameters and the physical quantities derived using these fitted parameters are given in Table \ref{tab:rad_spec}. The reduced $\chi^2$ values are given in Table \ref{tab:rad_spec}. Galaxy-wise results from the modelling procedure are given below:

\begin{figure}
\includegraphics[width=15cm]{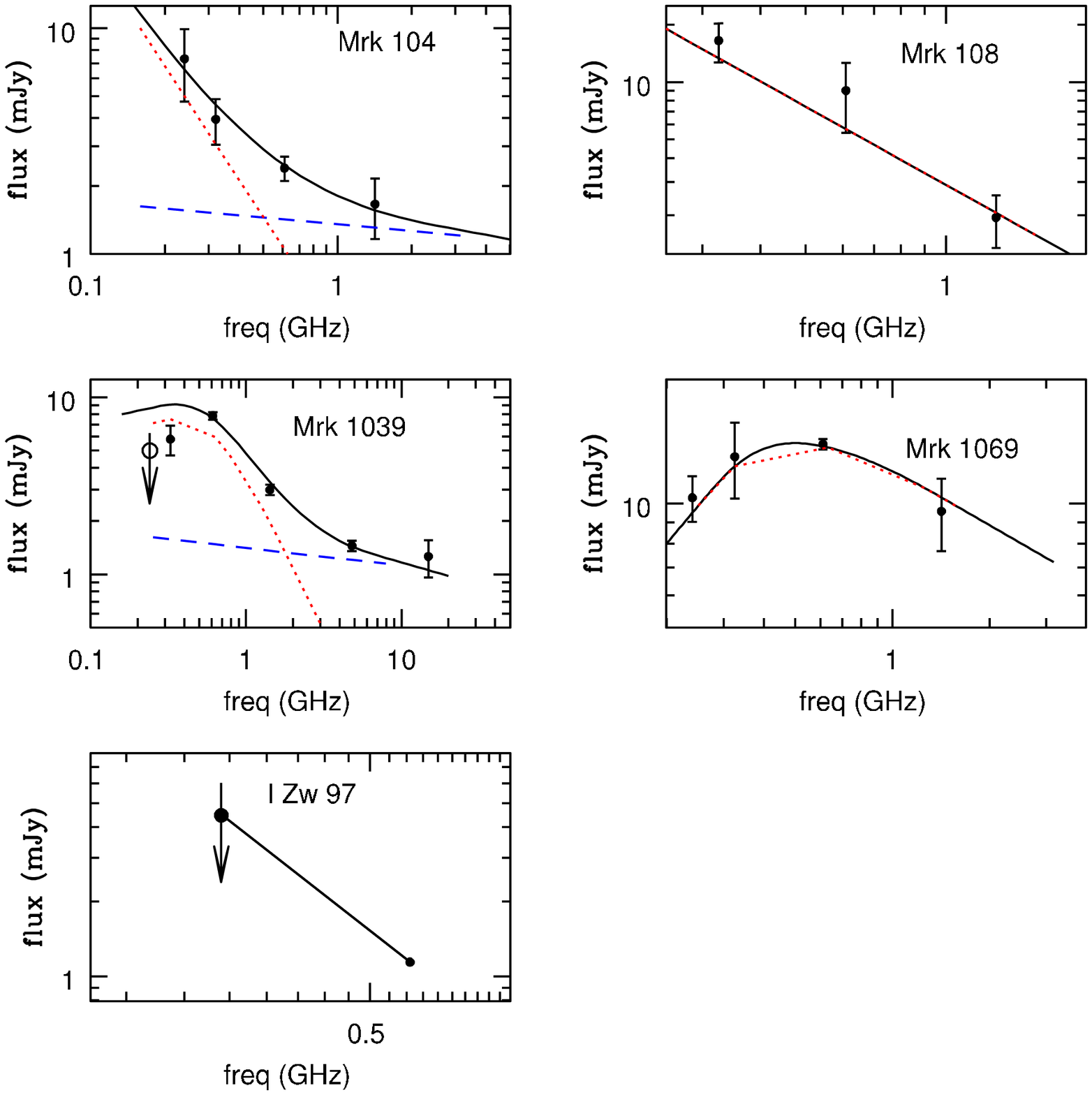}
\caption[\footnotesize The model fits to the observed radio spectra for four galaxies]{\footnotesize The 
model fits to the observed spectra of four galaxies. In each of the panels, the dashed line represents the thermal component, dotted line represents the non-thermal component (modified by free-free absorption in the case of Mrk 1039 and Mrk 1069).  The solid line represents the best fit to the observed points. The best fit model to the galaxy Mrk 104 is the combination of synchrotron emission and thermal bremstrahlung emission. Mrk 108 is best fit with a power law spectrum. Mrk 1039 is best fit assuming a combination of thermal and non-thermal emission, and the non-thermal emission is absorbed by the thermal gas mixed in the region. Mrk 1069 is best fit with synchrotron emission and thermal free-free absorption at lower frequencies. Table \ref{tab:rad_spec} gives the values of the fitted parameters.}
\label{fig:rad_spec}
\end{figure}

\begin{table*}[h!]\footnotesize
\begin{center}
\caption{\footnotesize Parameters of the model fits to the observed data.}
\label{tab:rad_spec}
\begin{tabular}{lllll}
\hline\hline 
\textbf{Parameters} & \textbf{Mrk 104} & \textbf{Mrk 108} & \textbf{Mrk 1039} & \textbf{Mrk 1069} \\
\hline 
\textbf{Model} & eqn 3 & eqn 2 & eqn 4 & eqn 1 \\
\textbf{${\chi^2}_{red}$} & 0.39 & 4.91 & 2.34 & 1.124 \\
\textbf{c$_1$,c$_2$} & --- & --- & $10.06\pm0.17$, $1.41\pm0.03$ & $98.18\pm2.78$ \\
\textbf{$\alpha$} & $-1.67\pm0.26$ & $-1.36\pm0.12$ & $-1.85\pm0.10$ & $-0.51\pm0.12$ \\
\textbf{$\tau_1$} & --- & --- & $0.42\pm0.02$ & $0.13\pm0.01$ \\
\textbf{EM from $\tau_1$} (pc cm$^{-6}$) & --- & --- & $(1.26\pm0.10)\times10^6$ & $(3.67\pm0.20)\times10^5$ \\
\textbf{$n(\textrm{\hii})$} cm$^{-3}$ & --- & --- & $\sim354 \ ^a$ & $\sim192 \ ^a$ \\
\textbf{fraction of {S$_{th}$}$_{1.4GHz}$} (\%) & 80\% & --- & 45\% & --- \\
\hline\hline 
\end{tabular}
\end{center}
$^a$ - Assuming a size of 10 pc for the emitting region.
\end{table*}

\textbf{Mrk 104:}  The best fit model (Figure \ref{fig:rad_spec}) for this galaxy is when the fitted emission consists of 
both thermal and non-thermal emission.  This is also obvious from the observed spectrum which shows that the emission at 1.4 GHz is more than expected from a power law. We estimate the thermal fraction at 1.4 GHz of $\sim 80\%$ and the spectral index, $\alpha$ of the synchrotron spectrum is $-1.67$ (Table \ref{tab:rad_spec}).  Compare this with a typical thermal fraction of $\le 10$\% and synchrotron spectral index of $\sim-0.7$ or $-0.8$ for normal spiral galaxies (\citealt{dale02}; \citealt{cond92}). However, no signature of free-free absorption towards the lower frequencies is seen down to 240 MHz.  The fraction of thermal emission in the galaxy is fairly large implying a young burst of star formation. \cite{ramya09} show that this galaxy has an abundant 4 Gyr population + $\sim500$ Myr population and young burst of star formation which is $\sim10-15$ Myr old.  Moreover they show that the major contribution to the total mass in this galaxy is the Gyr old population whereas the dominant contribution to the total light comes from the $\sim500$ Myr population.

Star formation rate (SFR) is calculated using the 610 MHz emission as given by equation 21 of \cite{cond92}. The spectral index $\alpha$ obtained from the fit is used to calculate the SFR. The SFR is $0.023\pm0.011$ \msyr for Mrk 104 (Table \ref{tab:fir_rad}). The SFR estimated using the FIR luminosity, following \cite{yun01} is $0.12\pm0.04$ \msyr. The SFR ($>10$ \msun) estimated from the H$\alpha$ flux is 0.04 \msyr \ and is in the range $0.0003-0.004$ \msyr \ for individual star forming regions \citep{ramya09}. The SFR calculated using H$\alpha$ flux is in reasonably good agreement with the SFR found using the 610 MHz emission. The parameter $q$ \citep{cond91} which is a measure of the starburst versus AGN nature of the radio emission is also estimated and tabulated. Normal galaxies which show the FIR-radio (1.4 GHz) correlation have a value of $q=2.35\pm0.2$ (\citealt{cond92}; \citealt{yun01}), while $q$ is $2.44\pm0.02$ (Table \ref{tab:fir_rad}) for Mrk 104 which is within the observed scatter for the correlation.  \\

\textbf{Mrk 108:} The galaxy located in the group Holmberg 124 is best fit (Figure \ref{fig:rad_spec}) with a power law spectrum. The spectral index of this non-thermal emission is $\sim-1.36$ (Table \ref{tab:rad_spec}) and we note that the possibility that the lower frequency flux density is overestimated due to contribution by the large companion NGC 2820 cannot be ruled out. If the error introduced by NGC 2820 is small and the steep index is real then it could be due to galaxy-wide star formation induced by tidal interaction. The SFR (refer Table \ref{tab:fir_rad}) calculated using the flux at 610 MHz is $0.051\pm0.036$ respectively. $q$ value could not be estimated due to the lack of IRAS FIR fluxes in the literature.\\

\begin{landscape}
\begin{table*}[h!]\footnotesize
\begin{center}
\caption{\footnotesize Radio-FIR properties for the five BCDs. 1$\sigma$ errors are quoted in the brackets.}
\label{tab:fir_rad}
\begin{tabular}{lllllllll}
\\
\hline\hline
\textbf{Galaxy} &  \textbf{log(L$_{1.4GHz}$(WHz$^{-1}$)} & {\textbf{FIR}}$^a$ & {\textbf{q}}$^a$ & {\textbf L$_{FIR}$}$^a$ & {\textbf SFR$_{FIR}$}$^a$ & {\textbf SFR$_{610MHz}$}$^b$ & {\textbf SFR$_{\textbf{H}\alpha}$($>10$ M$_\odot$)} & {\textbf SFR$_{H\alpha}$(knots)}\\
  ---      &           log(WHz$^{-1}$)             &  \textit{Wm$^{-2}$}   &  ---  & \textit{L$_\odot$} & \msyr & \msyr & \msyr & \msyr \\
\hline 
Mrk 104 & 20.29 &   1.73$\times10^{-14}$ & 2.51 & $5.35\times10^8$ & 0.12 &  0.023 & 0.041$^c$ & 0.0003-0.004   \\
--- & (0.20) &  (0.40$\times10^{-14}$)& (0.04) & ($2.02\times10^8$) & (0.04) &  (0.011) & --- & --- \\
Mrk 108 & 20.92 &  --- & --- & --- & --- &  0.051 & --- & --- \\
  ---& (0.11) & --- & --- & --- & --- &  (0.036) & --- & --- \\
Mrk 1039 &  20.48 & $8.17\times10^{-14}$ & 2.92 & $2.15\times10^9$ & 0.48 & 0.059 & 0.173$^d$ & 0.01-0.15\\
 --- & (0.08) & ($0.31\times10^{-14}$) & (0.07) & ($0.38\times10^9$) & (0.09) &  (0.016) & --- & ---\\
Mrk 1069 &  20.69 & $6.47\times10^{-14}$ & 2.32 &  $8.81\times10^8$ & 0.19 & 0.103 & 0.139$^d$ & 0.005-0.07 \\
 --- & (0.15) & ($0.92\times10^{-14}$) & (0.06) & ($0.20\times10^8$) & (0.04) & (0.026) & --- & ---\\
I Zw 97 & --- & $1.87\times10^{-14}$  &  --- &  $7.73\times10^8$ & 0.17 & 0.023 & 0.03$^c$ & 0.001-0.03\\
 --- & --- & ($0.22\times10^{-14}$) & --- & ($2.03\times10^8$) & (0.05) & (0.02) & --- & ---\\
\hline \hline
\end{tabular}
\end{center}
$^a$ - FIR data is obtained from IRAS Faint Source Catalog (source: NED) and the parameters are calculated using the equations given in \cite{yun01}. \\
$^b$ - The non-thermal fraction of the total 610 MHz flux obtained from the model is used to estimate the SFR using the equations given in \cite{cond92}. \\
$^c$ - The \ha SFR ($>10$ M$_\odot$) is calculated using the equation of \cite{kenni83} and quoted from \cite{ramya09}. \\
$^d$ - The H$\alpha$ SFR is quoted from \cite{ram10}.\\
\end{table*}
\end{landscape}

\textbf{Mrk 1039:} The observed emission of Mrk 1039 for which we have also analysed and included higher frequency data 
from the VLA archives shows a flattening at higher frequencies and turnover at lower observed frequencies. The observed 
emission is well fitted by a combination of thermal and non-thermal emission and the non-thermal emission is absorbed at
 lower frequencies by thermal gas  as shown in Figure \ref{fig:rad_spec}. The thermal and non-thermal emission are mixed.
 This is possible in a young ($\sim 10^{6-7}$ yr) starburst region which is also recording supernovae as the massive stars
 evolve but the relativistic electrons have not yet had time to diffuse through the galaxy. The spectrum is modelled using
 equation (4). The optical depth is $\tau_1$=$0.42$. This gives an emission measure (EM) of $1.26\times10^6$ pc~cm$^{-6}$
 for an assumed temperature of 10000 K. The \hii regions electron density is then $n_e\sim354$ cm$^{-3}$ assuming a 10\,pc
 size for the region. The parameters of the fit are given in Table \ref{tab:rad_spec}. Using this and following 
\cite{hunt04}, the total number of ionizing photons is calculated to be $9.82\times10^{52}$ photons/sec which corresponds
 to 9820 O7 V stars in that region. The H$\beta$ flux calculated using the thermal radio flux is $4.72\times10^{-13}$ erg
 cm$^{-2}$s$^{-1}$, which agrees with the values quoted by \cite{yin03} and \cite{huang99}. This implies that there is a
 young and intense star formation in this galaxy, \cite{huang99} have estimated an age of 4 Myr for starforming regions in
 the galaxy. The spectral index of the non-thermal emission is fairly steep, $\alpha\sim-1.85$ (Table \ref{tab:rad_spec}).
 We note that a type II supernova SN1985S has occurred to the south west of the centre of the galaxy
 (shown by a star in Figure \ref{fig:mrk1039_r}).  Although its location is close to the part of the galaxy from which
 radio emission is detected, no radio emission is seen to be coincident with it. The 610 and 325 MHz emission is clearly more 
extended than the image at 15 GHz obtained using VLA archival data (see Figure \ref{fig:mrk1039_r}~i,j,k) but does not
 encompass the supernova.

We estimate the $q$ parameter to be $2.90\pm0.03$ (Table \ref{tab:fir_rad}). A similar value for this galaxy was also 
obtained by \cite{yin03} ($q=2.94$). SFR calculated using FIR luminosity is $0.48\pm0.09$ \msyr. Using the 610 MHz 
emission, we estimate a SFR of $0.060\pm0.016$ \msyr \ for this galaxy. The SFR($>10~M_\odot$) quoted in the table is 
$0.173$ \msyr \ but the SFR for individual knots varies from 0.01--0.15 \msyr.

\textbf{Mrk 1069:} The observed spectrum of Mrk 1069 shows a clear turnover at lower frequencies and hence was modelled 
using a combination of non-thermal emission and free-free absorption at the lower frequency. Higher frequency archival
 data was not available for this galaxy. The best fit is shown in Figure \ref{fig:rad_spec}. The spectral index of the 
non-thermal emission is  $\sim-0.51$ and the optical depth of the absorbing thermal gas is $0.13$.  Assuming an electron 
temperature of $10^4$ K  implies a modest EM of $3.7\times10^5$ cm$^{-6}$~pc (Table \ref{tab:rad_spec}). For an \hii 
region of size 10 pc, this implies an electron density of about 192 cm$^{-3}$. The radio emission is intense near the
 optical centre of the galaxy and extends towards the north along the major axis as seen in our 610 MHz image. We also 
detect a bipolar outflow like feature extending along the minor axis of the galaxy in our 1.4 GHz image (Figure 
\ref{fig:mrk1069_r}~h) and hints of the same are visible at 610 MHz. Similar feature is not detected in H$\alpha$ 
(Figures \ref{fig:mrk1069_r}~j) but our images are not deep enough. The feature needs to be confirmed with higher
 frequency, higher resolution radio continuum data. We note that \cite{hunt05} have detected a similar outflow in the 
galaxy I Zw 18. We estimate  $q=2.32\pm0.06$ (Table \ref{tab:fir_rad}) which agrees with the FIR-radio correlation within
 errors. The SFR calculated for Mrk 1069 using the FIR luminosity is around $0.19\pm0.04$ \msyr \ and from 610 MHz flux 
$0.103\pm0.026$ \msyr.

\textbf{I Zw 97:} The galaxy, I Zw 97 is detected only at 610 MHz and is undetected in any other frequency. Due to this, 
the modelling of the spectra could not be done. The SFR calculated using the FIR luminosity is $0.17\pm0.05$ \msyr (Table
 \ref{tab:fir_rad}) and using 610 MHz flux $0.023\pm0.023$ \msyr. Using the H$\alpha$ flux quoted in \cite{ramya09}, we
 find the SFR is $0.001-0.03$ \msyr \ if we consider the production rate of stars of masses greater than $10 M_\odot$.

For all the five galaxies, the SFR calculated using FIR emission is higher than SFR calculated using the radio continuum 
emission at 610 MHz by a factor ranging from 2 to 10. The SFR calculated using radio emission is comparable to the SFR 
calculated for individual star forming regions from H$\alpha$ for Mrk 1039 and I Zw 97, whereas, the SFR for Mrk 104
 matches well with the SFR calculated from the total \ha emission.

\vspace{0.5cm}

We also examined the nature of the radio spectra of the BCDs studied in the radio continuum by \cite{kle91} and \cite{deeg93}.  \cite{kle91} have studied 23 galaxies and a smaller sample of seven galaxies from this has been studied at lower frequencies by \cite{deeg93}. In our sample of five, we notice that the galaxies in groups (i.e. Mrk 104 and Mrk 108) appear to show a power law spectrum to the lowest observed frequency whereas the relatively isolated galaxies (Mrk 1069, Mrk 1039) show a low frequency turnover. Although Mrk 1039 is in a group, we note that the other members are low mass dwarf galaxies unlike in the case of Mrk 104 and Mrk 108 which have massive spirals as members. We examine the behaviour exhibited by the \cite{deeg93} sample. We have tabulated the results in Table \ref{tab:glo_spec_inx}. The first column gives the name of the galaxy, the second the group name or notes its isolated status, the third the number of group members. The fourth and fifth columns list the radio spectral indices of the galaxy for two sets of frequencies, and the last column gives the reference to the information collected. For the galaxies whose membership is 
not determined, its status is kept as `single ?' and also the information regarding the nearby galaxy and possible partner is mentioned, as obtained from Hyperleda. Interestingly, the spectra of galaxies which are relatively isolated show flattening at the lower frequencies (less than 1 GHz) both in our sample and that of \cite{deeg93}.

Thus, three out of seven galaxies from \cite{deeg93} are isolated, and show flattening at the lowest frequency (325 MHz) that they observed. The galaxy Mrk 1069  which shows flattening at the lower frequency is isolated, while Mrk 1039, also showing flattening, is a member of a group having only dwarfs as group members. The flattening is probably related to the compact radio emitting, star forming regions. The localised emission could be a result of the trigger of star formation being relatively local in the case of isolated BCDs, following the scenario of stochastic self propagating star formation (SSPSF; \cite{gerola80}). Thus, from Table \ref{tab:glo_spec_inx} we can derive two inferences; (a) the overall spectrum of relatively isolated galaxies is flatter compared to BCDs in denser environments and (b) the spectrum turns over at low frequencies ($\nu\sim610$ MHz) for relatively isolated galaxies. However, these are preliminary results based on a small sample and should be confirmed with a study of a larger sample of BCDs in radio.

\begin{table*}[h!]\footnotesize
\begin{center}
\renewcommand{\arraystretch}{1.3}
\caption[\footnotesize Study of global spectral index of a sample of BCDs]{\footnotesize Spectral index of a sample of BCDs including our data and \cite{deeg93}; D93 data. Columns 2 and 3 are obtained from NED and Hyperleda.}
\label{tab:glo_spec_inx}
\begin{tabular}{llllll}
\\
\hline\hline 
{\bf Galaxy} & {\bf Group name} & {\bf No. Members} & {\bf ${\alpha_{0.3}}^{1.4}$}  & {\bf ${\alpha_{1.4}}^{4.8}$} & {\bf Reference} \\
\hline
II Zw 70 & GPair & 2 & $-0.62^{+0.10}_{-0.07}$ & $-0.43^{+0.14}_{-0.23}$ & D93 \\ 
Mrk 297  & GPair & 2 & $-0.57^{+0.01}_{-0.01}$ & $-0.99^{+0.09}_{-0.12}$ & D93 \\ 
Mrk 314  & LGG 469 & 8  & $-0.33^{+0.05}_{-0.06}$ & $-0.43^{+0.02}_{-0.03}$ & D93 \\ 
Mrk 527  & LGG 473 & 25 & $-0.56^{+0.18}_{-0.13}$ & $-0.34^{+0.05}_{+0.06}$  & D93 \\ 
\hline
III Zw 102 & Single ? & --- & $-0.26^{+0.001}_{-0.001}$ & $-0.52^{-0.001}_{-0.001}$ & D93 \\ 
II Zw 40   & Single & --- & $-0.16^{+0.06}_{-0.05}$ & $-0.29^{+0.05}_{-0.06}$ &  D93 \\ 
Haro 15    & Single & --- & $-0.10^{+0.009}_{-0.012}$ & $-0.29^{+0.03}_{-0.04}$ &  D93 \\ 
\hline
  &  &  & {\bf $\alpha_{0.3}^{0.6}$} & {\bf $\alpha_{0.6}^{1.4}$} &  \\ 
\hline
Mrk 104     & UZC-CG 94 & 3 & $-0.79^{+0.18}_{-0.12}$ & $-0.43^{+0.17}_{-0.26}$ & this paper \\ 
Mrk 108     & Holm 124 & 4 & $-0.96^{+0.20}_{-0.39}$ & $-1.77^{+0.16}_{-0.08}$ & this paper \\ 
Mrk 1039    & USGC S087 & 4 & $+0.48^{+0.25}_{-0.19}$ & $-1.11^{+0.02}_{-0.02}$ & this paper \\ 
Mrk 1069    & isolated & --- & $+0.12^{+0.33}_{-0.26}$ & $-0.43^{+0.17}_{-0.22}$ & this paper
\\ 
\hline\hline
\end{tabular}
\end{center}
\end{table*}

\subsection{\textsc{Atomic gas distribution}}
\label{subsec:atomic_gas}

The atomic gas is much more extended than the stellar extent, a feature generally seen in most late type galaxies and also most BCDs \citep{thuan04}. The typical peak \hi column density in these galaxies is $\sim~10^{21}$ cm$^{-2}$ at a resolution of 
$45\arcsec\times35\arcsec$ ( $\sim5.0\times5.0$ kpc). Hence most of the galaxy has lower \hi column densities. According to \cite{skill87}, \cite{tay94}, the star formation in dwarf galaxies occurs only above a constant threshold \hi column density of $N$(H{\sc~i})$~\sim~10^{21}$ cm$^{-2}$. However, column densities of  $10^{20}$ cm$^{-2}$ are fairly common in metal deficient dwarf galaxies (\citealt{ekta06}; \citealt{ekta08}; \citealt{begum08}, and references therein). A preliminary study of a small sample of galaxies from the FIGGS survey \citep{begum06} suggested that unlike brighter dwarfs, the faintest dwarf galaxies do not show well-defined threshold density \citep{begum08}. Thus, the requirement for the threshold column density may not be uniform.

The \hi masses (Table \ref{tab:obs_hi}) calculated here are in the range $10^8-10^9\, M_\odot$ similar to the ones quoted in \cite{thuan99} and \cite{thuan81}. A large discrepancy in Mrk 108 is noticed. Since Mrk 108 is located on the edges of  NGC 2820, our mass estimate is more accurate.

\begin{figure}[h!]
\centering
\includegraphics[width=15cm]{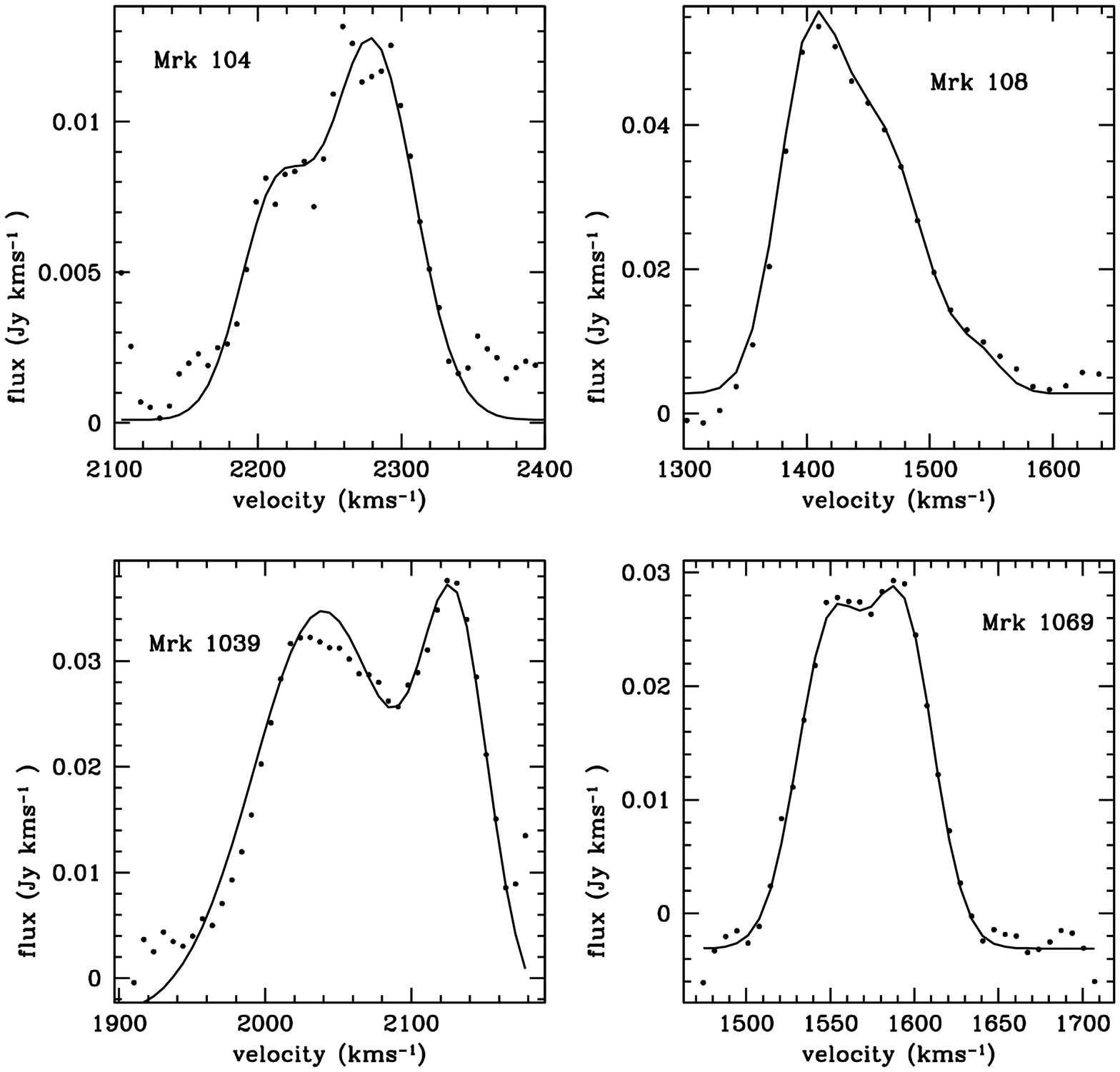}
\caption[\footnotesize The integrated \hi line profile of the galaxies Mrk 104, Mrk 108, Mrk 1039 and Mrk 1069]
{\footnotesize The integrated \hi line profile of the galaxies Mrk 104, Mrk 108, Mrk 1039 and Mrk 1069 are shown. The dots indicate the 
observed spectra and the solid line indicates the gaussian fits to the observed spectra. Note that while Mrk 104 and Mrk 108 show a profile 
consisting of more than one gaussian component, the spectra of Mrk 1039 and Mrk 1069 show the classic double-horned profile characteristic of a
 rotating disk viewed edge-on.}
\label{fig:line_prof}
\end{figure}

Figure \ref{fig:line_prof} shows the observed integrated line profiles of \hi emission for all the four galaxies. 
Wide double-horned profiles typical of edge-on rotating disks are observed for Mrk 1039 and Mrk 1069.  The widest 
line is seen from Mrk 1039 with a width at 50\% of peak of about 124 \kms. We fit the observed profiles with 
gaussian components. The \hi line profile (Figure \ref{fig:line_prof}) of Mrk 104 was best fit with two components
 centered at velocities 2211 and 2279 \kms \ with widths 57 and 69 \kms, respectively. The widths estimated here 
are smaller than the widths quoted in single dish observations (see Tables \ref{tab:para} and \ref{tab:obs_hi}) by 
\cite{thuan81} and \cite{thuan99}. The component at 2211 \kms \ is from the \hi cloud seen to the north of Mrk 104
 which has no optical counterpart. The velocity field of Mrk 104 is distorted as shown in Figure \ref{fig:mrk104_r}
(b) --- we also note that it is classified by NED as a peculiar galaxy.  The velocity field is receding on either 
side of the centre.  The \hi distribution across the galaxy shows a steep rise in the column density on the eastern
 side of the galaxy and diffused gas in the SW and NW.  Note that the closest projected part of the galaxy to this
 cloud has velocities of $\sim2260$ \kms. Thus either (1) the cloud is falling onto the galaxy from behind; a 
remnant of tidal interaction in the group or (2) the cloud has been pulled out of the galaxy. 

The nearest neighbour to Mrk 104 is the 
galaxy UGC 4906 whose integrated \hi profile \citep{spring05} shows it to be a rotating mildly 
lop-sided \hi disk. PGC 26253 (2297 \kms) 
$\sim17\arcmin.5$ (159 kpc) away towards south of Mrk 104 is the other member in the group. 
  A separate cloud-like entity is detected to the north of Mrk 104. The galaxy has a distorted velocity field. 
All of the above seem to suggest that Mrk 104 has experienced tidal forces. 
Using the peak velocity from the P-V diagram, we estimated the dynamical mass using
($V^2R/G$). The $M({\hi})/L_K$ and $M{\textrm {(dyn)}}/L_K$ ratios 
calculated are given in Table \ref{tab:obs_hi}. 
$M({\hi})/L_K$ is about $0.12 M_\odot/L_\odot$ and $M{\textrm {(dyn)}}/L_K$ is 
about $0.83 M_\odot/L_\odot$ for this galaxy and the width of the \hi line is around 69 \kms. Most of the 
parameters seem to suggest that the dark matter content is low and a tidal dwarf origin for Mrk 104 is possible. 

 \hi emission is detected from Mrk 108 (see Figure \ref{fig:mrk108_r}). The velocity field of Mrk 108 is distinct 
from its large neighbour NGC 2820 (Figure \ref{fig:mrk108_r}(c)). The line profile of Mrk 108 is best fit with two
components centered at 1406 and 1446 \kms \ with widths 52 and 117 \kms. The component with centre velocity at 
1446 \kms \ is contributed by NGC 2820. The emission from Mrk 108 is continuous to that from NGC 2820; hence we
 note that the mass from single dish observation by \cite{thuan81} is overestimated. The $M({\hi})/L_K$ ratio is 0.41 
and $M{\textrm{(dyn)}}/L_K$ is 2.86 adopting ($V^2R/G$) and the width of \hi line is around 52 \kms.  A possible 
tidal dwarf galaxy is detected in the west \citep{nim05} of NGC 2820. Mrk 108 has experienced tidal interaction 
due to the large neighbouring spirals. Thus the situation is similar to Mrk 104 and Mrk108 as a tidal dwarf 
galaxy cannot be ruled out. With masses being similar to those of tidal dwarf galaxy candidates \citep{wetzstein07} and being 
rotation dominated, the idea that a good number of BCDs in the groups might have originated in the tidal tails of larger galaxies is 
not an improbable scenario.

The observed \hi emission of Mrk 1039 is about 1.1 times the size of the optical galaxy and the \hi disk is warped 
(see Figure \ref{fig:mrk1039_r}(a)). No warp is visible in the underlying stellar disk.  The optical and kinematic centres 
seem to coincide.  The velocity field shows a smooth gradient across the galaxy with some distortions in the western half.
  The higher resolution map in Figure \ref{fig:mrk1039_r}(c) shows the higher column density regions on both edges of the 
galaxy disk. We note that radio continuum emission is coincident with the eastern half of the stellar disk hosting a 
strong \hii region (see Figure \ref{fig:mrk1039_r}(f)) and a \hi clump is also seen in the same region. High column 
density \hi is also associated with the stellar component in the western half of the galaxy from which no radio continuum
 has been detected. The P--V diagram through the major axis ($90^\circ$ east of north; see Figure \ref{fig:mrk1039_r}(d))
 also shows that the gas in the galaxy is mainly executing solid body rotation. The \hi line profile is double-horned (see
 Figure \ref{fig:line_prof}). The fitted parameters are given in Table \ref{tab:obs_hi}. The $M{({\hi})}/L_K$ ratio for
 Mrk 1039 is 0.4 $M_\odot/L_\odot$ and $M{\textrm {(dyn)}}/L_K$ is 1.47. The width of the \hi line is 124 \kms.  \\

The \hi distribution and velocity field of Mrk 1069 are shown in Figure \ref{fig:mrk1069_r}. The \hi disk 
($\sim185\arcsec$ diameter) is about six times the size of the optical disk ($D_{25} = \sim34\arcsec$) and higher column
 density gas is seen in the central and northern parts of the galaxy similar to the distribution of the radio continuum
 emission (see Figure \ref{fig:mrk1069_r}c,d). Diffuse \hi is seen in the southern parts of the galaxy. The galaxy shows a
 smooth gradient in the velocity field with minor distortions. The optical and kinematic centre appear to be displaced.
 This has also been observed in other BCDs by \citet{thuan04} who attributed it to the energy input into the ISM through
 violent starbursts. The nearest galaxy, a dwarf, UGCA 052, is located around $16\arcmin$ (96 kpc) to the south western
 side of this galaxy but the two galaxies have not been classified in the literature as being associated. The 
$M{({\hi})}/L_K$ ratio calculated for Mrk 1069 is 0.10 $M_\odot/L_\odot$ and the width of line emission is 61 \kms. The
 $M{\textrm {(dyn)}}/L_K$ is about 1.99.

\section{\textsc{Conclusion}}
\label{sec:con_rad}
The radio continuum observations at frequencies 610 MHz, 240 MHz, 325 MHz, 1.28 GHz and 1.4 GHz with \hi observations and analyses of five blue compact dwarf galaxies (BCDs) using the Giant Meterwave Radio Telescope (GMRT) are presented here. The \hi observations of four BCDs namely Mrk 104, Mrk 108, Mrk 1039 and Mrk 1069 have revealed their large \hi disks about 1.1 -- 3 times the optical size of the galaxy. The typical \hi masses range between $2\times10^8-10^9\, M_\odot$. These values are typical of blue compact dwarf galaxies.  We also detect a cloud close to the disk in Mrk 104 with no obvious optical counterpart and speculate that these might have influenced the current burst of star formation. Rotation is clearly noticed in the velocity maps of Mrk 108, Mrk 1039 and Mrk 1069. The \hi line profiles show multiple components. Mrk 104 has a \hi mass of $2\times10^8$, rotation velocity is 69 \kms \ and $M\textrm{(dyn)}/L_K\sim1$. Mrk 108 with total \hi mass is $1.6\times10^8\, M_\odot$, $M\textrm{(dyn)}/L_K\sim3$, rotation velocity $\sim52$ \kms \ and is situated very close to the galaxy NGC 2820 (a spiral galaxy). Mrk 104 and Mrk 108 are similar with the nearest group member being a large spiral. A tidal origin cannot be ruled out for these two galaxies. 

All the galaxies are detected in radio continuum at 325 MHz and 610 MHz and the emission at this frequency is dominated by  non-thermal emission. Out of the five galaxies, only two galaxies namely Mrk 104 and Mrk 1069 are detected at 240 MHz. I Zw 97 is detected for the first time in the radio; neither the NVSS survey nor the VLA FIRST survey detect it. Thermal and non-thermal separation is attempted for the four galaxies namely, Mrk 104, Mrk 108, Mrk 1039 and Mrk 1069. While the observed spectrum of Mrk 1069 can be explained by synchrotron spectrum absorbed by thermal gas at the lower frequencies, the observed spectrum of Mrk 104 could be explained as being due to combined thermal and non-thermal emissions. The observed spectrum of Mrk 108 is only due to pure non-thermal emission and the continuum emission in Mrk 1039 is explained as a combination of thermal and non-thermal emission with the non-thermal emission being absorbed at the lower frequencies by the thermal gas mixed in the region. We estimate a thermal fraction at 1.4 GHz of $\sim80\%$ and $\sim45\%$ of the total emission for Mrk 104 and Mrk 1039 which is much higher than the $\sim10\%$ generally seen in normal spiral galaxies \citep{cond92}. The emission in Mrk 1039, I Zw 97 and Mrk 1069 is confined to the \hii region and we note that the epoch of star formation in Mrk 1039 is considered to be fairly young $\sim 4$ Myr \citep{huang99}. 

 The SFR estimated from the observed emission at 610 MHz results in values in the range 0.01-0.1 \msyr. These are similar to the values of SFRs obtained for the individual star forming regions using H$\alpha$ fluxes \citep{ramya09} in case of Mrk 1039 and I Zw 97 whereas it matches with the SFR estimated from global \ha for Mrk 104. Emission detected at 610 MHz from Mrk 1039, Mrk 1069 and I Zw 97 is localized to the \hii regions seen in the H$\alpha$ images, which indicates that the SFRs obtained from the radio measurement represents the SFRs of a few individual \hii regions in these galaxies. On the other hand, the 610 MHz emission from Mrk 104 is seen to encompass the entire galaxy.

Combining our results with those of \cite{deeg93} --- the only other study of BCDs in low frequency radio continuum --- we find that on the average, the galaxies which are classified as isolated galaxies tend to show a somewhat flatter observed spectrum as compared to galaxies which are classified as being in groups. This, we interpret as due to a larger fraction of thermal emission mixed with the non-thermal in the former case, as one expects for young localized starbursts. However a flatter injection spectrum of a young localized starburst region could also explain this. We further note that the starburst in a galaxy in  tenuous environment is likely to be more 
localized. The flatter spectrum seen in isolated galaxies with localized star formation in the form of compact star forming regions and localized radio continuum emission, all suggest stochastic self-propagating star formation.

\bibliographystyle{apj}
\bibliography{ram_thesis}

\end{document}